\documentclass[10pt]{article}
\usepackage{amsfonts}
\usepackage{amsmath}
\usepackage{amssymb} 
\usepackage[compress]{cite}
\usepackage{bm} 
\usepackage{color}
\usepackage{comment}
\includecomment{comment}
\usepackage{dcolumn}
\usepackage{enumerate}
\usepackage[margin=1.2in]{geometry}                
\usepackage{graphicx}
\usepackage{epstopdf}   
\usepackage{mathrsfs}
\usepackage{mathtools} 
\usepackage{paralist}
\usepackage[parfill]{parskip}    
\usepackage{textcomp}
\usepackage{titlesec}
\usepackage{varioref}
\usepackage{hyperref} 
\allowdisplaybreaks
\hypersetup{
    bookmarks=false,         
    pdfstartview={FitH},    
    colorlinks=true,       
    linkcolor=blue,          
    citecolor=blue,        
    filecolor=blue,      
    urlcolor=magenta           
}
\DeclareFontFamily{OT1}{pzc}{}
\DeclareFontShape{OT1}{pzc}{m}{it}%
             {<-> s * [0.900] pzcmi7t}{}
\DeclareMathAlphabet{\mathscr}{OT1}{pzc}%
                                 {m}{it}
\labelformat{section}{Section #1} 
\labelformat{subsection}{Section #1} 
\labelformat{equation}{Eq.~(#1)} 
\labelformat{figure}{Fig.~#1} 
\labelformat{subfigure}{Fig.~\thefigure#1} 
\labelformat{table}{Tab.~#1} 
\titleclass{\subsubsubsection}{straight}[\subsection]

\newcounter{subsubsubsection}[subsubsection]
\renewcommand\thesubsubsubsection{\thesubsubsection.\arabic{subsubsubsection}}

\titleformat{\subsubsubsection}
{\normalfont\normalsize\bfseries}{\thesubsubsubsection}{1em}{}
\titlespacing*{\subsubsubsection}
{0pt}{3.25ex plus 1ex minus .2ex}{1.5ex plus .2ex}
\makeatletter
\renewcommand\paragraph{\@startsection{paragraph}{5}{\z@}%
	{3.25ex \@plus1ex \@minus.2ex}%
	{-1em}%
	{\normalfont\normalsize\bfseries}}
\renewcommand\subparagraph{\@startsection{subparagraph}{6}{\parindent}%
	{3.25ex \@plus1ex \@minus .2ex}%
	{-1em}%
	{\normalfont\normalsize\bfseries}}
\def\toclevel@subsubsubsection{4}
\def\toclevel@paragraph{5}
\def\toclevel@paragraph{6}
\def\l@subsubsubsection{\@dottedtocline{4}{7em}{4em}}
\def\l@paragraph{\@dottedtocline{5}{10em}{5em}}
\def\l@subparagraph{\@dottedtocline{6}{14em}{6em}}
\makeatother
\newcommand{\be}{\begin{equation}}
\newcommand{\ee}{\end{equation}}
\newcommand{\bea}{\begin{eqnarray}}
\newcommand{\eea}{\end{eqnarray}}
\newcommand{\nn}{\nonumber}

\newcommand{\sqg}{\sqrt{-g}}

\newcommand{\df}{\delta}

\newcommand{\sqq}{\sqrt{q}}

\newcommand{\ST}{\df \mathcal{A}_{_{\partial  \mathcal  V}}}

\newcommand\Mycomb[2][n]{\prescript{#1\mkern-0.5mu}{}C_{#2}}
\newcommand{\ts}{\textsuperscript}

\def\sap{see Appendix A.3 in arXiv version of \cite{Parattu:2015gga}}

\def\P[#1]#2{\Pi^{#1}_{\phantom{a}#2}}
\def\Pt[#1]#2{\tilde{\Pi}_{#1}^{\phantom{a}#2}}
\def\pdlr#1#2{\left(\frac{\partial#1}{ \partial#2}\right)}

\def\EH{Einstein-Hilbert }
\def\GB{Gauss-Bonnet }
\def\LL{Lanczos-Lovelock }

\title{Null boundary terms for Lanczos-Lovelock gravity}
\author{Sumanta Chakraborty\footnote{sumantac.physics@gmail.com}$~^{1}$ 
and
Krishnamohan Parattu\footnote{mailofkrishnamohan@gmail.com}$~^{2,3,4}$\\
{\small{$^{1}$School of Physical Sciences, Indian Association for the Cultivation of Science, Kolkata-700032, India}}\\
{$^{2}$\small{Instituto de Fisica, Pontificia Universidad Catolica de Valparaiso, Curauma, Valparaiso, Chile}}\\
{$^{3}$\small{Instituto de Ciencias Nucleares, Universidad Nacional Aut\'onoma de M\'exico, M\'exico D.F. 04510, M\'exico}}\\
{$^{4}$\small{Department of Physics, IIT Madras, Chennai - 600 036, India}}}
\begin{document}
\maketitle
\begin{abstract}
We derive boundary terms appropriate for the general Lanczos-Lovelock action on a null boundary, when Dirichlet boundary conditions are imposed. We believe that these boundary terms have been derived for the \emph{first} time in the literature. In this derivation, we rely \emph{only} on the structure of the boundary variation of the action for Lanczos-Lovelock gravity. We also provide the null boundary term for Gauss-Bonnet gravity separately.   
\end{abstract}

\section{Introduction} 
\label{Introduction}

Einstein-Hilbert action \cite{Einstein:1916cd} is the standard action for general relativity. It is a covariant action and is constructed out of the metric and its derivatives. On varying the Einstein-Hilbert action, we obtain Einstein's field equations, which are in very good agreement with experiments and observations \cite{Will:2014kxa}. The \EH action is unusual in that it is of second order in the derivatives of the dynamical variable of the theory, namely the metric, but has a special structure that allows the Einstein field equations to come out to be also of only second order. Since the action is of second order, one has to fix both the metric and its derivatives at the boundary surface (as the derivatives in directions tangential to the boundary are automatically fixed once we fix the metric on the boundary, we need to fix separately only the derivatives of the metric in directions non-tangential to the boundary) to kill the boundary contribution in the variation of the action. This is too much for the second order equations of motion, and there will be no solutions for most choices of the boundary conditions. We say that the action is \emph{not} well-posed (see \cite{Hawking:1980gf,Dyer:2008hb} and Chapter 6 in \cite{gravitation}). The ready remedy is to add boundary terms, since they modify the boundary conditions without affecting the equations of motion. A suitable boundary term needs to be added to the Einstein-Hilbert action to cancel the variation of the non-tangential derivatives of the metric on the boundary hypersurface in the variation of the action. Various proposals for such boundary terms exist in the literature but the standard one is the Gibbons-Hawking-York (GHY) boundary term \cite{Gibbons:1976ue,York:1972sj} (see also \cite{Katz:1985,Katz:1996ym}). But the GHY term has the deficiency that it requires the normalized normal and non-degenerate determinant of the boundary metric, and hence can be constructed directly only for a non-null surface. A proposal for a null boundary term was first provided in \cite{Barth:Phd}, but it was derived in too special a case to be of much use. Much later, a null boundary term was suggested in \cite{Neiman:2012fx}, albeit for the case of a null foliation. A completely independent derivation of the null boundary term for the Einstein-Hilbert action without restricting to a foliation was presented in \cite{Parattu:2015gga}, and further refined in \cite{Parattu:2016trq,Chakraborty:2016yna,Lehner:2016vdi,Hopfmuller:2016scf,Jubb:2016qzt,Aghapour:2018icu}. The references \cite{Parattu:2016trq} and \cite{Jubb:2016qzt} provide treatments that do not distinguish between null and non-null boundaries (in this respect, also see \cite{Aghapour:2018icu}). When one considers non-smooth boundaries, one has to add corner terms to the gravitational action in addition to the boundary terms. Such corner terms have been explored for both null and non-null boundaries \cite{Hartle:1981cf,Farhi:1989yr,Brill:1991rw,Hayward:1993my,Brill:1994mb,Lehner:2016vdi,Jubb:2016qzt}. 

The null boundary term has found numerous applications in various contexts. There has been quite a bit of work in the area of AdS/CFT correspondence, where the relationship between quantum complexity of states in the boundary field theory and bulk gravitational action has been explored \cite{Lehner:2016vdi,Reynolds:2017lwq,Kim:2017lrw,Reynolds:2016rvl,Carmi:2016wjl,Chapman:2016hwi,Yang:2016awy,Ben-Ami:2016qex,Fareghbal:2018ngr,Auzzi:2018pbc,Alishahiha:2018tep,Bolognesi:2018ion,Reynolds:2017jfs,Kim:2017qrq,Carmi:2017jqz,Yang:2017nfn} in connection with the ``complexity=action'' conjecture \cite{Brown:2015bva,Brown:2015lvg} (see \cite{Eling:2016qvx} for work in another direction in the area of Ads/CFT correspondence). There has also been work on formulating de Sitter spacetimes in the framework of string theory \cite{Maltz:2016max,Maltz:2016iaw}. The right form of the boundary term is also important for the causal set approach to quantum gravity \cite{Buck:2015oaa}. The topic of degrees of freedom of gravity on a null surface has also been explored \cite{Parattu:2015gga,Hopfmuller:2016scf,Aghapour:2018icu}. 

A well-known issue with the theory of general relativity is the presence of singularities, where predictability of the theory breaks down, in the context of black hole and cosmological spacetimes. One possible way to cure this issue is by supplementing the Einstein-Hilbert action with higher curvature corrections. However, diffeomorphism invariance alone is not strong enough to reasonably constrain the higher curvature candidate terms. It turns out that if one imposes an additional criterion, that the associated gravitational field equations be second order in derivatives of the metric, a very unique set of Lagrangians, including the \EH Lagrangian, gets selected. These Lagrangians are known as Lanczos-Lovelock Lagrangians \cite{Lanczos:1938sf,Lovelock:1971yv,Dadhich:2008df,Padmanabhan:2013xyr,Deruelle:2003ck,Chakraborty:2017bcg}. The Lanczos-Lovelock Lagrangians are perhaps the most well-motivated extensions of the \EH Lagrangian. The actions for the theories in the \LL series are all second order in the derivatives of the metric, but with the equations of motion being also second order. Thus, even the \LL theories do \emph{not} have a well-posed variational problem and one must add appropriate boundary terms. The non-null boundary term for the Gauss-Bonnet Lagrangian, the first non-trivial correction to the \EH Lagrangian that is quadratic in the curvature tensor, was derived by Bunch \cite{bunch1981surface}. For general Lanczos-Lovelock theories, the non-null boundary terms were derived by Myers \cite{Myers:1987yn} (see also \cite{Davis:2002gn,Yale:2011dq,Miskovic:2007mg,Deruelle:2017xel,Deruelle:2018vtt}). However, the structure of the full boundary variation on a non-null boundary, including the total surface derivative term and the Dirichlet variation term that tells us the degrees of freedom to be fixed on the boundary and the corresponding conjugate momenta, was not known for \LL theories. We have recently derived the full structure of the non-null boundary variation starting directly from the variation of the \LL action \cite{Chakraborty:2017zep} without using any topological arguments. Subsequently, the corner terms that one must add to the \LL Lagrangian in the case of non-smooth boundaries were presented in \cite{Cano:2018ckq}. The corner terms were derived for non-null boundaries and then it was argued that the corner terms on null boundaries can be obtained as a straightforward extensions. But, to our knowledge, there have been no proposals in the literature for the boundary terms associated with the general \LL Lagrangian when the boundary surface is null in nature. In this paper, we fill this gap by explicitly deriving the boundary terms that one must add on a null boundary to make the variational principle for \LL gravity well-posed.  

The procedure that we will follow will be the same as that adopted in \cite{Chakraborty:2017zep}. We will start from the boundary variation of the \LL action on a null boundary and then judiciously manipulate the terms to arrive at the desired boundary term. Thus, the action itself will provide the appropriate boundary terms to be added to make it well-posed (see \cite{Padmanabhan:2014BT}, which is perhaps the earliest work to make use of this technique). Note that one may add any term without off-the-boundary derivatives of the metric (defined appropriately for null boundaries) to these boundary terms without affecting the well-posedness of the action. We attempt in this work to provide a minimal set of tensorial terms by throwing away any term without off-the-boundary derivatives of the metric. 

The paper is organized as follows: We start with a very brief overview of \LL theories in \ref{sec:LL_intro}. Then, we provide the various projections of the Riemann tensor near a null surface, required for our derivation, in \ref{sec:Proj_Riem_Intro}. The formulae from these two sections are applied subsequently in \ref{Null_Boundary_LL} to derive the null boundary terms. After doing some preliminary simplification of the null boundary variation for \LL theories in \ref{sec:LL_init_simp}, we warm up by deriving the null boundary term for the Gauss-Bonnet case in \ref{sec:GB}. Then, in \ref{sec:gen_LL}, we derive the boundary term for general \LL gravity as the main result of this paper. We run a consistency check on this result in \ref{sec:LL_to_EH_GB} by obtaining the \EH and \GB null boundary terms as special cases of the general result. The paper ends with a brief conclusion in \ref{sec:conclusion}. Most of the background material required for the computations have been included in the appendices.

\textit{Notations and Conventions:} We use the mostly positive metric signature $(-,+,+,+,\dots)$ and the fundamental constants $G$ and $c$ have been set to unity. The Latin indices, $a,b,\ldots$, run over all the space-time indices, and are hence summed over all the $D$ values when the spacetime is $D$-dimensional. Greek indices, $\alpha ,\beta ,\ldots$, will be used for indices restricted to a codimension-$1$ hypersurface and upper case Latin symbols, $A,B,\ldots$, will be used for indices corresponding to a codimension-$2$ hypersurface. We shall use the notation $R^a_{bcd}$ to refer to the Riemann tensor of the $D$-dimensional manifold, defined through the commutation relation $[\nabla _{b},\nabla _{c}]v^{a}=R^{a}_{~ibc}v^{i}$, and $~^{(D-2)}R^a_{bcd}$ to refer to the Riemann tensor of the codimension-$2$ surfaces on the null surface orthogonal to the auxiliary null vector. The $D$-dimensional covariant derivative will be denoted by $\nabla_a$ while the covariant derivative on the codimension-$2$ surfaces will be denoted by $D_a$.
\section{A short summary of \LL gravity}
\label{sec:LL_intro} 

In this section, we shall briefly revisit the basic geometrical structure of \LL gravity, defining relevant tensors and discussing the results pertinent to our forthcoming calculations. An interested reader may consult \cite{Padmanabhan:2013xyr} for further details and clarifications. The Lagrangian for \LL gravity is constructed as a sum over homogeneous polynomials of the Riemann curvature tensor of various orders, under the restriction that the equations of motion are second order in derivatives of the metric. For a spacetime of a particular dimension, only the terms up to a particular order are non-zero. For even number of dimensions, the highest order term is a total derivative and hence of no effect on the dynamics. Higher order terms of this series become effective as the spacetime dimension increases. For example, the Lanczos-Lovelock Lagrangian in four dimensions consists of the cosmological constant term, the Einstein-Hilbert term and the \GB term. The \GB term, which is the additional term present 
here with respect to the usual Lagrangian for general relativity, is topological in four dimensions, but can lead to non-trivial effects on the dynamics in five and higher dimensions. Thus, it is customary to consider \LL gravity in $D$ spacetime dimensions, where $D$ is greater than four, since otherwise the dynamics is effectively equal to that of general relativity with a cosmological constant term. But we are interested in the well-posedness of the action itself, and not just the dynamics. So, from our perspective, the Gauss-Bonnet term is important even in its role as a total derivative in four dimensions.

The most general action of the \LL gravity in a spacetime volume $\mathcal{V}$ in $D$ spacetime dimensions is given by
\begin{equation} \label{Lovelock_Action}
16\pi\mathcal{A}=\int d^{D}x\sqrt{-g}~L=\sum_{m=0}^{m_{\rm max}}c_m \int_{\mathcal{V}} d^{D}x~\sqrt{-g}L_m~;\quad L_m\equiv\frac{1}{2^{m}}\delta ^{a_{1}b_{1}\dots a_{m}b_{m}}_{c_{1}d_{1}\dots c_{m}d_{m}}R^{c_{1}d_{1}}_{a_{1}b_{1}}\dots R^{c_{m}d_{m}}_{a_{m}b_{m}}~,
\end{equation}
where $m_{\rm max}=[D/2]$, i.e., the greatest integer less than or equal to $D/2$ and $c_{m}$'s are dimensionful coupling parameters. When $D=2m$, the $m$\ts{th} order term will be a total derivative and will not contribute to the equations of motion. The $m=0$ term in the above expression is essentially the cosmological constant term, while the $m=1$ term presents the standard Einstein-Hilbert action for the choice $c_1=1$. The object $\delta ^{a_{1}b_{1}\dots a_{m}b_{m}}_{c_{1}d_{1}\dots c_{m}d_{m}}$ is the determinant of a matrix made out of delta functions and is referred to in the literature as the completely antisymmetric determinant tensor \cite{Padmanabhan:2013xyr} or the generalized Kronecker delta \cite{lovelock1989tensors,frankel2011geometry}. Another tensor playing a very important role in the context of \LL theories is the \textit{entropy tensor} $P^{ab}_{cd}$. Starting from the definition of the \LL Lagrangian as presented in \ref{Lovelock_Action},
the entropy tensor takes the following form:
\begin{equation}
P^{ab}_{cd} \equiv \pdlr{L}{R_{ab}^{cd}}_{g_{ij}}=\sum _{m}c_{m}\frac{m}{2^{m}}\delta ^{aba_{1}b_{1}\dots a_{m-1}b_{m-1}}_{cdc_{1}d_{1}\dots c_{m-1}d_{m-1}}R^{c_{1}d_{1}}_{a_{1}b_{1}}\dots R^{c_{m-1}d_{m-1}}_{a_{m-1}b_{m-1}}
\equiv \sum _{m}c_{m}P^{ab}_{cd~~(m)}~. \label{P_def}
\end{equation}
It is called the entropy tensor since it enters the expression for the Wald entropy. Using the last expression, we can write down the $m$\ts{th} order \LL Lagrangian as a product of the $m$\ts{th} order entropy tensor, $P^{ab}_{cd~(m)}$, and the Riemann tensor:
\begin{equation}
L_m =\frac{1}{m}P^{ab}_{cd\ (m)} R^{cd}_{ab}~.
\end{equation}
The entropy tensor $P^{abcd}$, and also $P^{ab}_{cd~(m)}$, satisfies all the algebraic symmetries of the Riemann curvature tensor. It, as well as every $P^{ab}_{cd~(m)}$, also has the additional property that its divergence on any index is zero:
\begin{equation}\label{div_P}
\nabla_{a}P^{abcd}=0~.
\end{equation}
This concludes our discussion of the essential geometrical structure and identities associated with the general \LL theory. Before we proceed, let us briefly comment on the two most important terms in the series for the \LL action, namely the \EH (the usual action for general relativity without the cosmological constant term) and the \GB terms. The first non-trivial case corresponds to $m=1$, the \EH action, for which the tensor $P^{ab}_{cd~(1)}$ takes the following form:
\begin{equation}\label{P_EH}
P^{ab}_{cd\ (1)}=\frac{1}{2}\left(\delta ^{a}_{c}\delta ^{b}_{d}-\delta ^{a}_{d}\delta ^{b}_{c}\right),
\end{equation}
and the Lagrangian, as expected, is given by
\begin{equation} \label{L_EH}
L_1 =P^{ab}_{cd\ (1)} R^{cd}_{ab}=R~. 
\end{equation}
The next term in the \LL Lagrangian, for $m=2$, is the the \GB term, with the following expression for the entropy tensor:
\begin{equation}\label{P_Gauss_Bonnet}
P^{ab}_{cd\ (2)}=2\Big[R^{ab}_{cd}+G^{b}_{c}\delta ^{a}_{d}-G^{a}_{c}\df^{b}_{d}+R^{a}_{d}\df^{b}_{c}-R^{b}_{d}\delta ^{a}_{c}\Big]~.
\end{equation}
The Lagrangian is
\begin{equation}\label{L_Gauss_Bonnet}
L_2 = \frac{1}{2} P^{ab}_{cd\ (2)} R^{cd}_{ab} = R^2-4 R^{ab}R_{ab}+R^{abcd}R_{abcd}~.
\end{equation}
We will require these expressions when we derive the null boundary term for \GB Lagrangian in \ref{sec:GB}. Having listed the results in \LL gravity required for our work, we proceed forward to present the decompositions of the Riemann tensor near a null surface in the next section.
\section{Projections of the Riemann tensor near a null surface}\label{sec:Proj_Riem_Intro}

In this section, we shall provide expressions for the various projections of the Riemann tensor near a null surface that will be used in our derivation of the \LL boundary term near a null boundary. The derivations of the identities provided in this section are given in a detailed fashion in \ref{app:R_decomp_null}. (For more detailed discussions on the geometrical quantities associated with null surface that are used in these derivations, we refer the reader to Appendix A.3 in the arXiv version of \cite{Parattu:2015gga})

The identities and the results presented here are valid under the following setup. We start with a null surface in a $D$-dimensional spacetime and parametrise it by $\phi(x)=\textrm{constant}$, for some scalar function $\phi(x)$. (For the purpose of this paper, we shall assume that we have taken a null foliation, such that the normal vector $\ell_{a}$ satisfies $\ell_{a}\ell^{a}=0$ everywhere and not just on the null boundary that we are considering.) 

A null normal $\ell _{a}$ to the surface will have the form $\ell_a=A(x)\nabla_{a}\phi(x)$, for some scalar function $A(x)$. As is standard (see, e.g., \cite{Carter:1979,Poisson}), we choose an auxiliary null vector $k^a$ such that $k^a\ell_a=-1$. This allows us to define the induced metric on the null surface as
\begin{equation}\label{q_def-1}
q_{ab}=g_{ab}+\ell_a k_b+k_a \ell_b~.
\end{equation}
This is a two-metric on a two-surface orthogonal to $\ell^a$ and $k^a$. It turns out that it is advantageous to assume $k_{a}$ to be hyper-surface orthogonal, i.e., $k_{a}=B(x)\nabla _{a}\psi(x)$, for some scalar function $\psi(x)$ and $B(x)$. As shown in \ref{appsubsec:ind_met}, this condition is sufficient (but does not appear to be necessary) to ensure that we can define an appropriate covariant derivative $D_{a}$ on the null surface without introducing torsion. These assumptions on the null normal $\ell_{a}$ and the auxiliary null vector $k_{a}$ imply a bunch of relations that will be useful for our calculations. The ``acceleration" of the null normal satisfies the following equation: 
\begin{equation}\label{def_kappa}
\ell^c \nabla_c \ell^d=\kappa \ell^d~,
\end{equation}
where $\kappa$ may be called the \textit{non-affinity coefficient}. The direct formula for $\kappa$ is
\begin{equation}\label{formula_kappa}
\kappa = \ell^a \partial_a \ln A -\frac{k^a}{2} \partial_a\left(\ell^b \ell_b\right)~.
\end{equation}
Since the null normal has vanishing norm everywhere, we have $\nabla _{a}(\ell ^{b}\ell _{b})=0$ and hence the non-affinity coefficient associated with the null normal becomes
\begin{equation}\label{rel_kappa_kappa_t}
\kappa = \ell^a \partial_a \ln A~.
\end{equation}
The second fundamental form associated with the null surface is
\begin{equation}\label{theta_def-main}
\Theta_{cd}\equiv q^m_c q^k_d \nabla_m \ell_k~,
\end{equation}
while the corresponding quantity for the null vector $k^a$, the second fundamental form for the foliation by $\Psi$ with $\ell^a$ chosen as the auxiliary null vector to $k_a$, is
\begin{equation}\label{psi_def-main}
\Psi_{cd}\equiv q^m_c q^n_d \nabla_m k_n~.
\end{equation}
Both $\Theta_{cd}$ and $\Psi_{cd}$ are symmetric. Finally,  we have the following formula for the variation of $\ell_a$ when the metric is varied:
\begin{equation}\label{dl_0}
\delta \ell _{a}=\ell _{a}\delta \ln A~.
\end{equation}
This can be easily proved as follows: $\delta \ell _{a}=A\delta \left(\partial_{a}\phi\right)+\partial_{a}\phi \delta A= A\partial_{a}\left(\delta\phi\right)+\ell _{a}\delta \ln A=\ell _{a}\delta \ln A$, where the term $A\partial_{a}\left(\delta\phi\right)$ identically vanishes since $\phi$ is a scalar independent of the metric that does not vary under metric variations. Thus, after variation, $\ell_a+\df \ell_a=\left(1+\df \ln A\right)\ell_a=\left(1+\df \ln A\right)A\partial_a \phi$ is still normal to the $\phi=\textrm{constant}$ surfaces. In order to see how $A$ will vary under metric variations, consider a coordinate system with $\phi$ and $\psi$ as two of the coordinates. In such a coordinate system, the conditions $g^{ab}\ell_a \ell_b=0$, $g^{ab}k_a k_b=0$ and $g^{ab}\ell_a k_b=-1$ turn into the following conditions on the inverse metric, respectively:
\begin{align}
g^{\phi \phi}=0,\quad g^{\psi \psi}=0,\quad AB g^{\phi \psi}=-1~;  
\end{align}  
where we have used the fact that $\ell_a$ and $k_a$ have to be non-zero and hence $A$ and $B$ are non-zero. The first two conditions do not constrain $A$ or $B$, but the third condition specifies their product in terms of the metric as
\begin{equation} \label{AB_in_metric}
AB=\frac{-1}{g^{\phi \psi}}~.
\end{equation}
Considering variations, it is clear that enforcing $\delta\left(\bm{\ell}.\bm{\ell}\right)=0$ and $\delta\left(\bm{k}.\bm{k}\right)=0$ are restrictions on the metric, unlike the non-null case where $\delta\left(\bm{n}.\bm{n}\right)=0$ can be achieved by varying the normalization factor of the normal appropriately. But the constraint $\delta\left(\bm{\ell}.\bm{k}\right)=0$ can be achieved by tweaking the product $AB$ appropriately. For the purpose of this paper, we shall \textit{assume} $\delta\left(\bm{\ell}.\bm{\ell}\right)=0$ and $\delta\left(\bm{k}.\bm{k}\right)=0$ everywhere, because otherwise all the expressions and equations we have for null surfaces will be meaningless under variation, and we shall also \textit{impose} $\delta\left(\bm{\ell}.\bm{k}\right)=0$ everywhere. [In our earlier paper deriving the null boundary term for general relativity \cite{Parattu:2015gga}, we had assumed $\delta\left(\bm{\ell}.\bm{\ell}\right)=0$ only on the null boundary and $\phi$ was not assumed to be a null foliation. We hope to remove these extra restrictions that we have imposed in this paper, for ease of calculation, in a later work.] We shall also assume $B$ to be a constant independent of the metric. To simplify matters, we shall equate it to $1$:
\begin{equation}
B=1~,
\end{equation}
which provides $A$ and the variation of $A$ in terms of the metric degrees of freedom as
\begin{equation}
A=\frac{-1}{g^{\phi \psi}}\Rightarrow \df A=\frac{\df\left(g^{\phi \psi}\right)}{\left(g^{\phi \psi}\right)^{2}}~.
\end{equation}
This is also a restriction we hope to remove in a future work.

Having described the set-up under which we will carry out the derivation of null boundary terms in the context of \LL gravity, we will now consider the projections of the Riemann tensor on the null surface that we will have occasion to use. First, we shall consider the case when all the indices of the $D$-dimensional Riemann tensor are projected on the null surface, which is  given by (for a detailed derivation, see \ref{Gauss-Codazzi-alt} in \ref{Projection_Riem_App})
\begin{align} \label{Gauss-Codazzi-main}
R^{ab}_{cd}~q^{p}_{a}q^{q}_{b}q^{c}_{m}q^{d}_{n}=~^{(D-2)}R^{pq}_{mn}
+\Theta ^{p}_{m}\Psi ^{q}_{n}+\Psi ^{p}_{m}\Theta ^{q}_{n}
-\Theta ^{p}_{n}\Psi ^{q}_{m}-\Psi ^{p}_{n}\Theta ^{q}_{m}~,
\end{align}
where $\Theta^{a}_{b}$ and $\Psi ^{c}_{d}$ are the second fundamental forms associated with the null normal $\ell_{a}$ and the auxiliary null vector $k_{a}$, respectively. When three indices of the Riemann tensor are projected on the null surface while the remaining index is contracted with $\ell^a$, one ends up with the following result (derived in \ref{R_3q} in \ref{Projection_Riem_App}):
\begin{align}\label{R_3q-main}
R^{ab}_{cd}~\ell _{a}q^{m}_{b}q^{c}_{p}q^{d}_{q}=D_{q}\Theta ^{m}_{p}-D_{p}\Theta ^{m}_{q}
+\left(q^{i}_{q}\Theta ^{m}_{p}-q^{i}_{p}\Theta ^{m}_{q}\right)k^{j}\nabla _{i}\ell _{j}~.
\end{align}
And finally, when two indices of the Riemann tensor are projected on the null surface and the other two are contracted with the null normal $\ell^a$, we get
\begin{align}\label{qlqlR-5-main}
R^{ab}_{cd}~\ell _{a}\ell ^{c}q^{m}_{b}q^{d}_{n}=-q^{m}_{i}q^{j}_{n}\pounds _{\ell}\Theta ^{i}_{j}
+\kappa \Theta ^{m}_{n}-\Theta ^{m}_{i}\Theta ^{i}_{n}~.
\end{align}
For a derivation of the above equation, we refer the reader to \ref{qlqlR-5} in \ref{Projection_Riem_App}. Note that the left-hand side of this equation is a symmetric tensor and each of the terms on the right-hand side are also symmetric. These are all the projections we can have without making use of $k^a$, since a term with three indices of the Riemann tensor contracted with the null normal $\ell_a$ would identically vanish due to the symmetries of the Riemann tensor. 

\section{The null boundary term for \LL gravity}
\label{Null_Boundary_LL}

\subsection{Manipulating the boundary variation in general \LL gravity} \label{sec:LL_init_simp}
                             	 
In this section, we shall start with the total divergence term in the variation of the action of general \LL gravity, express it as the boundary variation on a null boundary and separate out the total variation term to be cancelled by the addition of a boundary term. The expression for the total divergence term in the action of \LL gravity, following \cite{gravitation}, is
\begin{equation}
\ST = \int_{\mathcal{V}} d^{D}x \sqg \nabla_j \left[2P^{ibjd}\nabla_b \left(\df g_{di}\right)- 2  \left(\nabla_cP^{ijcd}\right) \df g_{di}\right]~.
\end{equation}
Enforcing \ref{div_P}, i.e., the result that the contracted covariant derivative of the entropy tensor $P^{abcd}$ identically vanishes, one can eliminate the second term appearing in the above expression. We use the divergence theorem to convert the remaining expression from a volume integral to a boundary integral on the null boundary with normal $\ell_a$, leading to
\begin{equation}
\ST = \int_{\mathcal{\partial V}} d^{D-1}x ~2 \left(\frac{\sqrt{-g}}{A}\right) \ell_j P^{ibjd}\nabla_b \left(\df g_{di}\right)~,\label{ST_1}
\end{equation} 
where the form of the volume element used is for $\ell_a=A\nabla_a \phi$ with $\phi$ taken as one of the coordinates \cite{Parattu:2015gga}. In \ref{app:g_in_q}, we have derived the following relation between $\sqg$ and the determinant $q$ of the metric on the codimension-$2$ surfaces:
\begin{align}\label{sqg_in_q_main}
\sqg=A\sqq~.
\end{align}
Using this relation in \ref{ST_1} we obtain
\begin{equation}
\ST = \int_{\mathcal{\partial V}} d^{D-1}x ~2 \sqq \ell_j P^{ibjd}\nabla_b \left(\df g_{di}\right)~.\label{ST_2}
\end{equation} 
Let us lose the integral and just write the integrand appearing above:
\begin{align}
\sqq Q[\ell_c]&\equiv 2 \sqq ~\ell_c P^{abcd}\nabla_b \left(\delta g_{ad}\right) 
\nn 
\\
&=2 \sqq \left[\nabla_b \left(P^{abcd}\ell_c \df g_{ad}\right)-P^{abcd}\left(\nabla_b \ell_c\right) \delta g_{ad}\right]~. \label{start}
\end{align}
The last term is a Dirichlet variation term (a term with only the variations of the dynamical variables, and not of their derivatives) and it will be killed when we fix the metric on the boundary. (There is the question of which components of the metric are to be fixed on the boundary for a consistent variational principle. Let us assume for now that, just as in the non-null case \cite{Chakraborty:2017zep}, the decomposition of the boundary term will automatically select the relevant components and ensure consistency.) Since this does not contribute to the boundary term, let us ignore it and write
\begin{align}
\sqq Q[\ell_c]&=2 \sqq \nabla_b \left(P^{abcd}\ell_c \df g_{ad}\right)+\dots~. \label{start-1}
\end{align}
From now on, any term deemed not relevant for the boundary term will be dumped into the `$\dots$'. Decomposing the first term using the projector $\P[a]{b}=\df^a_b + k^a \ell_b$ (see \cite{Parattu:2016trq}), we obtain
\begin{align}
2 \sqq\nabla_b \left(P^{abcd}\ell_c \df g_{ad}\right)&=2 \frac{\sqg}{A}\nabla_b \left(P^{abcd}\ell_c \df g_{ad}\right)= 2 \sqg\nabla_b \left(\frac{1}{A}P^{abcd}\ell_c \df g_{ad}\right)+ \dots \nn\\
&=2 \sqrt{-g} \nabla_e \left(\frac{1}{A}\P[e]{b}P^{abcd}\ell_c \df g_{ad}\right)-2 \sqrt{-g} \nabla_e \left(\frac{1}{A}k^e \ell_b P^{abcd}\ell_c \df g_{ad}\right)+\dots \nn \\
&=2 \partial_e \left(\frac{\sqrt{-g}}{A}\P[e]{b}P^{abcd}\ell_c \df g_{ad}\right)-2 \sqrt{-g} \nabla_e \left(\frac{1}{A}k^e \ell_b P^{abcd}\ell_c \df g_{ad}\right)+\dots~. \label{Eq1}
\end{align}
Since $\P[e]{b}\ell_e=0$, the first term above is a surface total derivative term on the null boundary. This term has variations of the derivatives of the metric, but all these derivatives are tangential to the boundary and hence will disappear when the metric is fixed all along the boundary. Thus, this term does not contribute to the boundary term. Relegating it also to the oblivion of the three dots, we obtain
\begin{align}
 \sqq Q[\ell_c]&=-2 \sqq \nabla_e \left(k^e \ell_b P^{abcd}\ell_c \df g_{ad}\right)+\dots~. \label{start-3}
\end{align}
Note that, as we have done above, we can take $A$ inside and outside of the derivative as we see fit, since the extra terms are anyway proportional to $\delta g_{ab}$ and hence do not contribute to the boundary term. Taking the covariant derivative in \ref{start-3} inside, we obtain
\begin{align}
\sqq Q[\ell_c]&= -2 \sqq \nabla_e \left(k^e \ell_b P^{abcd}\ell_c\right) \df g_{ad}-2 \sqq  k^e \ell_b P^{abcd}\ell_c \nabla_e\df g_{ad}+\dots~.
\end{align}
Again, we spy a Dirichlet variation term in the above expression. Relegating it to the dots, we can write
\begin{align}
\sqq  Q[\ell_c]&=-2 \sqq  k^e \ell_b P^{abcd}\ell_c \nabla_e\df g_{ad}+\dots~. \label{start-4}
\end{align}
Next, we need to bring the $\delta$ out of the covariant derivative in $\nabla _{e}\delta g_{ad}$. We can manipulate this term as follows:
\begin{align}
\nabla_e \df g_{ad}&=\partial_e \df g_{ad}-\Gamma^{f}_{ea}\df g_{fd}-\Gamma^{f}_{ed}\df g_{af} \nn \\
&=\partial_e \df g_{ad}-\df\left(\Gamma^{f}_{ea} g_{fd}\right)-\df\left(\Gamma^{f}_{ed} g_{af}\right)+g_{fd}\df\Gamma^{f}_{ea} +g_{af}\df \Gamma^{f}_{ed} \nn \\
&= \df \left(\nabla_e g_{ad}\right)+g_{fd}\df\Gamma^{f}_{ea} +g_{af}\df \Gamma^{f}_{ed}=g_{fd}\df\Gamma^{f}_{ea} +g_{af}\df \Gamma^{f}_{ed}~. \label{dgad}
\end{align}
We have converted the variations of the metric and its derivatives to variations of the Christoffel symbols. In \ref{start-4}, the index $e$ above is contracted with the auxiliary null vector $k^{e}$. We shall use the following identity to transform these terms: 
\begin{align}
k^e\df\Gamma^{f}_{ea}= \df \left(\nabla_a k^f\right)-\nabla_a \left(\df k^f\right)~. \label{kdG}
\end{align}
Using \ref{dgad} and \ref{kdG} in \ref{start-4}, we obtain
\begin{align}
-2 \sqq k^e \ell_b P^{abcd}\ell_c \nabla_e\df g_{ad} &=2 \sqq  k^e P^{abdc}\ell_b\ell_c \left(g_{fd}\df\Gamma^{f}_{ea} +g_{af}\df \Gamma^{f}_{ed}\right) 
\nn 
\\
&=4 \sqq  k^e P^{abdc}\ell_b\ell_c g_{fd}\df\Gamma^{f}_{ea} 
\nn 
\\
&=4 \sqq  P^{abdc}\ell_b\ell_c g_{fd}\left[\df \left(\nabla_a k^f\right)-\nabla_a \left(\df k^f\right)\right]~. \label{dt}
\end{align}
Decomposing the second term involving $\nabla _{a}\delta k^{f}$ above,
\begin{align}
-4 \sqq P^{abdc}\ell_b\ell_c g_{fd}\nabla_a \left(\df k^f\right)=&-4 \sqq P^{abdc}\ell_b\ell_c g_{fd}\left(\P[e]{a}-k^e \ell_a\right)\nabla_e \left[\df k^f\right] 
\nn 
\\
=&-4 \sqq P^{abdc}\ell_b\ell_c g_{fd}\P[e]{a}\nabla_e \left[\df k^f\right] 
\nn 
\\
=&-4 \sqq \nabla_e \left[\P[e]{a}P^{abdc}\ell_b\ell_c g_{fd}\df k^f\right]
\nonumber
\\
&+4 \sqq  \nabla_e \left[P^{abdc}\ell_b\ell_c g_{fd}\P[e]{a}\right]\df k^f~, \label{Eq4}
\end{align}
where we have used the antisymmetry of indices $a,b$ in $P^{abcd}$ in the second step. The second term involving $\delta k^{f}$ is a Dirichlet variation term, since our assumption $k_a=\nabla_{a}\psi$ means that $\delta k^{f}=k_a\delta g^{fa}$. 
As in \ref{Eq1}, the first term can be converted to a total derivative term on the boundary surface by taking the $\sqq$ inside the covariant derivative at the cost of appearance of a Dirichlet variation term. Thus, consigning all the terms in \ref{Eq4} to the dots, we have
\begin{align}
\sqq Q[\ell_c]&=4 \sqq  P^{abdc}\ell_b\ell_c g_{fd}\df \left(\nabla_a k^f\right)+\dots
 \nn
 \\
&=4 \sqq  P^{abcd}\ell_a\ell_c g_{fd}\df \left(\nabla_b k^f\right)+\dots
\nn
\\
&=4 \sqq P^{ab}_{cd}\ell_a\ell^c \left(\tilde{\Pi}_f^{\phantom{a}d}-\ell^d k_f\right) \left(\Pi^m_{\phantom{a}b}-k^m\ell_b \right)\df \left(\nabla_m k^f\right)+\dots
\nn
\\
&=4 \sqq P^{ab}_{cd}\ell_a\ell^c \tilde{\Pi}_f^{\phantom{a}d} \Pi^m_{\phantom{a}b}\df \left(\nabla_m k^f\right)+\dots~, \label{start-5}
\end{align}
where we have introduced the notation $\tilde{\Pi}_a^{\phantom{a}b}=\df^b_a+ k_a \ell^b$. This is just $\P[a]{b}$ with the first index lowered and the second one raised. But we have added the tilde so that it will be easier to recognize this without looking minutely at index placements. 

Dropping the $\sqq$ factor for the time being for ease, the boundary variation can be further manipulated as follows:
\begin{align}
Q[\ell_c]=&4P^{ab}_{cd}\ell _{a}\ell ^{c}\Pi ^{m}_{\phantom{a}b}\tilde{\Pi}^{\phantom{a}d}_{n}\delta \left(\nabla _{m}k^{n}\right) +\dots
\nonumber
\\
=&4P^{ab}_{cd}\ell _{a}\ell ^{c}\left(q^{m}_{b}-\ell ^{m}k_{b}\right)\left(q^{d}_{n}-\ell _{n}k^{d}\right)\delta \left(\nabla _{m}k^{n}\right)+\dots
\nonumber
\\
=&4P^{ab}_{cd}\ell _{a}\ell ^{c}q^{m}_{b}q^{d}_{n}\delta \left(\nabla _{m}k^{n} \right)
-4P^{ab}_{cd}\ell _{a}\ell ^{c}q^{m}_{b}\ell _{n}k^{d}\delta \left(\nabla _{m}k^{n}\right)
-4P^{ab}_{cd}\ell _{a}\ell ^{c}\ell ^{m}k_{b}q^{d}_{n}\delta \left(\nabla _{m}k^{n}\right)
\nonumber
\\
&+4P^{ab}_{cd}\ell _{a}\ell ^{c}\ell ^{m}k_{b}k^{d}\ell_{n}\delta \left(\nabla _{m}k^{n}\right)+\dots
\nonumber
\\
=&4P^{ab}_{cd}\ell _{a}\ell ^{c}\delta \Psi ^{d}_{b}-4P^{ab}_{cd}\ell _{a}\ell ^{c}k^{d}q^{m}_{b}
\delta \left(\ell _{n}\nabla _{m}k^{n}\right)
-4P^{ab}_{cd}\ell _{a}\ell ^{c}k_{b}q^{d}_{n}\delta \left(\ell ^{m}\nabla _{m}k^{n}\right)
\nonumber
\\
&+4P^{ab}_{cd}\ell _{a}\ell ^{c}k_{b}k^{d}\delta \left(\ell ^{m}\ell_{n}\nabla _{m}k^{n}\right) +\dots
\nonumber
\\
=&4P^{ab}_{cd}\ell _{a}\ell ^{c}\delta \Psi ^{d}_{b}+4P^{ab}_{cd}\ell _{a}\ell ^{c}k^{d}q^{m}_{b}
\delta \left(k^{p}\nabla _{m}\ell _{p}\right)
+4P^{ab}_{cd}\ell _{a}\ell ^{c}k_{b}q^{d}_{n}\delta \left(k^{p}\nabla ^{n}\ell _{p}\right)
\nonumber
\\
&+4P^{ab}_{cd}\ell _{a}\ell ^{c}k_{b}k^{d}\delta \kappa +\dots 
\nn
\\
=&4P^{ab}_{cd}\ell _{a}\ell ^{c}\delta \Psi ^{d}_{b}
+8P^{ab}_{cd}\ell _{a}\ell ^{c}k^{d}q^{m}_{b}\delta \left(k^{p}\nabla _{m}\ell _{p}\right)
+4P^{ab}_{cd}\ell _{a}\ell ^{c}k_{b}k^{d}\delta \kappa+\dots~. \label{BT_LL-interm}
\end{align}
In the above derivation, we have replaced $\Pi$ and $\tilde{\Pi}$ in terms of the induced metric $q^{a}_{b}$ and simplified the resultant expressions. Any terms involving metric variations ($\df \ell^a$, $\df k^a$, $\df q^a_b$, $\df g^{ab}$, etc.) have been banished to the anonymity of the three dots and we have made use of the equations \ref{def_kappa} and \ref{psi_def-main}. Further, we have used the assumption that $\ell_a k^a=-1$ everywhere and even under variations to write
\begin{align}
\ell_n \nabla_m k^n=\nabla_m \left(\ell_nk^n\right)-k^n\nabla_m \ell_n =\partial_m \left(\ell_nk^n\right)-k^n\nabla_m \ell_n=-k^n\nabla_m \ell_n~.
\end{align}
The simplification of the term $-4P^{ab}_{cd}\ell _{a}\ell ^{c}k_{b}q^{d}_{n}\delta \left(\ell ^{m}\nabla _{m}k^{n}\right)$ was performed using
\begin{align}
\ell ^{m}\nabla _{m}k^{n}
=\ell ^{m}\nabla _{m}\nabla^n \psi
= \ell ^{m}\nabla^n \nabla_{m} \psi = \ell ^{m}\nabla^n k_m= - k^{m}\nabla^n \ell_m~,
\end{align}
to write
\begin{align}
-4P^{ab}_{cd}\ell _{a}\ell ^{c}k_{b}q^{d}_{n}\delta \left(\ell ^{m}\nabla _{m}k^{n}\right)= 4P^{ab}_{cd}\ell _{a}\ell ^{c}k_{b}q^{d}_{n}\delta \left( k^{m}\nabla^n \ell_m\right)+ \dots~,
\end{align}
which was then combined with another term using the symmetries of $P^{abcd}$ to obtain \ref{BT_LL-interm}.

We will use \ref{BT_LL-interm} to evaluate the structure of the boundary term for the general \LL Lagrangian. But first, as a warmup and due to its significance as the first higher order correction to Einstein gravity, we consider the case of the \GB term.
\subsection{Specializing to Gauss-Bonnet gravity} \label{sec:GB}

Before deriving the boundary term for general \LL gravity, let us briefly discuss the boundary term structure associated with Gauss-Bonnet gravity, i.e., with the $m=2$ term in the \LL series. The expression for the entropy tensor $P^{ab}_{cd}$ in the context of \GB gravity was given in \ref{P_Gauss_Bonnet}, which we reproduce below for convenience:
\begin{align}
P^{ab}_{cd\ (2)}=2\left(R^{ab}_{cd}+G^{b}_{c}\delta ^{a}_{d}-G^{a}_{c}\delta ^{b}_{d}+R^{a}_{d}\delta ^{b}_{c}-R^{b}_{d}\delta ^{a}_{c}\right)~.
\end{align}
Using the above formula for the entropy tensor and \ref{BT_LL-interm}, we shall derive the boundary term in \GB gravity. We shall simplify each of the terms in \ref{BT_LL-interm} in turn. The first among them is
\begin{align}
Q^{(1)}_{\textrm{GB}}[\ell_c]\equiv &4 P^{ab}_{cd\ (2)}\ell _{a}\ell ^{c}\delta \Psi ^{d}_{b}
\nonumber
\\
=&4P^{ab}_{cd\ (2)}\ell _{a}\ell ^{c}q^{m}_{b}q^{d}_{n}\delta \left(\nabla _{m}k^{n}\right)+\dots
\nonumber
\\
=&8 \left(R^{ab}_{cd}\ell _{a}\ell ^{c}q^{m}_{b}q^{d}_{n}-G^{a}_{c}\ell _{a}\ell ^{c}q^{m}_{n}\right)
\delta \left(\nabla _{m}k^{n}\right)+\dots
\nonumber
\\
=&8 \left(R^{ab}_{cd}\ell _{a}\ell ^{c}q^{m}_{b}q^{d}_{n}-R^{a}_{c}\ell _{a}\ell ^{c}q^{m}_{n}\right)
\delta \left(\nabla _{m}k^{n}\right)+\dots
\nonumber
\\
=&8\left[\left(-q^{m}_{a}q^{b}_{n}\pounds _{\ell}\Theta ^{a}_{b}+\kappa \Theta ^{m}_{n}-\Theta ^{m}_{p}\Theta ^{p}_{n}\right) 
-\left(\kappa \Theta -\frac{d\Theta}{d\lambda}-\Theta ^{a}_{b}\Theta ^{b}_{a}\right)q^{m}_{n}\right]\delta \left(\nabla _{m}k^{n}\right)+\dots 
\nn 
\\
=& 8\left(-q^{m}_{a}q^{b}_{n}\pounds _{\ell}\Theta ^{a}_{b}+\kappa \Theta ^{m}_{n}-\Theta ^{m}_{p}\Theta ^{p}_{n}\right)\delta \Psi _{m}^{n}-8\left(\kappa \Theta -\frac{d\Theta}{d\lambda}-\Theta ^{a}_{b}\Theta ^{b}_{a}\right)\delta \Psi+\dots~, \label{QGB_1}
\end{align}
where we have used \ref{qlqlR-5-main} to substitute for $R^{ab}_{cd}\ell _{a}\ell ^{c}q^{m}_{b}q^{d}_{n}$ and also used 
\begin{align}
R^{a}_{c}\ell _{a}\ell ^{c}=\kappa \Theta -\frac{d\Theta}{d\lambda}-\Theta^{a}_{b}\Theta^{b}_{a}~. \label{null_Raych}
\end{align}
Here we have introduced a parameter $\lambda$ such that $\ell^c \nabla_c \lambda=1$ to write $\ell^c \nabla_c \Theta=(d\Theta/d\lambda)$. \ref{null_Raych} is the null Raychaudhuri equation \cite{Poisson}, but with non-affine parametrisation \cite{Chakraborty:2015hna}. In fact, we can derive this equation easily from the projections of the Riemann tensor. We have given this derivation in \ref{app:null_raych}.

The next term in \ref{BT_LL-interm} is
\begin{align}
Q^{(2)}_{\textrm{GB}}[\ell_c]\equiv &8P^{ab}_{cd\ (2)}\ell _{a}\ell ^{c}k^{d}q^{m}_{b}\delta \left(k^{p}\nabla _{m}\ell _{p}\right)
\nonumber
\\
=&16\left(R^{ab}_{cd}\ell _{a}\ell ^{c}k^{d}q^{m}_{b}-G^{b}_{c}\ell ^{c}q^{m}_{b}\right)
\delta \left(k^{p}\nabla _{m}\ell _{p}\right)
\nonumber
\\
=&16 R^{ab}_{cd}\ell ^{c}q^{m}_{b}\left(\ell _{a}k^{d}+\delta ^{d}_{a}\right)
\delta \left(k^{p}\nabla _{m}\ell _{p}\right)
\nonumber
\\
=&16 R^{ab}_{cd}\ell ^{c}q^{m}_{b}\left(q^{d}_{a}-k_{a}\ell ^{d}\right)
\delta \left(k^{p}\nabla _{m}\ell _{p}\right)
\nonumber
\\
=&-16 R^{ab}_{cd}\ell ^{c}q^{m}_{a}q^{d}_{b}\delta \left(k^{p}\nabla _{m}\ell _{p}\right)
\nonumber
\\
=&-16\left[-D^{m}\Theta+D_{p}\Theta ^{mp}-\left(\Theta q^{am}-\Theta ^{ma}\right)k^{p}\nabla _{a}\ell _{p}
\right]\delta \left(k^{q}\nabla _{m}\ell _{q}\right)~,\label{QGB_2}
\end{align}
where we have used a contracted version of \ref{R_3q-main} in arriving at the last step. The remaining term in \ref{BT_LL-interm} has the following structure
\begin{align}
Q^{(3)}_{\textrm{GB}}[\ell_c]&\equiv 4P^{ab}_{cd\ (2)}\ell _{a}\ell ^{c}k_{b}k^{d}\delta \kappa
=8\left(R^{ab}_{cd}\ell _{a}\ell ^{c}k_{b}k^{d}-G^{b}_{c}\ell ^{c}k_{b}
-R^{a}_{d}\ell _{a}k^{d}\right)\delta \kappa
\nonumber
\\
&=8\left(R^{ab}_{cd}\ell _{a}\ell ^{c}k_{b}k^{d}-2R^{a}_{d}\ell _{a}k^{d}-\frac{1}{2}R\right)\delta \kappa
=-4R^{ab}_{cd}q^{c}_{a}q^{d}_{b}\delta \kappa
\nonumber
\\
&=-4\left(~^{(D-2)}R+2\Theta \Psi -2\Theta _{ab}\Psi ^{ab}\right)\delta \kappa~,\label{QGB_3}
\end{align}
where, in the penultimate step, we have used the following result:
\begin{align}
	R^{ab}_{cd}q^{c}_{a}q^{d}_{b}=R+4R^{a}_{b}\ell _{a}k^{b}-2R^{ab}_{cd}\ell _{a}\ell ^{c}k_{b}k^{d}~,
\end{align}
and contracted version of \ref{Gauss-Codazzi-main} in the ultimate step. Finally, combining \ref{QGB_1}, \ref{QGB_2} and \ref{QGB_3}, we arrive at
\begin{align}
Q_{\textrm{GB}}[\ell_c]\equiv& Q^{(1)}_{\textrm{GB}}[\ell_c]+Q^{(2)}_{\textrm{GB}}[\ell_c]+Q^{(3)}_{\textrm{GB}}[\ell_c] \nn\\
=&-8\left(q^{m}_{p}q^{q}_{n}\pounds _{\ell}\Theta ^{n}_{m}\right)\delta \Psi ^{p}_{q}+8\kappa \Theta ^{m}_{n}\delta \Psi ^{n}_{m}-8\Theta ^{m}_{p}\Theta ^{p}_{n}\delta \Psi ^{n}_{m}-8\kappa \Theta \delta \Psi +8\frac{d\Theta}{d\lambda}\delta \Psi 
\nonumber
\\
&+8\Theta ^{a}_{b}\Theta ^{b}_{a}\delta \Psi + 16D^{m}\Theta \delta \left(k^{q}\nabla _{m}\ell _{q}\right)-16 D_{p}\Theta ^{mp}\delta \left(k^{q}\nabla _{m}\ell _{q}\right)
\nonumber
\\
&+16\left(k^{q}\nabla _{a}\ell _{q}\right)\left(\Theta q^{am}-\Theta ^{am}\right)\delta \left(k^{p}\nabla _{m}\ell _{p}\right)-4~^{(D-2)}R\delta \kappa -8\Theta \Psi \delta \kappa +8\Theta _{ab}\Psi ^{ab}\delta \kappa + \dots 
\nn 
\\
=&-8\left(q^{m}_{p}q^{q}_{n}\pounds _{\ell}\Theta ^{n}_{m}\right)\delta \Psi ^{p}_{q}+8 \Theta ^{m}_{n}\delta \left(\kappa\Psi ^{n}_{m}\right)-8\Theta ^{m}_{p}\Theta ^{p}_{n}\delta \Psi ^{n}_{m}-8 \Theta \delta \left(\kappa\Psi\right) +8\frac{d\Theta}{d\lambda}\delta \Psi 
\nonumber
\\
&+8\Theta ^{a}_{b}\Theta ^{b}_{a}\delta \Psi + 16D^{m}\Theta \delta \left(k^{q}\nabla _{m}\ell _{q}\right)-16 D_{p}\Theta ^{mp}\delta \left(k^{q}\nabla _{m}\ell _{q}\right)
\nonumber
\\
&+8\left(\Theta q^{am}-\Theta ^{am}\right)\delta \left[\left(k^{q}\nabla _{a}\ell _{q}\right)\left(k^{p}\nabla _{m}\ell _{p}\right)\right]-4~^{(D-2)}R\delta \kappa + \dots~,
\end{align}
where we have again thrown some metric variations to the dots. All these terms have variations of the derivatives of the metric orthogonal to the null surface, i.e., derivatives along $k^a$. Taking the delta commonly outside and eliminating the terms with only variations of the metric or its surface derivatives on the boundary to the group of terms represented by the dots, we obtain
\begin{align}
Q_{\textrm{GB}}[\ell_c]=&\df \left[-8\Psi ^{p}_{q}\pounds _{\ell}\Theta ^{q}_{p}+8\kappa \Theta ^{p}_{q}\Psi ^{q}_{p}-8\Theta ^{a}_{c}\Theta ^{c}_{b}\Psi ^{b}_{a}-8\kappa \Theta \Psi +8\Psi \frac{d\Theta}{d\lambda}+8\Psi \Theta ^{a}_{b}\Theta ^{b}_{a}
\right.\nonumber
\\
&\phantom{\df \left[\right.}+16\left(D^{m}\Theta -D^{a}\Theta ^{m}_{a}\right)\left(k^{q}\nabla _{m}\ell _{q}\right)
+8\left(k^{q}\nabla _{m}\ell _{q}\right)\left(k^{p}\nabla _{a}\ell_{p}\right)\left(\Theta q^{ma}-\Theta ^{ma}\right)
\nonumber
\\
&\phantom{\df \left[\right.}\left.-4\kappa~^{(D-2)}R\right]+\dots~,
\end{align}
where we have used the fact that $\Theta^{ab}$, $D_c$ acting on $\Theta^{ab}$, $\Theta$, $q^{m}_{p}q^{q}_{n}\pounds _{\ell}\Theta ^{n}_{m}$, $d\Theta/d\lambda$ and $~^{(D-2)}R$ do not have derivatives of the metric along $k^a$. (From \ref{R_d-2_def}, one can see that $~^{(D-2)}R$ contains only derivatives tangential to the null surface.) Introducing back the $\sqq$ which we had dropped before \ref{BT_LL-interm}, we have
\begin{align}
\sqq Q_{\textrm{GB}}[\ell_c]=&\df \left\{\sqq  \left[-8\Psi ^{p}_{q}\pounds _{\ell}\Theta ^{q}_{p}+8\kappa \Theta ^{p}_{q}\Psi ^{q}_{p}-8\Theta ^{a}_{c}\Theta ^{c}_{b}\Psi ^{b}_{a}-8\kappa \Theta \Psi +8\Psi \frac{d\Theta}{d\lambda}+8\Psi \Theta ^{a}_{b}\Theta ^{b}_{a}
\right.\right. \nonumber
\\
&\phantom{\df \left\{\sqq\left[\right.\right.}+16\left(D^{m}\Theta -D^{a}\Theta ^{m}_{a}\right)\left(k^{q}\nabla _{m}\ell _{q}\right)
+8\left(k^{q}\nabla _{m}\ell _{q}\right)\left(k^{p}\nabla _{a}\ell_{p}\right)\left(\Theta q^{ma}
-\Theta ^{ma}\right)
\nonumber
\\
&\phantom{\df \left\{\sqq\left[\right.\right.}\left.\left.-4\kappa~^{(D-2)}R\right]\right\}+\dots~,
\end{align}
where the variations of $\sqq$ have been removed to the dots. Therefore, the boundary term to be added to cancel off the normal derivatives of the metric in the boundary variation for \GB gravity is
\begin{align}\label{BTGB_pre}
\sqq~\mathcal{B}_{\textrm{\tiny GB}}=&4\sqq\left[2\Psi ^{p}_{q}\pounds _{\ell}\Theta ^{q}_{p}-2\kappa \Theta ^{p}_{q}\Psi ^{q}_{p}+2\Theta ^{a}_{c}\Theta ^{c}_{b}\Psi ^{b}_{a}+2\kappa \Theta \Psi -2\Psi \frac{d\Theta}{d\lambda}-2\Psi \Theta ^{a}_{b}\Theta ^{b}_{a}
\right. \nonumber
\\
&\phantom{\sqq\left[\right.}-4\left(D^{m}\Theta -D^{a}\Theta ^{m}_{a}\right)\left(k^{q}\nabla _{m}\ell _{q}\right)
-2\left(k^{q}\nabla _{m}\ell _{q}\right)\left(k^{p}\nabla _{a}\ell _{p}\right)\left(\Theta q^{ma}
-\Theta ^{ma}\right)
\nonumber
\\
&\phantom{\sqq\left[\right.}\left.+\kappa~^{(D-2)}R\right]~.
\end{align}
Grouping terms with $\kappa$, we can rewrite this boundary term as
\begin{align}\label{BTGB}
\sqq~\mathcal{B}_{\textrm{\tiny GB}}=&4\sqq\bigg[\kappa\left(~^{(D-2)}R-2 \Theta ^{p}_{q}\Psi ^{q}_{p}+2 \Theta \Psi\right)+2\Psi ^{p}_{q}\pounds _{\ell}\Theta ^{q}_{p}-2\Psi \frac{d\Theta}{d\lambda} +2\Theta ^{a}_{c}\Theta ^{c}_{b}\Psi ^{b}_{a} -2\Psi \Theta ^{a}_{b}\Theta ^{b}_{a}
 \nonumber
\\
&\qquad \qquad \quad -4\left(D^{m}\Theta -D^{a}\Theta ^{m}_{a}\right)\left(k^{q}\nabla _{m}\ell _{q}\right)
-2\left(k^{q}\nabla _{m}\ell _{q}\right)\left(k^{p}\nabla _{a}\ell_{p}\right)\left(\Theta q^{ma}
-\Theta ^{ma}\right)\bigg]~.
\end{align}
This finishes our computation of the null boundary term in \GB gravity. This boundary term is considerably larger than the null boundary term for Einstein-Hilbert action \cite{Neiman:2012fx,Parattu:2015gga,Lehner:2016vdi}. The derivatives of the metric in directions non-tangential to the null boundary are present in the tensors $\kappa,\Psi_{ab}$ and $k^{p}\nabla _{a}\ell_{p}$, while only $\kappa$ was present in the \EH null boundary term. 

We will now take up our main task, that of finding out the boundary term for general \LL gravity.
\subsection{Boundary term for general \LL gravity}\label{sec:gen_LL}

In this section, we shall derive the null boundary term for general \LL gravity by proceeding on a path similar to the one used in the last section to successfully derive the null boundary term in the context of \GB gravity.

We shall derive the null boundary term for $m^{\textrm th}$ order \LL gravity. The null boundary term for any \LL action can be obtained as a suitable sum of these terms. The expression of the entropy tensor $P^{ab}_{cd}$ for $m^{\textrm th}$ order \LL gravity takes the following form (see \ref{P_def}):
\begin{equation}\label{P-m-def-1}
P^{ab}_{cd\ (m)}=\frac{m}{2^{m}}\delta ^{aba_{1}b_{1}\dots a_{m-1}b_{m-1}}_{cdc_{1}d_{1}\dots c_{m-1}d_{m-1}}R^{c_{1}d_{1}}_{a_{1}b_{1}}\dots R^{c_{m-1}d_{m-1}}_{a_{m-1}b_{m-1}}~.
\end{equation}
To make the index notation less cumbersome, we shall do the evaluation for $(m+1)^{th}$ order \LL gravity, since then we will have $m$ instead of $(m-1)$ in the last set of indices presented above and the expressions will be slightly shorter. 
\subsubsection{Projecting and decomposing the Riemann tensors in the boundary variation}

In \ref{BT_LL-interm}, we start with the last term, involving $\delta \kappa$, which is the simplest to work with. We multiply this term by $[2^{m+1}/(m+1)]$ to remove the corresponding factor from $P^{ab}_{cd\ (m+1)}$ to obtain
\begin{align}
\frac{2^{m+1}}{m+1}Q^{(3)}_{m+1}[\ell_c]\equiv&\frac{2^{m+1}}{m+1} \left[4P^{ab}_{cd\ (m+1)}\ell _{a}\ell ^{c}k_{b}k^{d}\delta \kappa\right]
\nonumber
\\
=&4\delta \kappa ~\ell _{a}\ell ^{c}k_{b}k^{d}~\delta ^{abp_{1}q_{1}\dots p_{m}q_{m}}_{cdr_{1}s_{1}\dots r_{m}s_{m}}R^{r_{1}s_{1}}_{p_{1}q_{1}}\dots R^{r_{m}s_{m}}_{p_{m}q_{m}}
\nonumber
\\
=&4\delta \kappa ~\ell _{a}\ell ^{c}k_{b}k^{d}~\delta ^{abp_{1}q_{1}\dots p_{m}q_{m}}_{cdr_{1}s_{1}\dots r_{m}s_{m}}\left(q^{a_{1}}_{p_{1}}q^{b_{1}}_{q_{1}}q^{r_{1}}_{c_{1}} q^{s_{1}}_{d_{1}}R^{c_{1}d_{1}}_{a_{1}b_{1}}\right)\dots \left(q^{a_{m}}_{p_{m}}q^{b_{m}}_{q_{m}}q^{r_{m}}_{c_{m}} q^{s_{m}}_{d_{m}}R^{c_{m}d_{m}}_{a_{m}b_{m}}\right)
\nonumber
\\
=&4\delta \kappa ~\ell _{a}\ell ^{c}k_{b}k^{d}~\delta ^{abp_{1}q_{1}\dots p_{m}q_{m}}_{cdr_{1}s_{1}\dots r_{m}s_{m}}\left(~^{(D-2)}R^{r_{1}s_{1}}_{p_{1}q_{1}}+4\Theta ^{r_{1}}_{p_{1}}\Psi ^{s_{1}}_{q_{1}}\right)\dots 
\left(~^{(D-2)}R^{r_{m}s_{m}}_{p_{m}q_{m}}+4\Theta ^{r_{m}}_{p_{m}}\Psi ^{s_{m}}_{q_{m}}\right)
\nonumber
\\
=&4\delta \kappa ~\ell _{a}\ell ^{c}k_{b}k^{d}~\delta ^{abp_{1}q_{1}\dots p_{m}q_{m}}_{cdr_{1}s_{1}\dots r_{m}s_{m}}
\nonumber
\\
&\times \sum _{t=0}^{m}\Mycomb[m]{t}~4^{t}\left(~^{(D-2)}R^{r_{1}s_{1}}_{p_{1}q_{1}}
\dots ~^{(D-2)}R^{r_{m-t}s_{m-t}}_{p_{m-t}q_{m-t}}\right)\left(\Theta ^{r_{m-t+1}}_{p_{m-t+1}}\Psi ^{s_{m-t+1}}_{q_{m-t+1}}\dots \Theta ^{r_{m}}_{p_{m}}\Psi ^{s_{m}}_{q_{m}}\right)~, \label{BT_LL-3}
\end{align}
where, in the third step, we have written the Kronecker delta as $\delta ^{a}_{b}=q^{a}_{b}-\ell ^{a}k_{b}-\ell _{b}k^{a}$ and have used the antisymmetry of the determinant tensor. In the fourth line, we have made use of \ref{Gauss-Codazzi-main} to write down the projected $D$-dimensional Riemann tensor in terms of the induced Riemann tensor and $\Theta_{ab}$ and $\Psi_{ab}$. In the next step, we have expanded out the product of the $m$ factors. Each term in the sum involves $m$ factors, the factors present being either the components of the induced Riemann tensor or factors of the form $4\Theta^a_b\Psi^c_d$. All terms having the same number of Riemann tensor factors are equivalent due to the symmetries of $\delta ^{abp_{1}q_{1}\dots p_{m}q_{m}}_{cdr_{1}s_{1}\dots r_{m}s_{m}}$. The binomial coefficient $\Mycomb[m]{t}$ represents the number of ways of choosing $k$ factors out of the $m$ factors present. It is to be understood that the indices are to be taken only from the pool $(p_1,q_1,r_1,s_1)$ to $(p_{m},q_{m},r_{m},s_{m})$, with each index appearing only once. For example, the $t=0$ term is built from the induced Riemann tensor alone, while the $t=m$ term depends solely on factors inheriting $\Theta^a _b$ and $\Psi^a_b$. 

Next, we tackle the second term in \ref{BT_LL-interm} involving the variation of $k^{p}\nabla _{m}\ell_{p}$, again multiplying by $[2^{m+1}/(m+1)]$:
\begin{align}
	\frac{2^{m+1}}{m+1}Q^{(2)}_{m+1}[\ell_c]\equiv &\frac{2^{m+1}}{m+1} \left[8P^{ab}_{cd\ (m+1)}\ell _{a}\ell ^{c}k^{d}q^{n}_{b}\delta \left(k^{p}\nabla _{n}\ell _{p}\right)\right]
	\nonumber
	\\
	=&8\ell _{a}\ell ^{c}k^{d}q^{n}_{b}\delta \left(k^{p}\nabla _{n}\ell _{p}\right)
	~\delta ^{abp_{1}q_{1}\dots p_{m}q_{m}}_{cdr_{1}s_{1}\dots r_{m}s_{m}}
	R_{p_{1}q_{1}}^{r_{1}s_{1}}\dots R_{p_{m}q_{m}}^{r_{m}s_{m}}
	\nonumber
	\\
	=&8\ell _{a}\ell ^{c}k^{d}q^{n}_{b}\delta \left(k^{p}\nabla _{n}\ell _{p}\right)
	~\delta ^{abp_{1}q_{1}\dots p_{m}q_{m}}_{cdr_{1}s_{1}\dots r_{m}s_{m}}\left(q^{a_{1}}_{p_{1}}q^{b_{1}}_{q_{1}}q^{r_{1}}_{c_{1}} q^{s_{1}}_{d_{1}}R^{c_{1}d_{1}}_{a_{1}b_{1}}\right)\dots \left(q^{a_{m}}_{p_{m}}q^{b_{m}}_{q_{m}}q^{r_{m}}_{c_{m}} q^{s_{m}}_{d_{m}}R^{c_{m}d_{m}}_{a_{m}b_{m}}\right)
	\nonumber
	\\
	&-16m\ell _{a}\ell ^{c}k^{d}q^{n}_{b}\delta \left(k^{p}\nabla _{n}\ell _{p}\right)
	~\delta ^{abp_{1}q_{1}p_{2}q_{2}\dots p_{m}q_{m}}_{cdr_{1}s_{1}r_{2}s_{2}\dots r_{m}s_{m}}k_{p_{1}}
	\left(\ell ^{a_{1}}q^{b_{1}}_{q_{1}}q^{r_{1}}_{c_{1}}q^{s_{1}}_{d_{1}}R^{c_{1}d_{1}}_{a_{1}b_{1}}\right)
	\nonumber
	\\
	&\phantom{-}\times \left(q^{a_{2}}_{p_{2}}q^{b_{2}}_{q_{2}}q^{r_{2}}_{c_{2}} q^{s_{2}}_{d_{2}}R^{c_{2}d_{2}}_{a_{2}b_{2}}\right)\dots \left(q^{a_{m}}_{p_{m}}q^{b_{m}}_{q_{m}}q^{r_{m}}_{c_{m}} q^{s_{m}}_{d_{m}}R^{c_{m}d_{m}}_{a_{m}b_{m}}\right)~,\label{Q-2}
\end{align}
where, again, the decomposition of the Kronecker delta as $\delta ^{a}_{b}=q^{a}_{b}-\ell ^{a}k_{b}-\ell _{b}k^{a}$ and the symmetry properties of the determinant tensor were used. Since none of the upper indices of the determinant tensor are a priori contracted with a $k_a$, terms of the form $k_{b}\ell^c$ arising from the expansion of $\delta ^{c}_{b}$ survive when the index $b$ is contracted with the determinant tensor and the index $c$ is contracted with a lower index of one of the Riemann tensors, with all the other Riemann indices present, of that particular Riemann tensor and of others in the product, contracted with $q^a_b$. All such terms are equivalent due to the symmetries of the determinant tensor. Since there are two lower indices per Riemann tensor and there are $m$ Riemann tensors to choose from, we get an overall factor of $2m$ for this term. We shall first simplify the first term in the \ref{Q-2}, where all the indices of the Riemann tensors are contracted with $q$'s:
\begin{align}
	\frac{2^{m+1}}{m+1}Q^{(2,a)}_{m+1}[\ell_c]\equiv&8\ell _{a}\ell ^{c}k^{d}q^{n}_{b}\delta \left(k^{p}\nabla _{n}\ell _{p}\right)
	\delta ^{abp_{1}q_{1}\dots p_{m}q_{m}}_{cdr_{1}s_{1}\dots r_{m}s_{m}}\left(q^{a_{1}}_{p_{1}}q^{b_{1}}_{q_{1}}q^{r_{1}}_{c_{1}} q^{s_{1}}_{d_{1}}R^{c_{1}d_{1}}_{a_{1}b_{1}}\right)\dots \left(q^{a_{m}}_{p_{m}}q^{b_{m}}_{q_{m}}q^{r_{m}}_{c_{m}} q^{s_{m}}_{d_{m}}R^{c_{m}d_{m}}_{a_{m}b_{m}}\right)
	\nonumber
	\\
	=&8\ell _{a}\ell ^{c}k^{d}q^{n}_{b}\delta \left(k^{p}\nabla _{n}\ell _{p}\right)
	\delta ^{abp_{1}q_{1}\dots p_{m}q_{m}}_{cdr_{1}s_{1}\dots r_{m}s_{m}}
	\nonumber
	\\
	&\times \sum _{t=0}^{m}\Mycomb[m]{t}~4^{t}\left(^{(D-2)}R^{r_{1}s_{1}}_{p_{1}q_{1}}
	\dots ~^{(D-2)}R^{r_{m-t}s_{m-t}}_{p_{m-t}q_{m-t}}\right)\left(\Theta ^{r_{m-t+1}}_{p_{m-t+1}}\Psi ^{s_{m-t+1}}_{q_{m-t+1}}\dots \Theta ^{r_{m}}_{p_{m}}\Psi ^{s_{m}}_{q_{m}}\right)~, \label{BT_LL-2-a}
\end{align}
where we have made use of \ref{Gauss-Codazzi-main} and expanded just as in \ref{BT_LL-3}. Going on to the second term in \ref{Q-2}, we have
\begin{align}
	\frac{2^{m+1}}{m+1}Q^{(2,b)}_{m+1}[\ell_c]\equiv&-16m\ell _{a}\ell ^{c}k^{d}q^{n}_{b}\delta \left(k^{p}\nabla _{n}\ell _{p}\right)
	~\delta ^{abp_{1}q_{1}p_{2}q_{2}\dots p_{m}q_{m}}_{cdr_{1}s_{1}r_{2}s_{2}\dots r_{m}s_{m}}k_{p_{1}}
	\left(\ell ^{a_{1}}q^{b_{1}}_{q_{1}}q^{r_{1}}_{c_{1}}q^{s_{1}}_{d_{1}}R^{c_{1}d_{1}}_{a_{1}b_{1}}\right)
	\nonumber
	\\
	&\times \left(q^{a_{2}}_{p_{2}}q^{b_{2}}_{q_{2}}q^{r_{2}}_{c_{2}} q^{s_{2}}_{d_{2}}R^{c_{2}d_{2}}_{a_{2}b_{2}}\right)\dots \left(q^{a_{m}}_{p_{m}}q^{b_{m}}_{q_{m}}q^{r_{m}}_{c_{m}} q^{s_{m}}_{d_{m}}R^{c_{m}d_{m}}_{a_{m}b_{m}}\right)
	\nonumber
	\\
	=&-32m\ell _{a}\ell ^{c}k_{b}k^{d}\delta \left(q^{n}_{p_{1}}k^{p}\nabla _{n}\ell _{p}\right)
	~\delta ^{abp_{1}q_{1}p_{2}q_{2}\dots p_{m}q_{m}}_{cdr_{1}s_{1}r_{2}s_{2}\dots r_{m}s_{m}}
	\times \left\{D^{r_{1}}\Theta ^{s_{1}}_{q_{1}}+\left(k^{q}\nabla _{j}\ell _{q}\right)q^{jr_{1}}
	\Theta ^{s_{1}}_{q_{1}}\right\}
	\nonumber
	\\
	&\times \sum _{t=0}^{m-1}\Mycomb[m-1]{t}~4^{t}\left(~^{(D-2)}R^{r_{2}s_{2}}_{p_{2}q_{2}}
	\dots ~^{(D-2)}R^{r_{m-t}s_{m-t}}_{p_{m-t}q_{m-t}}\right)
	\nonumber
	\\
	&\phantom{~ \sum _{t=1}^{m}4^{t}\Mycomb[m-1]{t-1}}\times \left(\Theta ^{r_{m-t+1}}_{p_{m-t+1}}\Psi ^{s_{m-t+1}}_{q_{m-t+1}}\dots \Theta ^{r_{m}}_{p_{m}}\Psi ^{s_{m}}_{q_{m}}\right) \nn\\
	&+\dots~.
	 \label{BT_LL-2-b}
\end{align}
Here we have made use of \ref{Gauss-Codazzi-main}, again, as well as \ref{R_3q-main}, and thrown away some metric variations into the dots. The summation has appeared in the same way as in \ref{BT_LL-2-a}, with the difference that there are only $m-1$ factors here. 

Finally, we shall tackle the first term of \ref{BT_LL-interm}, which, after multiplying with $[2^{m+1}/(m+1)]$, is
\begin{align}
	\frac{2^{m+1}}{m+1}Q^{(1)}_{m+1}[\ell_c]\equiv&\frac{2^{m+1}}{m+1}\left(4P^{ab}_{cd\ (m+1)}\ell _{a}\ell ^{c}\delta \Psi ^{d}_{b}\right)
	\nonumber
	\\
	=&4\ell _{a}\ell ^{c}\delta \Psi ^{d}_{b}~\delta ^{abp_{1}q_{1}\dots p_{m}q_{m}}_{cdr_{1}s_{1}\dots r_{m}s_{m}}
	R_{p_{1}q_{1}}^{r_{1}s_{1}}\dots R_{p_{m}q_{m}}^{r_{m}s_{m}}
	\nonumber
	\\
	=&4\ell _{a}\ell ^{c}\delta \Psi ^{d}_{b}~\delta ^{abp_{1}q_{1}\dots p_{m}q_{m}}_{cdr_{1}s_{1}\dots r_{m}s_{m}}\left(q^{a_{1}}_{p_{1}}q^{b_{1}}_{q_{1}}q^{r_{1}}_{c_{1}} q^{s_{1}}_{d_{1}}R^{c_{1}d_{1}}_{a_{1}b_{1}}\right)\dots \left(q^{a_{m}}_{p_{m}}q^{b_{m}}_{q_{m}}q^{r_{m}}_{c_{m}} q^{s_{m}}_{d_{m}}R^{c_{m}d_{m}}_{a_{m}b_{m}}\right)
	\nonumber
	\\
	&-16m\ell _{a}\ell ^{c}\delta \Psi ^{d}_{b}~\delta ^{abp_{1}q_{1}\dots p_{m}q_{m}}_{cdr_{1}s_{1}\dots r_{m}s_{m}}k_{p_{1}}\left(\ell ^{a_{1}}q^{b_{1}}_{q_{1}}q^{r_{1}}_{c_{1}}q^{s_{1}}_{d_{1}}R^{c_{1}d_{1}}_{a_{1}b_{1}}\right)
	\nonumber
	\\
	&\phantom{-}\times \left(q^{a_{2}}_{p_{2}}q^{b_{2}}_{q_{2}}q^{r_{2}}_{c_{2}} q^{s_{2}}_{d_{2}}R^{c_{2}d_{2}}_{a_{2}b_{2}}\right)\dots \left(q^{a_{m}}_{p_{m}}q^{b_{m}}_{q_{m}}q^{r_{m}}_{c_{m}} q^{s_{m}}_{d_{m}}R^{c_{m}d_{m}}_{a_{m}b_{m}}\right)
	\nonumber
	\\
	&+32\Mycomb[m]{2}~\ell _{a}\ell ^{c}\delta \Psi ^{d}_{b}~\delta ^{abp_{1}q_{1}\dots p_{m}q_{m}}_{cdr_{1}s_{1}\dots r_{m}s_{m}}\left(\ell ^{a_{1}}q^{b_{1}}_{q_{1}}q^{r_{1}}_{c_{1}}q^{s_{1}}_{d_{1}}R^{c_{1}d_{1}}_{a_{1}b_{1}}\right)k_{p_{1}}
	 \left(\ell _{c_{2}}q^{b_{2}}_{q_{2}}q^{a_{2}}_{p_{2}}q^{s_{2}}_{d_{2}}R^{c_{2}d_{2}}_{a_{2}b_{2}}\right)k^{r_{2}}
	\nonumber
	\\
	&\phantom{+}\times \left(q^{a_{3}}_{p_{3}}q^{b_{3}}_{q_{3}}q^{r_{3}}_{c_{3}} q^{s_{3}}_{d_{3}}R^{c_{3}d_{3}}_{a_{3}b_{3}}\right)\dots \left(q^{a_{m}}_{p_{m}}q^{b_{m}}_{q_{m}}q^{r_{m}}_{c_{m}} q^{s_{m}}_{d_{m}}R^{c_{m}d_{m}}_{a_{m}b_{m}}\right)
	\nonumber
	\\
	&+16m\ell _{a}\ell ^{c}\delta \Psi ^{d}_{b}~\delta ^{abp_{1}q_{1}\dots p_{m}q_{m}}_{cdr_{1}s_{1}\dots r_{m}s_{m}}\left(\ell ^{a_{1}} q^{b_{1}}_{q_{1}}\ell _{c_{1}}q^{s_{1}}_{d_{1}}R^{c_{1}d_{1}}_{a_{1}b_{1}}\right)k_{p_{1}}k^{r_{1}}
	\nonumber
	\\
	&\phantom{+}\times\left(q^{a_{2}}_{p_{2}}q^{b_{2}}_{q_{2}}q^{r_{2}}_{c_{2}} q^{s_{2}}_{d_{2}}R^{c_{2}d_{2}}_{a_{2}b_{2}}\right)\dots \left(q^{a_{m}}_{p_{m}}q^{b_{m}}_{q_{m}}q^{r_{m}}_{c_{m}} q^{s_{m}}_{d_{m}}R^{c_{m}d_{m}}_{a_{m}b_{m}}\right)\nn\\
	&+\dots~, \label{BT_LL_1}
\end{align}
where the dots at the end indicate that we have removed some metric variation terms. Here also, we have expressed the Kronecker delta in terms of the induced metric $q^{a}_{b}$ and the null vectors $\ell_{a}$ and $k_{a}$. In this case, we get four non-zero terms on expansion: (a) a term with \textit{all} the indices of \textit{all} the Riemann tensors contracted with the induced metric; (b) a term with \textit{one} index of just \textit{one} Riemann tensor contracted with the null normal $\ell_{a}$ and all other Riemann tensor indices contracted with the induced metric $q^{a}_{b}$; (c) a term with \emph{one} index each of \emph{two} Riemann tensors contracted with $\ell_a$ and the rest with $q^{a}_{b}$ and (d) a term with \emph{two} indices of \emph{one} Riemann tensor contracted with $\ell_{a}$, with the other indices contracted with $q^{a}_{b}$.  The second term is obtained in a manner similar to the second term in \ref{Q-2}. But here there is no $k_a$ contracted a priori with the lower or upper indices of the determinant tensor. Therefore, terms with factors of the form $k^{r}\left(\ell_{c} q^{b}_{q}q^{a}_{p}q^{s}_{d}R^{cd}_{ab}\right)$ survive in addition to terms of the form $k_{p}\left(\ell ^{a}q^{b}_{q}q^{r}_{c}q^{s}_{d}R^{cd}_{ab}\right)$, where the open indices are contracted with the determinant tensor. But these two types of terms are equivalent up to metric variations, as can be shown by lowering all upper indices and raising all lower indices. In proving this equivalence, one would have to throw away some metric variations to the dots while raising and lowering indices in $\df \Psi^d_b$ and make use of the fact that $\Psi^d_b$ is a symmetric tensor. So we obtain an additional factor of two for this term in comparison with \ref{Q-2}, giving an overall factor of $4m$ on summing all such equivalent terms. The third term arises out of the possibility of having one Riemann tensor factor contracted as $k^{r}\left(\ell_{c} q^{b}_{q}q^{a}_{p}q^{s}_{d}R^{cd}_{ab}\right)$ and another contracted as $k_{p}\left(\ell ^{a}q^{b}_{q}q^{r}_{c}q^{s}_{d}R^{cd}_{ab}\right)$. Symmetries of the Riemann tensor ensure that all such terms are equivalent. There are $\Mycomb[m]{2}$ ways of choosing two out of the $m$ Riemann tensors. Once two are chosen, one of the two upper indices of the first Riemann tensor can be contracted with $\ell_c$ while one of  the two upper indices of the other are contracted with $\ell^a$, giving four possibilities. Similarly, there are four possibilities using the lower indices of the first and the upper indices of the second. In total, thus, there are $8$ terms arising out of any pair of Riemann tensors. Thus, the total number of such terms formed is $8\Mycomb[m]{2}=4m(m-1)$. The last term has one Riemann tensor contracted in the form $\ell ^{a} q^{b}_{q}\ell _{c}q^{s}_{d}R^{cd}_{ab}k_{p}k^{r}$. Again, all such terms are equivalent. There are $m$ Riemann tensor factors to choose from, with each Riemann tensor offering two lower indices as options to contract with $\ell ^{a}$ and two upper indices as options for contracting with $\ell_{c}$, giving four combinations in total per Riemann tensor. Thus, we obtain an overall factor of $4m$.

We shall now simplify the terms in \ref{BT_LL_1} one by one. We start with the first term, which has all the indices of all the Riemann tensors contracted with $q^{a}_{b}$. Making use of \ref{Gauss-Codazzi-main} and noting that this term has the same structure we encountered in \ref{BT_LL-3}, we can immediately write
\begin{align}
	\frac{2^{m+1}}{m+1}Q^{(1,a)}_{m+1}[\ell_c]\equiv&4\ell _{a}\ell ^{c}\delta \Psi ^{d}_{b}~\delta ^{abp_{1}q_{1}\dots p_{m}q_{m}}_{cdr_{1}s_{1}\dots r_{m}s_{m}}\left(q^{a_{1}}_{p_{1}}q^{b_{1}}_{q_{1}}q^{r_{1}}_{c_{1}} q^{s_{1}}_{d_{1}}R^{c_{1}d_{1}}_{a_{1}b_{1}}\right)\dots \left(q^{a_{m}}_{p_{m}}q^{b_{m}}_{q_{m}}q^{r_{m}}_{c_{m}} q^{s_{m}}_{d_{m}}R^{c_{m}d_{m}}_{a_{m}b_{m}}\right)
	\nonumber
	\\
	=&4\ell _{a}\ell ^{c}\delta \Psi ^{d}_{b}~\delta ^{abp_{1}q_{1}\dots p_{m}q_{m}}_{cdr_{1}s_{1}\dots r_{m}s_{m}}
	\nonumber
	\\
	&\times \sum _{t=0}^{m}\Mycomb[m]{t}~4^{t}\left(~^{(D-2)}R^{r_{1}s_{1}}_{p_{1}q_{1}}
	\dots ~^{(D-2)}R^{r_{m-t}s_{m-t}}_{p_{m-t}q_{m-t}}\right)\left(\Theta ^{r_{m-t+1}}_{p_{m-t+1}}\Psi ^{s_{m-t+1}}_{q_{m-t+1}}\dots \Theta ^{r_{m}}_{p_{m}}\Psi ^{s_{m}}_{q_{m}}\right)~.\label{BT_LL-1-a}
\end{align}
Next, we simplify the second term in  \ref{BT_LL_1}:
\begin{align}
	\frac{2^{m+1}}{m+1}Q^{(1,b)}_{m+1}[\ell_c]\equiv&-16m\ell _{a}\ell ^{c}\delta \Psi ^{d}_{b}~\delta ^{abp_{1}q_{1}\dots p_{m}q_{m}}_{cdr_{1}s_{1}\dots r_{m}s_{m}}k_{p_{1}}\left(\ell ^{a_{1}}q^{b_{1}}_{q_{1}}q^{r_{1}}_{c_{1}}q^{s_{1}}_{d_{1}}R^{c_{1}d_{1}}_{a_{1}b_{1}}\right)
	\nonumber
	\\
	&\times \left(q^{a_{2}}_{p_{2}}q^{b_{2}}_{q_{2}}q^{r_{2}}_{c_{2}} q^{s_{2}}_{d_{2}}R^{c_{2}d_{2}}_{a_{2}b_{2}}\right)\dots \left(q^{a_{m}}_{p_{m}}q^{b_{m}}_{q_{m}}q^{r_{m}}_{c_{m}} q^{s_{m}}_{d_{m}}R^{c_{m}d_{m}}_{a_{m}b_{m}}\right)
	\nonumber
	\\
	=&-16m \ell _{a}\ell ^{c}k^{d}\delta \Psi ^{r_{1}}_{p_{1}}~\delta ^{abp_{1}q_{1}\dots p_{m}q_{m}}_{cdr_{1}s_{1}\dots r_{m}s_{m}}\left(q^{a_{2}}_{p_{2}}q^{b_{2}}_{q_{2}}q^{r_{2}}_{c_{2}} q^{s_{2}}_{d_{2}}R^{c_{2}d_{2}}_{a_{2}b_{2}}\right)\dots \left(q^{a_{m}}_{p_{m}}q^{b_{m}}_{q_{m}}q^{r_{m}}_{c_{m}} q^{s_{m}}_{d_{m}}R^{c_{m}d_{m}}_{a_{m}b_{m}}\right)
	\nonumber
	\\
	&\times \left(\ell _{c_{1}}q^{b_{1}}_{q_{1}}q^{a_{1}}_{b}q^{s_{1}}_{d_{1}}R^{c_{1}d_{1}}_{a_{1}b_{1}}\right)
	\nonumber
	\\
	=&32m \ell _{a}\ell ^{c}k^{d}\delta \Psi ^{r_{1}}_{p_{1}}~\delta ^{abp_{1}q_{1}\dots p_{m}q_{m}}_{cdr_{1}s_{1}\dots r_{m}s_{m}}\left(D_{b}\Theta ^{s_{1}}_{q_{1}}+q^{j}_{b}\Theta ^{s_{1}}_{q_{1}}~k^{p}\nabla _{j}\ell _{p} \right)
	\nonumber
	\\
	&\times \sum _{t=0}^{m-1}\Mycomb[m-1]{t}~4^{t}\left(~^{(D-2)}R^{r_{2}s_{2}}_{p_{2}q_{2}}
	\dots ~^{(D-2)}R^{r_{m-t}s_{m-t}}_{p_{m-t}q_{m-t}}\right)
	\nonumber
	\\
	&\phantom{\times \sum _{t=1}^{m}4^{t}\Mycomb[m-1]{t}}\times \left(\Theta ^{r_{m-t+1}}_{p_{m-t+1}}\Psi ^{s_{m-t+1}}_{q_{m-t+1}}\dots \Theta ^{r_{m}}_{p_{m}}\Psi ^{s_{m}}_{q_{m}}\right) \nn \\
	&+\dots~. \label{BT_LL-1-b}
\end{align}
To obtain the second line above, we have lowered all the upper indices and raised all the lower indices, while throwing away metric variations to the dots. We have also exchanged some indices using the symmetries of the determinant tensor. The last step was obtained making use of \ref{Gauss-Codazzi-main} and \ref{R_3q-main} and simplifying in a manner similar to \ref{BT_LL-2-b}.  
Using the same two equations, we can simplify the third term in \ref{BT_LL_1} as well. We have
\begin{align}
	\frac{2^{m+1}}{m+1}Q^{(1,c)}_{m+1}[\ell_c]\equiv&32\Mycomb[m]{2}~\ell _{a}\ell ^{c}\delta \Psi ^{d}_{b}~\delta ^{abp_{1}q_{1}p_{2}q_{2}\dots p_{m}q_{m}}_{cdr_{1}s_{1}r_{2}s_{2}\dots r_{m}s_{m}}\left(\ell ^{a_{1}}q^{b_{1}}_{q_{1}}q^{r_{1}}_{c_{1}}q^{s_{1}}_{d_{1}}R^{c_{1}d_{1}}_{a_{1}b_{1}}\right)k_{p_{1}}
	 \left(\ell _{c_{2}}q^{b_{2}}_{q_{2}}q^{a_{2}}_{p_{2}}q^{s_{2}}_{d_{2}}R^{c_{2}d_{2}}_{a_{2}b_{2}}\right)k^{r_{2}}
	\nonumber
	\\
	&\times \left(q^{a_{3}}_{p_{3}}q^{b_{3}}_{q_{3}}q^{r_{3}}_{c_{3}} q^{s_{3}}_{d_{3}}R^{c_{3}d_{3}}_{a_{3}b_{3}}\right)\dots \left(q^{a_{m}}_{p_{m}}q^{b_{m}}_{q_{m}}q^{r_{m}}_{c_{m}} q^{s_{m}}_{d_{m}}R^{c_{m}d_{m}}_{a_{m}b_{m}}\right)
	\nonumber
	\\
	=&-16m\left(m-1\right)\ell _{a}\ell ^{c}k_{b}k^{d}\delta \Psi ^{r_{1}}_{p_{1}}~\delta ^{abp_{1}q_{1}p_{2}q_{2}\dots p_{m}q_{m}}_{cdr_{1}s_{1}r_{2}s_{2}\dots r_{m}s_{m}}
	\left(\ell ^{a_{1}}q^{b_{1}}_{q_{1}}q^{r_{2}}_{c_{1}}q^{s_{1}}_{d_{1}}R^{c_{1}d_{1}}_{a_{1}b_{1}}\right)
	\left(\ell _{c_{2}}q^{b_{2}}_{q_{2}}q^{a_{2}}_{p_{2}}q^{s_{2}}_{d_{2}}R^{c_{2}d_{2}}_{a_{2}b_{2}}\right)
	\nonumber
	\\
	&\times \left(q^{a_{3}}_{p_{3}}q^{b_{3}}_{q_{3}}q^{r_{3}}_{c_{3}} q^{s_{3}}_{d_{3}}R^{c_{3}d_{3}}_{a_{3}b_{3}}\right)\dots \left(q^{a_{m}}_{p_{m}}q^{b_{m}}_{q_{m}}q^{r_{m}}_{c_{m}} q^{s_{m}}_{d_{m}}R^{c_{m}d_{m}}_{a_{m}b_{m}}\right)
	\nonumber
	\\
	=&-64m\left(m-1\right)\ell _{a}\ell ^{c}k_{b}k^{d}\delta \Psi ^{r_{1}}_{p_{1}}~\delta ^{abp_{1}q_{1}p_{2}q_{2}\dots p_{m}q_{m}}_{cdr_{1}s_{1}r_{2}s_{2}\dots r_{m}s_{m}}
	\left[D^{r_{2}}\Theta ^{s_{1}}_{q_{1}}+\left(k^{p}\nabla _{j}\ell _{p}\right)q^{jr_{2}}\Theta ^{s_{1}}_{q_{1}}\right]
	\nonumber
	\\
	&\times \left[D_{p_{2}}\Theta ^{s_{2}}_{q_{2}}+\left(k^{q}\nabla _{i}\ell _{q}\right)q^{i}_{p_{2}}\Theta ^{s_{2}}_{q_{2}}\right]
	\nonumber
	\\
	&\times \sum _{t=0}^{m-2}\Mycomb[m-2]{t}~4^{t}\left(~^{(D-2)}R^{r_{3}s_{3}}_{p_{3}q_{3}}
	\dots ~^{(D-2)}R^{r_{m-t}s_{m-t}}_{p_{m-t}q_{m-t}}\right)
	\left(\Theta ^{r_{m-t+1}}_{p_{m-t+1}}\Psi ^{s_{m-t+1}}_{q_{m-t+1}}\dots \Theta ^{r_{m}}_{p_{m}}\Psi ^{s_{m}}_{q_{m}}\right)~. \label{BT_LL-1-c}
\end{align}
We have used the symmetries of the determinant tensor to exchange some indices in the second step and have expanded the product as a sum in the last step as before.

Finally, we turn to the last term in \ref{BT_LL_1}. This can be simplified in the following manner:
\begin{align}
	\frac{2^{m+1}}{m+1}Q^{(1,d)}_{m+1}[\ell_c]\equiv&16m\ell _{a}\ell ^{c}\delta \Psi ^{d}_{b}~\delta ^{abp_{1}q_{1}\dots p_{m}q_{m}}_{cdr_{1}s_{1}\dots r_{m}s_{m}}\left(\ell ^{a_{1}} q^{b_{1}}_{q_{1}}\ell _{c_{1}}q^{s_{1}}_{d_{1}}R^{c_{1}d_{1}}_{a_{1}b_{1}}\right)k_{p_{1}}k^{r_{1}}
	\nonumber
	\\
	&\left(q^{a_{2}}_{p_{2}}q^{b_{2}}_{q_{2}}q^{r_{2}}_{c_{2}} q^{s_{2}}_{d_{2}}R^{c_{2}d_{2}}_{a_{2}b_{2}}\right)\dots \left(q^{a_{m}}_{p_{m}}q^{b_{m}}_{q_{m}}q^{r_{m}}_{c_{m}} q^{s_{m}}_{d_{m}}R^{c_{m}d_{m}}_{a_{m}b_{m}}\right)
	\nonumber
	\\
	=&16m \ell _{a}\ell ^{c}k_{b}k^{d}\delta \Psi ^{r_{1}}_{p_{1}}~\delta ^{abp_{1}q_{1}\dots p_{m}q_{m}}_{cdr_{1}s_{1}\dots r_{m}s_{m}}\left(-q^{s_{1}}_{i}q^{j}_{q_{1}}\pounds _{\ell}\Theta ^{i}_{j}+\kappa \Theta ^{s_{1}}_{q_{1}}-\Theta ^{s_{1}}_{i}\Theta ^{i}_{q_{1}}\right)
	\nonumber
	\\
	&\times \sum _{t=0}^{m-1}\Mycomb[m-1]{t}~4^{t}\left(~^{(D-2)}R^{r_{2}s_{2}}_{p_{2}q_{2}}
	\dots ~^{(D-2)}R^{r_{m-t}s_{m-t}}_{p_{m-t}q_{m-t}}\right)
	 \left(\Theta ^{r_{m-t+1}}_{p_{m-t+1}}\Psi ^{s_{m-t+1}}_{q_{m-t+1}}\dots \Theta ^{r_{m}}_{p_{m}}\Psi ^{s_{m}}_{q_{m}}\right)~, \label{BT_LL-1-d}
\end{align}
which was obtained using \ref{Gauss-Codazzi-main} and \ref{qlqlR-5-main}. This completes our job of writing all the relevant terms in the boundary variation by decomposing them using the projections of the Riemann tensor near the null surface given in \ref{Gauss-Codazzi-main}, \ref{R_3q-main} and \ref{qlqlR-5-main}. We will now regroup these terms appropriately.
\subsubsection{Grouping terms to obtain the total variation terms}

Thus, having written down the appropriate decomposition of all the terms, our next step would be to separate out the total variation terms while throwing away metric variations and their surface derivatives. For this purpose, it is advantageous to group certain terms together. First, note that the only terms involving the non-affinity parameter $\kappa$, which involves the derivatives of the metric along $k^a$, are in \ref{BT_LL-3} and \ref{BT_LL-1-d}. In \ref{BT_LL-3}, $\kappa$ appears as a variation, while in \ref{BT_LL-1-d} it appears outside the variation. Grouping these two terms together results in the following expression:
\begin{align}
	\frac{2^{m+1}}{m+1}&Q^{(3)}_{m+1}[\ell_c]+\frac{2^{m+1}}{m+1}Q^{(1,d)}_{m+1}[\ell_c]
	\nn \\
	=&4\delta \kappa ~\ell _{a}\ell ^{c}k_{b}k^{d}~\delta ^{abp_{1}q_{1}\dots p_{m}q_{m}}_{cdr_{1}s_{1}\dots r_{m}s_{m}}
	\nonumber
	\\
	&\times \sum _{t=0}^{m}\Mycomb[m]{t}~4^{t}\left(~^{(D-2)}R^{r_{1}s_{1}}_{p_{1}q_{1}}
	\dots ~^{(D-2)}R^{r_{m-t}s_{m-t}}_{p_{m-t}q_{m-t}}\right)\left(\Theta ^{r_{m-t+1}}_{p_{m-t+1}}\Psi ^{s_{m-t+1}}_{q_{m-t+1}}\dots \Theta ^{r_{m}}_{p_{m}}\Psi ^{s_{m}}_{q_{m}}\right)
	\nonumber
	\\
	&+16m \ell _{a}\ell ^{c}k_{b}k^{d}\delta \Psi ^{r_{1}}_{p_{1}}~\delta ^{abp_{1}q_{1}\dots p_{m}q_{m}}_{cdr_{1}s_{1}\dots r_{m}s_{m}}\left(-q^{s_{1}}_{i}q^{j}_{q_{1}}\pounds _{\ell}\Theta ^{i}_{j}+\kappa \Theta ^{s_{1}}_{q_{1}}-\Theta ^{s_{1}}_{i}\Theta ^{i}_{q_{1}}\right)
	\nonumber
	\\
	&\phantom{+}\times \sum _{t=0}^{m-1}\Mycomb[m-1]{t}~4^{t}\left(~^{(D-2)}R^{r_{2}s_{2}}_{p_{2}q_{2}}
	\dots ~^{(D-2)}R^{r_{m-t}s_{m-t}}_{p_{m-t}q_{m-t}}\right)
	\left(\Theta ^{r_{m-t+1}}_{p_{m-t+1}}\Psi ^{s_{m-t+1}}_{q_{m-t+1}}\dots \Theta ^{r_{m}}_{p_{m}}\Psi ^{s_{m}}_{q_{m}}\right)
	\nonumber
	\\
	=&4\delta \kappa ~\ell _{a}\ell ^{c}k_{b}k^{d}~\delta ^{abp_{1}q_{1}\dots p_{m}q_{m}}_{cdr_{1}s_{1}\dots r_{m}s_{m}}
	\nonumber
	\\
	&\times \sum _{t=0}^{m}\Mycomb[m]{t}~4^{t}\left(~^{(D-2)}R^{r_{1}s_{1}}_{p_{1}q_{1}}
	\dots ~^{(D-2)}R^{r_{m-t}s_{m-t}}_{p_{m-t}q_{m-t}}\right)\left(\Theta ^{r_{m-t+1}}_{p_{m-t+1}}\Psi ^{s_{m-t+1}}_{q_{m-t+1}}\dots \Theta ^{r_{m}}_{p_{m}}\Psi ^{s_{m}}_{q_{m}}\right)
	\nonumber
	\\
	&+4\kappa \ell _{a}\ell ^{c}k_{b}k^{d}\delta ^{abp_{1}q_{1}\dots p_{m}q_{m}}_{cdr_{1}s_{1}\dots r_{m}s_{m}}
	\times \sum _{t=0}^{m-1}m\Mycomb[m-1]{t}~4^{t+1}\left(~^{(D-2)}R^{r_{2}s_{2}}_{p_{2}q_{2}}
	\dots ~^{(D-2)}R^{r_{m-t}s_{m-t}}_{p_{m-t}q_{m-t}}\right)
	\nonumber
	\\
	&\phantom{4\kappa \ell _{a}\ell ^{c}k_{b}k^{d}\delta ^{abp_{1}q_{1}\dots p_{m}q_{m}}_{cdr_{1}s_{1}\dots r_{m}s_{m}}
		\times \sum _{t=0}^{m-1}m}
	\times\left(\Theta ^{s_{1}}_{q_{1}} \delta \Psi ^{r_{1}}_{p_{1}}\right)\left(\Theta ^{r_{m-t+1}}_{p_{m-t+1}}\Psi ^{s_{m-t+1}}_{q_{m-t+1}}\dots \Theta ^{r_{m}}_{p_{m}}\Psi ^{s_{m}}_{q_{m}}\right)
	\nonumber
	\\
	&+4\ell _{a}\ell ^{c}k_{b}k^{d}\delta \Psi ^{r_{1}}_{p_{1}}~\delta ^{abp_{1}q_{1}\dots p_{m}q_{m}}_{cdr_{1}s_{1}\dots r_{m}s_{m}}\left(-q^{s_{1}}_{i}q^{j}_{q_{1}}\pounds _{\ell}\Theta ^{i}_{j}-\Theta ^{s_{1}}_{i}\Theta ^{i}_{q_{1}}\right)
	\nonumber
	\\
	&\phantom{+}\times \sum _{t=0}^{m-1}m\Mycomb[m-1]{t}~4^{t+1}\left(~^{(D-2)}R^{r_{2}s_{2}}_{p_{2}q_{2}}
	\dots ~^{(D-2)}R^{r_{m-t}s_{m-t}}_{p_{m-t}q_{m-t}}\right)
	\left(\Theta ^{r_{m-t+1}}_{p_{m-t+1}}\Psi ^{s_{m-t+1}}_{q_{m-t+1}}\dots \Theta ^{r_{m}}_{p_{m}}\Psi ^{s_{m}}_{q_{m}}\right)
	\nonumber
	\\
	=&\delta \Bigg[4\kappa \ell _{a}\ell ^{c}k_{b}k^{d}\delta ^{abp_{1}q_{1}\dots p_{m}q_{m}}_{cdr_{1}s_{1}\dots r_{m}s_{m}}
	 \sum _{t=0}^{m}\Mycomb[m]{t}~4^{t}\left(~^{(D-2)}R^{r_{1}s_{1}}_{p_{1}q_{1}}
	\dots ~^{(D-2)}R^{r_{m-t}s_{m-t}}_{p_{m-t}q_{m-t}}\right) \nn \\
	&\phantom{\delta \Bigg[4\kappa \ell _{a}\ell ^{c}k_{b}k^{d}\delta ^{abp_{1}q_{1}\dots p_{m}q_{m}}_{cdr_{1}s_{1}\dots r_{m}s_{m}}
		\times \sum _{t=0}^{m}\Mycomb[m]{t}}\times\left(\Theta ^{r_{m-t+1}}_{p_{m-t+1}}\Psi ^{s_{m-t+1}}_{q_{m-t+1}}\dots \Theta ^{r_{m}}_{p_{m}}\Psi ^{s_{m}}_{q_{m}}\right)\Bigg]
	\nonumber
	\\
	&-\delta \Bigg[4\ell _{a}\ell ^{c}k_{b}k^{d}~\delta ^{abp_{1}q_{1}\dots p_{m}q_{m}}_{cdr_{1}s_{1}\dots r_{m}s_{m}}\left(q^{s_{1}}_{i}q^{j}_{q_{1}}\pounds _{\ell}\Theta ^{i}_{j}+\Theta ^{s_{1}}_{i}\Theta ^{i}_{q_{1}}\right)
	\nonumber
	\\
	&\quad \quad \times \sum _{t=0}^{m-1}\Mycomb[m]{t+1}~4^{t+1}\left(~^{(D-2)}R^{r_{2}s_{2}}_{p_{2}q_{2}}
	\dots ~^{(D-2)}R^{r_{m-t}s_{m-t}}_{p_{m-t}q_{m-t}}\right)
	 \left(\Theta ^{r_{m-t+1}}_{p_{m-t+1}}\Psi ^{s_{m-t+1}}_{q_{m-t+1}}\dots \Theta ^{r_{m}}_{p_{m}}\Psi ^{s_{m}}_{q_{m}}\right)\Psi ^{r_{1}}_{p_{1}}\Bigg] \nn \\
	&+\dots~, \label{Q_fin_1}
\end{align}
where we have separated out the total variation term by removing the metric variations and variations of the surface derivatives of the metric to the dots, using the facts that only $\kappa$ and $\Psi^{a}_b$ among the tensors present above have derivatives of the metric along $k^a$ and that a $\delta$ on any $\Psi^{a}_b$ is equivalent to a $\delta$ on any other $\Psi^{a}_b$ due to the symmetries of the determinant tensor. Also,
\begin{align}
&\delta ^{abp_{1}q_{1}\dots p_{m}q_{m}}_{cdr_{1}s_{1}\dots r_{m}s_{m}}
\delta \Bigg[\sum _{t=0}^{m}\Mycomb[m]{t}~4^{t}\left(~^{(D-2)}R^{r_{1}s_{1}}_{p_{1}q_{1}}
\dots ~^{(D-2)}R^{r_{m-t}s_{m-t}}_{p_{m-t}q_{m-t}}\right)
\left(\Theta ^{r_{m-t+1}}_{p_{m-t+1}}\Psi ^{s_{m-t+1}}_{q_{m-t+1}}\dots \Theta ^{r_{m}}_{p_{m}}\Psi ^{s_{m}}_{q_{m}}\right)\Bigg] \nn \\
&=\delta ^{abp_{1}q_{1}\dots p_{m}q_{m}}_{cdr_{1}s_{1}\dots r_{m}s_{m}}
\delta \Bigg[\sum _{t=1}^{m}\Mycomb[m]{t}~4^{t}\left(~^{(D-2)}R^{r_{1}s_{1}}_{p_{1}q_{1}}
\dots ~^{(D-2)}R^{r_{m-t}s_{m-t}}_{p_{m-t}q_{m-t}}\right)
\left(\Theta ^{r_{m-t+1}}_{p_{m-t+1}}\Psi ^{s_{m-t+1}}_{q_{m-t+1}}\dots \Theta ^{r_{m}}_{p_{m}}\Psi ^{s_{m}}_{q_{m}}\right)\Bigg]\nn\\
&\phantom{=}+\dots \nn \\
&=\delta ^{abp_{1}q_{1}\dots p_{m}q_{m}}_{cdr_{1}s_{1}\dots r_{m}s_{m}}
 \Bigg[\sum _{t=1}^{m}\Mycomb[m]{t}~4^{t}\left(~^{(D-2)}R^{r_{1}s_{1}}_{p_{1}q_{1}}
\dots ~^{(D-2)}R^{r_{m-t}s_{m-t}}_{p_{m-t}q_{m-t}}\right)
\delta\left(\Theta ^{r_{m-t+1}}_{p_{m-t+1}}\Psi ^{s_{m-t+1}}_{q_{m-t+1}}\dots \Theta ^{r_{m}}_{p_{m}}\Psi ^{s_{m}}_{q_{m}}\right)\Bigg]\nn\\
&\phantom{=}+\dots \nn\\
&=\delta ^{abp_{1}q_{1}\dots p_{m}q_{m}}_{cdr_{1}s_{1}\dots r_{m}s_{m}}
\Bigg[\sum _{t=1}^{m}\Mycomb[m]{t}~4^{t}t\left(~^{(D-2)}R^{r_{1}s_{1}}_{p_{1}q_{1}}
\dots ~^{(D-2)}R^{r_{m-t}s_{m-t}}_{p_{m-t}q_{m-t}}\right)
\left(\Theta ^{r_{m-t+1}}_{p_{m-t+1}}\Psi ^{s_{m-t+1}}_{q_{m-t+1}}\dots \Theta ^{r_{m}}_{p_{m}}\delta\Psi ^{s_{m}}_{q_{m}}\right)\Bigg]\nn\\
&\phantom{=}+\dots \nn\\
&=\delta ^{abp_{1}q_{1}\dots p_{m}q_{m}}_{cdr_{1}s_{1}\dots r_{m}s_{m}}
\Bigg[\sum _{t'=0}^{m-1}\Mycomb[m]{t'+1}~4^{t'+1}\left(t'+1\right)\left(~^{(D-2)}R^{r_{1}s_{1}}_{p_{1}q_{1}}
\dots ~^{(D-2)}R^{r_{m-t'-1}s_{m-t'-1}}_{p_{m-t'-1}q_{m-t'-1}}\right)\nn\\
&\phantom{=\delta ^{abp_{1}q_{1}\dots p_{m}q_{m}}_{cdr_{1}s_{1}\dots r_{m}s_{m}}
	\Bigg[\sum _{t'=0}^{m-1}\Mycomb[m]{t'+1}4^{t'+1}\left(t'+1\right)}\times\left(\Theta ^{r_{m-t'}}_{p_{m-t'}}\Psi ^{s_{m-t'}}_{q_{m-t'}}\dots \Theta ^{r_{m}}_{p_{m}}\delta\Psi ^{s_{m}}_{q_{m}}\right)\Bigg]\nn\\
&\phantom{=}+\dots \nn\\
&=\delta ^{abp_{1}q_{1}\dots p_{m}q_{m}}_{cdr_{1}s_{1}\dots r_{m}s_{m}}
\Bigg[\sum _{t'=0}^{m-1}m\Mycomb[m-1]{t'}~4^{t'+1}\left(\Theta^{s_{1}}_{q_{1}} \delta\Psi^{r_{1}}_{p_{1}} \right)\left(~^{(D-2)}R^{r_{2}s_{2}}_{p_{2}q_{2}}
\dots ~^{(D-2)}R^{r_{m-t'}s_{m-t'}}_{p_{m-t'}q_{m-t'}}\right)\nn\\
&\phantom{=\delta ^{abp_{1}q_{1}\dots p_{m}q_{m}}_{cdr_{1}s_{1}\dots r_{m}s_{m}}
	\Bigg[\sum _{t'=0}^{m-1}\Mycomb[m]{t'+1}4^{t'+1}\left(t'+1\right)}\times\left(\Theta ^{r_{m-t'+1}}_{p_{m-t'+1}}\Psi ^{s_{m-t'+1}}_{q_{m-t'+1}}\dots \Theta ^{r_{m}}_{p_{m}}\Psi ^{s_{m}}_{q_{m}}\right)\Bigg]\nn\\
&\phantom{=}+\dots~,
\end{align}
which is how the $\delta\Psi^{r_{1}}_{p_{1}}$ term appears from the first term in \ref{Q_fin_1}.

Progressing forward, let us consider the term presented in \ref{BT_LL-1-a}. One can convert this term to a total variation in a straightforward manner as follows:
\begin{align}
		\frac{2^{m+1}}{m+1}Q^{(1,a)}_{m+1}[\ell_c]=&4\ell _{a}\ell ^{c}\delta \Psi ^{d}_{b}~\delta ^{abp_{1}q_{1}\dots p_{m}q_{m}}_{cdr_{1}s_{1}\dots r_{m}s_{m}}
	\nonumber
	\\
	&\times \sum _{t=0}^{m}\Mycomb[m]{t}~4^{t}\left(~^{(D-2)}R^{r_{1}s_{1}}_{p_{1}q_{1}}
	\dots ~^{(D-2)}R^{r_{m-t}s_{m-t}}_{p_{m-t}q_{m-t}}\right)\left(\Theta ^{r_{m-t+1}}_{p_{m-t+1}}\Psi ^{s_{m-t+1}}_{q_{m-t+1}}\dots \Theta ^{r_{m}}_{p_{m}}\Psi ^{s_{m}}_{q_{m}}\right)
	\nonumber
	\\
	=&\delta \Bigg[4 \ell _{a}\ell ^{c}\Psi ^{d}_{b}~\delta ^{abp_{1}q_{1}\dots p_{m}q_{m}}_{cdr_{1}s_{1}\dots r_{m}s_{m}}
	\nonumber
	\\
	&\quad\times \sum _{t=0}^{m}4^{t}\frac{\Mycomb[m]{t}}{t+1}\left(~^{(D-2)}R^{r_{1}s_{1}}_{p_{1}q_{1}}
	\dots ~^{(D-2)}R^{r_{m-t}s_{m-t}}_{p_{m-t}q_{m-t}}\right)\left(\Theta ^{r_{m-t+1}}_{p_{m-t+1}}\Psi ^{s_{m-t+1}}_{q_{m-t+1}}\dots \Theta ^{r_{m}}_{p_{m}}\Psi ^{s_{m}}_{q_{m}}\right)\Bigg]\nn\\
	&+\dots~,\label{Q_fin_2}
\end{align}
where we have removed several terms involving variations of the metric and its surface derivatives into the dots. However, it turns out that the above variation is in fact zero. This is because if we expand the antisymmetric determinant tensor, the null normals $\ell_{a}$ and $\ell_{c}$ will be contracted either among themselves or with the other tensors present. All such contractions result in zero and thus we will receive a vanishing contribution from the above expression to the structure of the null boundary term.

For our next set of simplifications, we shall group the variations in \ref{BT_LL-2-a} and \ref{BT_LL-1-b} together. This leads to 
\begin{align}
	\frac{2^{m+1}}{m+1}&Q^{(2,a)}_{m+1}[\ell_c]+\frac{2^{m+1}}{m+1}Q^{(1,b)}_{m+1}[\ell_c]\nn\\
	=&8\ell _{a}\ell ^{c}k^{d}q^{j}_{b}\delta \left(k^{p}\nabla _{j}\ell_{p}\right)
	\delta ^{abp_{1}q_{1}\dots p_{m}q_{m}}_{cdr_{1}s_{1}\dots r_{m}s_{m}}
	\nonumber
	\\
	&\times \sum _{t=0}^{m}\Mycomb[m]{t}~4^{t}\left(~^{(D-2)}R^{r_{1}s_{1}}_{p_{1}q_{1}}
	\dots ~^{(D-2)}R^{r_{m-t}s_{m-t}}_{p_{m-t}q_{m-t}}\right)\left(\Theta ^{r_{m-t+1}}_{p_{m-t+1}}\Psi ^{s_{m-t+1}}_{q_{m-t+1}}\dots \Theta ^{r_{m}}_{p_{m}}\Psi ^{s_{m}}_{q_{m}}\right)
	\nonumber
	\\
	&+32m \ell _{a}\ell ^{c}k^{d}\delta \Psi ^{r_{1}}_{p_{1}}~\delta ^{abp_{1}q_{1}\dots p_{m}q_{m}}_{cdr_{1}s_{1}\dots r_{m}s_{m}}\left(D_{b}\Theta ^{s_{1}}_{q_{1}}+q^{j}_{b}\Theta ^{s_{1}}_{q_{1}}~k^{p}\nabla _{j}\ell _{p} \right)
	\nonumber
	\\
	&\times \sum _{t=0}^{m-1}\Mycomb[m-1]{t}~4^{t}\left(~^{(D-2)}R^{r_{2}s_{2}}_{p_{2}q_{2}}
	\dots ~^{(D-2)}R^{r_{m-t}s_{m-t}}_{p_{m-t}q_{m-t}}\right)
	\nonumber
	\\
	&\phantom{\times \sum _{t=1}^{m}4^{t}\Mycomb[m-1]{t}}\times \left(\Theta ^{r_{m-t+1}}_{p_{m-t+1}}\Psi ^{s_{m-t+1}}_{q_{m-t+1}}\dots \Theta ^{r_{m}}_{p_{m}}\Psi ^{s_{m}}_{q_{m}}\right)
	\nonumber
	\\
	=&\delta \Bigg[8\ell _{a}\ell ^{c}k^{d}q^{j}_{b}\left(k^{p}\nabla _{j}\ell _{p}\right)
	\delta ^{abp_{1}q_{1}\dots p_{m}q_{m}}_{cdr_{1}s_{1}\dots r_{m}s_{m}}
	\nonumber
	\\
	&\quad \times \sum _{t=0}^{m}\Mycomb[m]{t}~4^{t}\left(~^{(D-2)}R^{r_{1}s_{1}}_{p_{1}q_{1}}
	\dots ~^{(D-2)}R^{r_{m-t}s_{m-t}}_{p_{m-t}q_{m-t}}\right)\left(\Theta ^{r_{m-t+1}}_{p_{m-t+1}}\Psi ^{s_{m-t+1}}_{q_{m-t+1}}\dots \Theta ^{r_{m}}_{p_{m}}\Psi ^{s_{m}}_{q_{m}}\right)\Bigg]
	\nonumber
	\\
	&+\delta \Bigg[8\ell _{a}\ell ^{c}k^{d}\left(D_{b}\Theta ^{s_{1}}_{q_{1}}\right)
	\delta ^{abp_{1}q_{1}\dots p_{m}q_{m}}_{cdr_{1}s_{1}\dots r_{m}s_{m}}
	\nonumber
	\\
	&\quad\quad\times \sum _{t=0}^{m-1}\Mycomb[m]{t+1}~4^{t+1}\left(~^{(D-2)}R^{r_{2}s_{2}}_{p_{2}q_{2}}
	\dots ~^{(D-2)}R^{r_{m-t}s_{m-t}}_{p_{m-t}q_{m-t}}\right)
	\times \left(\Theta ^{r_{m-t+1}}_{p_{m-t+1}}\Psi ^{s_{m-t+1}}_{q_{m-t+1}}\dots \Theta ^{r_{m}}_{p_{m}}\Psi ^{s_{m}}_{q_{m}}\Psi ^{r_{1}}_{p_{1}}\right)\Bigg] \nn\\
	&+\dots~, \label{Q_fin_3}
\end{align}
where the dots again stand for terms with only variations of the metric and its surface derivatives. Here too, the derivatives along $k^a$ are present in only two of the tensors involved, $k^{p}\nabla _{a}\ell _{p}$ and $\Psi ^{a}_{b}$ being the tensors this time around. 

The final grouping that we need to do is of the terms in \ref{BT_LL-2-b} and \ref{BT_LL-1-c}. These terms are also dependent on $k^{p}\nabla _{a}\ell _{p}$, but in a quadratic manner. The simplification in this case involves a little more work, but the steps are analogous to those of our previous simplifications. The terms are
\begin{align}
	\frac{2^{m+1}}{m+1}&Q^{(2,b)}_{m+1}[\ell_c]+\frac{2^{m+1}}{m+1}Q^{(1,c)}_{m+1}[\ell_c]\nn\\
	=&-32m\ell _{a}\ell ^{c}k_{b}k^{d}\delta \left(q^{i}_{p_{1}}k^{p}\nabla _{i}\ell _{p}\right)
	~\delta ^{abp_{1}q_{1}p_{2}q_{2}\dots p_{m}q_{m}}_{cdr_{1}s_{1}r_{2}s_{2}\dots r_{m}s_{m}}
        \left\{D^{r_{1}}\Theta ^{s_{1}}_{q_{1}}+\left(k^{q}\nabla _{j}\ell _{q}\right)q^{jr_{1}}
	\Theta ^{s_{1}}_{q_{1}}\right\}
	\nonumber
	\\
	&~~\times \sum _{t=0}^{m-1}\Mycomb[m-1]{t}~4^{t}\left(~^{(D-2)}R^{r_{2}s_{2}}_{p_{2}q_{2}}
	\dots ~^{(D-2)}R^{r_{m-t}s_{m-t}}_{p_{m-t}q_{m-t}}\right)
	\left(\Theta ^{r_{m-t+1}}_{p_{m-t+1}}\Psi ^{s_{m-t+1}}_{q_{m-t+1}}\dots \Theta ^{r_{m}}_{p_{m}}\Psi ^{s_{m}}_{q_{m}}\right)
	\nonumber
	\\
	&-64m\left(m-1\right)\ell _{a}\ell ^{c}k_{b}k^{d}\delta \Psi ^{r_{1}}_{p_{1}}~\delta ^{abp_{1}q_{1}p_{2}q_{2}\dots p_{m}q_{m}}_{cdr_{1}s_{1}r_{2}s_{2}\dots r_{m}s_{m}}
	\left\{D^{r_{2}}\Theta ^{s_{1}}_{q_{1}}+\left(k^{p}\nabla _{i}\ell _{p}\right)q^{ir_{2}}\Theta ^{s_{1}}_{q_{1}}\right\} \nn\\
	&~~\times\left\{D_{p_{2}}\Theta ^{s_{2}}_{q_{2}}+\left(k^{q}\nabla _{j}\ell _{q}\right)q^{j}_{p_{2}}\Theta ^{s_{2}}_{q_{2}}\right\}
	\nonumber
	\\
	&~~\times \sum _{t=0}^{m-2}\Mycomb[m-2]{t}~4^{t}\left(~^{(D-2)}R^{r_{3}s_{3}}_{p_{3}q_{3}}
	\dots ~^{(D-2)}R^{r_{m-t}s_{m-t}}_{p_{m-t}q_{m-t}}\right)
	\times \left(\Theta ^{r_{m-t+1}}_{p_{m-t+1}}\Psi ^{s_{m-t+1}}_{q_{m-t+1}}\dots \Theta ^{r_{m}}_{p_{m}}\Psi ^{s_{m}}_{q_{m}}\right)
	\nonumber
	\\
	=&-32m\ell _{a}\ell ^{c}k_{b}k^{d}
	~\delta ^{abp_{1}q_{1}p_{2}q_{2}\dots p_{m}q_{m}}_{cdr_{1}s_{1}r_{2}s_{2}\dots r_{m}s_{m}}
	\left[\left(D^{r_{1}}\Theta ^{s_{1}}_{q_{1}}\right)\delta \left(q^{i}_{p_{1}}k^{p}\nabla _{i}\ell _{p}\right)+\Theta ^{s_{1}}_{q_{1}}q^{jr_{1}}\left(k^{q}\nabla _{j}\ell _{q}\right)
	\delta \left(q^{i}_{p_{1}}k^{p}\nabla _{i}\ell _{p}\right)\right]
	\nonumber
	\\
	&~~\times \sum _{t=0}^{m-1}\Mycomb[m-1]{t}~4^{t}\left(~^{(D-2)}R^{r_{2}s_{2}}_{p_{2}q_{2}}
	\dots ~^{(D-2)}R^{r_{m-t}s_{m-t}}_{p_{m-t}q_{m-t}}\right)
	\left(\Theta ^{r_{m-t+1}}_{p_{m-t+1}}\Psi ^{s_{m-t+1}}_{q_{m-t+1}}\dots \Theta ^{r_{m}}_{p_{m}}\Psi ^{s_{m}}_{q_{m}}\right)
	\nonumber
	\\
	&-64m\left(m-1\right)\ell _{a}\ell ^{c}k_{b}k^{d}~\delta ^{abp_{1}q_{1}p_{2}q_{2}\dots p_{m}q_{m}}_{cdr_{1}s_{1}r_{2}s_{2}\dots r_{m}s_{m}}
	 \nn\\
	&~~\times 
	\left[\left(D^{r_{2}}\Theta ^{s_{1}}_{q_{1}}\right)\left(D_{p_{2}}\Theta ^{s_{2}}_{q_{2}}\right)
	      +\left(k^{p}\nabla _{i}\ell _{p}\right)q^{ir_{2}}\Theta ^{s_{1}}_{q_{1}} \left(k^{q}\nabla _{j}\ell _{q}\right)q^{j}_{p_{2}}\Theta ^{s_{2}}_{q_{2}}  
	      +2 \left(D^{r_{2}}\Theta ^{s_{1}}_{q_{1}}\right)\left(k^{q}\nabla _{j}\ell _{q}\right)q^{j}_{p_{2}}\Theta ^{s_{2}}_{q_{2}} \right]\delta \Psi ^{r_{1}}_{p_{1}}
	\nonumber
	\\
	&~~\times \sum _{t=0}^{m-2}\Mycomb[m-2]{t}~4^{t}\left(~^{(D-2)}R^{r_{3}s_{3}}_{p_{3}q_{3}}
	\dots ~^{(D-2)}R^{r_{m-t}s_{m-t}}_{p_{m-t}q_{m-t}}\right)
	\times \left(\Theta ^{r_{m-t+1}}_{p_{m-t+1}}\Psi ^{s_{m-t+1}}_{q_{m-t+1}}\dots \Theta ^{r_{m}}_{p_{m}}\Psi ^{s_{m}}_{q_{m}}\right)
	\nonumber
	\\
	=&-\delta \Bigg[4m\ell _{a}\ell ^{c}k_{b}k^{d}~\delta ^{abp_{1}q_{1}p_{2}q_{2}\dots p_{m}q_{m}}_{cdr_{1}s_{1}r_{2}s_{2}\dots r_{m}s_{m}}\left(q^{i}_{p_{1}}k^{p}\nabla _{i}\ell _{p}\right)\left(q^{jr_{1}}k^{q}\nabla _{j}\ell _{q}\right)\Theta ^{s_{1}}_{q_{1}}
	\nonumber
	\\
	&\quad\times \sum _{t=0}^{m-1}\Mycomb[m-1]{t}~4^{t+1}\left(~^{(D-2)}R^{r_{2}s_{2}}_{p_{2}q_{2}}
	\dots ~^{(D-2)}R^{r_{m-t}s_{m-t}}_{p_{m-t}q_{m-t}}\right)\times \left(\Theta ^{r_{m-t+1}}_{p_{m-t+1}}\Psi ^{s_{m-t+1}}_{q_{m-t+1}}\dots \Theta ^{r_{m}}_{p_{m}}\Psi ^{s_{m}}_{q_{m}}\right)\Bigg]
	\nonumber
	\\
	&-\delta \Bigg[4m\ell _{a}\ell ^{c}k_{b}k^{d}~\delta ^{abp_{1}q_{1}p_{2}q_{2}\dots p_{m}q_{m}}_{cdr_{1}s_{1}r_{2}s_{2}\dots r_{m}s_{m}}\left(D_{p_{2}}\Theta ^{s_{2}}_{q_{2}}\right)\left(D^{r_{2}}\Theta ^{s_{1}}_{q_{1}}\right)
	\nonumber
	\\
	&\quad \quad \times \sum _{t=0}^{m-2}\Mycomb[m-1]{t+1}4^{t+2}\left(~^{(D-2)}R^{r_{3}s_{3}}_{p_{3}q_{3}}
	\dots ~^{(D-2)}R^{r_{m-t}s_{m-t}}_{p_{m-t}q_{m-t}}\right) \nn \\
	&\qquad \qquad \qquad \qquad \quad \; \;~ \times \left(\Theta ^{r_{m-t+1}}_{p_{m-t+1}}\Psi ^{s_{m-t+1}}_{q_{m-t+1}}\dots \Theta ^{r_{m}}_{p_{m}}\Psi ^{s_{m}}_{q_{m}}\Psi ^{r_{1}}_{p_{1}}\right)\Bigg]
	\nonumber
	\\
	&-\delta \Bigg[8m\ell _{a}\ell ^{c}k_{b}k^{d}~\delta ^{abp_{1}q_{1}p_{2}q_{2}\dots p_{m}q_{m}}_{cdr_{1}s_{1}r_{2}s_{2}\dots r_{m}s_{m}}\left(q^{j}_{p_{1}}k^{p}\nabla _{j}\ell _{p}\right)\left(D^{r_{1}}\Theta ^{s_{1}}_{q_{1}}\right)
	\nonumber
	\\
	&\quad\quad\times \sum _{t=0}^{m-1}\Mycomb[m-1]{t}~4^{t+1}\left(~^{(D-2)}R^{r_{2}s_{2}}_{p_{2}q_{2}}
	\dots ~^{(D-2)}R^{r_{m-t}s_{m-t}}_{p_{m-t}q_{m-t}}\right)\times \left(\Theta ^{r_{m-t+1}}_{p_{m-t+1}}\Psi ^{s_{m-t+1}}_{q_{m-t+1}}\dots \Theta ^{r_{m}}_{p_{m}}\Psi ^{s_{m}}_{q_{m}}\right)\Bigg]\nn\\
	&+\dots~, \label{Q_fin_4}
\end{align}
where we have again thrown away metric variations and their surface derivatives. In the second step, we have expanded out the products and used the result that the two cross terms arising in the product in the $\delta \Psi ^{r_{1}}_{p_{1}}$ term, i.e., the terms with $D^{r_{2}}\Theta ^{s_{1}}_{q_{1}}\left(k^{q}\nabla _{j}\ell _{q}\right)q^{j}_{p_{2}}\Theta ^{s_{2}}_{q_{2}}$ and $D_{p_{2}}\Theta ^{s_{2}}_{q_{2}}\left(k^{p}\nabla _{i}\ell _{p}\right)q^{ir_{2}}\Theta ^{s_{1}}_{q_{1}}$, are equivalent up to metric variations. This equivalence can be seen by using the symmetry properties of the determinant tensor first to exchange the indices on the $\Theta$s and then raising and lowering all indices present, while throwing away metric variations, to match the positions of the indices $p_2$ and $r_2$. As one can see, there are five different terms that occur after the expansion of the products. These are the terms containing the following expressions:
\begin{inparaenum}[i)]
\item $\left(D^{r_{1}}\Theta ^{s_{1}}_{q_{1}}\right)\delta \left(q^{i}_{p_{1}}k^{p}\nabla _{i}\ell _{p}\right)$, 
\item $\Theta ^{s_{1}}_{q_{1}}q^{jr_{1}}\left(k^{q}\nabla _{j}\ell _{q}\right)
\delta \left(q^{i}_{p_{1}}k^{p}\nabla _{i}\ell _{p}\right)$, 
\item $\left(D^{r_{2}}\Theta ^{s_{1}}_{q_{1}}\right)\left(D_{p_{2}}\Theta ^{s_{2}}_{q_{2}}\right)\delta \Psi ^{r_{1}}_{p_{1}}$, 
\item $\left(k^{p}\nabla _{i}\ell _{p}\right)q^{ir_{2}}\Theta ^{s_{1}}_{q_{1}} \left(k^{q}\nabla _{j}\ell _{q}\right)q^{j}_{p_{2}}\Theta ^{s_{2}}_{q_{2}}\delta \Psi ^{r_{1}}_{p_{1}}$ 
and finally
\item $ \left(D^{r_{2}}\Theta ^{s_{1}}_{q_{1}}\right)\left(k^{q}\nabla _{j}\ell _{q}\right)q^{j}_{p_{2}}\Theta ^{s_{2}}_{q_{2}}\delta \Psi ^{r_{1}}_{p_{1}}$
\end{inparaenum}. \\

Let us examine how these terms arise from the variations given in the last step of \ref{Q_fin_4}. The easiest variation to tackle in \ref{Q_fin_4} is the second one. Since $\left(D_{p_{2}}\Theta ^{s_{2}}_{q_{2}}\right)\left(D^{r_{2}}\Theta ^{s_{1}}_{q_{1}}\right)$ does not contain any derivatives along $k^a$, the only normal derivatives present in this term are in the $\Psi$s. There are $(t+1)$ $\Psi$s present, all of them being equivalent due to the symmetries of the determinant tensor. Thus, we obtain the term iii) in our list above, the one containing $\left(D^{r_{2}}\Theta ^{s_{1}}_{q_{1}}\right)\left(D_{p_{2}}\Theta ^{s_{2}}_{q_{2}}\right)\delta \Psi ^{r_{1}}_{p_{1}}$, from this variation. Next, we consider the last variation term in \ref{Q_fin_4}. Here, the derivatives along $k^a$ are present in $k^{p}\nabla _{j}\ell _{p}$ and the $\Psi$s. The variation acting on $k^{p}\nabla _{j}\ell _{p}$ gives us the term i) in our list above, while variation acting on the $\Psi$s produces the term v) in the list. We are left with the terms ii) and iv) in our list. These come from the first variation in \ref{Q_fin_4}, where again the derivatives along $k^a$ are present in $k^{p}\nabla _{j}\ell _{p}$, now present twice, and the $\Psi$s. Thus, we have verified that the variation terms in \ref{Q_fin_4} correctly provide all the terms we started with.
\subsubsection{Obtaining the boundary term}

We have obtained all the results that we require to write down an explicit expression for the null boundary term. Putting together \ref{Q_fin_1}, \ref{Q_fin_2}, \ref{Q_fin_3} and \ref{Q_fin_4} (where \ref{Q_fin_2} does not make any contribution to the boundary term, as we have stated before), and adding the factor of $\sqq$ back (which can be taken inside the $\delta$ at the cost of metric variation terms), we obtain the total structure of the boundary variation for Lanczos-Lovelock gravity as
\begin{align}
	\frac{2^{m+1}}{m+1}&\sqq Q_{m+1}[\ell_c] \nn\\
	\equiv&\frac{2^{m+1}}{m+1}\left(\sqq Q^{(1)}_{m+1}[\ell_c]+\sqq Q^{(2)}_{m+1}[\ell_c]+\sqq Q^{(3)}_{m+1}[\ell_c]\right)
	\nn \\
	=&-\delta \Bigg[8\sqq m\ell _{a}\ell ^{c}k_{b}k^{d}~\delta ^{abp_{1}q_{1}p_{2}q_{2}\dots p_{m}q_{m}}_{cdr_{1}s_{1}r_{2}s_{2}\dots r_{m}s_{m}}\left(q^{j}_{p_{1}}k^{p}\nabla _{j}\ell _{p}\right)\left(D^{r_{1}}\Theta ^{s_{1}}_{q_{1}}\right)
	\nonumber
	\\
	&\quad\quad\times \sum _{t=0}^{m-1}\Mycomb[m-1]{t}~4^{t+1}\left(~^{(D-2)}R^{r_{2}s_{2}}_{p_{2}q_{2}}
	\dots ~^{(D-2)}R^{r_{m-t}s_{m-t}}_{p_{m-t}q_{m-t}}\right)\times \left(\Theta ^{r_{m-t+1}}_{p_{m-t+1}}\Psi ^{s_{m-t+1}}_{q_{m-t+1}}\dots \Theta ^{r_{m}}_{p_{m}}\Psi ^{s_{m}}_{q_{m}}\right)\Bigg]
	\nn
	\\
	&-\delta \Bigg[4\sqq m\ell _{a}\ell ^{c}k_{b}k^{d}~\delta ^{abp_{1}q_{1}p_{2}q_{2}\dots p_{m}q_{m}}_{cdr_{1}s_{1}r_{2}s_{2}\dots r_{m}s_{m}}\left(D_{p_{2}}\Theta ^{s_{2}}_{q_{2}}\right)\left(D^{r_{2}}\Theta ^{s_{1}}_{q_{1}}\right)
	\nonumber
	\\
	&\quad \quad \times \sum _{t=0}^{m-2}\Mycomb[m-1]{t+1}4^{t+2}\left(~^{(D-2)}R^{r_{3}s_{3}}_{p_{3}q_{3}}
	\dots ~^{(D-2)}R^{r_{m-t}s_{m-t}}_{p_{m-t}q_{m-t}}\right) \nn \\
	&\qquad \qquad \qquad \qquad \quad \; \;~ \times \left(\Theta ^{r_{m-t+1}}_{p_{m-t+1}}\Psi ^{s_{m-t+1}}_{q_{m-t+1}}\dots \Theta ^{r_{m}}_{p_{m}}\Psi ^{s_{m}}_{q_{m}}\Psi ^{r_{1}}_{p_{1}}\right)\Bigg]
	\nonumber
	\\
	&-\delta \Bigg[4\sqq m\ell _{a}\ell ^{c}k_{b}k^{d}~\delta ^{abp_{1}q_{1}p_{2}q_{2}\dots p_{m}q_{m}}_{cdr_{1}s_{1}r_{2}s_{2}\dots r_{m}s_{m}}\left(q^{i}_{p_{1}}k^{p}\nabla _{i}\ell _{p}\right)\left(q^{jr_{1}}k^{q}\nabla _{j}\ell _{q}\right)\Theta ^{s_{1}}_{q_{1}}
	\nonumber
	\\
	&\quad\times \sum _{t=0}^{m-1}\Mycomb[m-1]{t}~4^{t+1}\left(~^{(D-2)}R^{r_{2}s_{2}}_{p_{2}q_{2}}
	\dots ~^{(D-2)}R^{r_{m-t}s_{m-t}}_{p_{m-t}q_{m-t}}\right)\times \left(\Theta ^{r_{m-t+1}}_{p_{m-t+1}}\Psi ^{s_{m-t+1}}_{q_{m-t+1}}\dots \Theta ^{r_{m}}_{p_{m}}\Psi ^{s_{m}}_{q_{m}}\right)\Bigg]
	\nonumber
	\\
	&+\delta \Bigg[8\sqq\ell _{a}\ell ^{c}k^{d}\left(D_{b}\Theta ^{s_{1}}_{q_{1}}\right)
	\delta ^{abp_{1}q_{1}\dots p_{m}q_{m}}_{cdr_{1}s_{1}\dots r_{m}s_{m}}
	\nonumber
	\\
	&\quad\quad\times \sum _{t=0}^{m-1}\Mycomb[m]{t+1}~4^{t+1}\left(~^{(D-2)}R^{r_{2}s_{2}}_{p_{2}q_{2}}
	\dots ~^{(D-2)}R^{r_{m-t}s_{m-t}}_{p_{m-t}q_{m-t}}\right)
	\times \left(\Theta ^{r_{m-t+1}}_{p_{m-t+1}}\Psi ^{s_{m-t+1}}_{q_{m-t+1}}\dots \Theta ^{r_{m}}_{p_{m}}\Psi ^{s_{m}}_{q_{m}}\Psi ^{r_{1}}_{p_{1}}\right)\Bigg]
	\nonumber
	\\
	&+\delta \Bigg[8\sqq\ell _{a}\ell ^{c}k^{d}q^{j}_{b}\left(k^{p}\nabla _{j}\ell _{p}\right)
	\delta ^{abp_{1}q_{1}\dots p_{m}q_{m}}_{cdr_{1}s_{1}\dots r_{m}s_{m}}
	\nonumber
	\\
	&\quad \times \sum _{t=0}^{m}\Mycomb[m]{t}~4^{t}\left(~^{(D-2)}R^{r_{1}s_{1}}_{p_{1}q_{1}}
	\dots ~^{(D-2)}R^{r_{m-t}s_{m-t}}_{p_{m-t}q_{m-t}}\right)\left(\Theta ^{r_{m-t+1}}_{p_{m-t+1}}\Psi ^{s_{m-t+1}}_{q_{m-t+1}}\dots \Theta ^{r_{m}}_{p_{m}}\Psi ^{s_{m}}_{q_{m}}\right)\Bigg]
	\nn
	\\
	&-\delta \Bigg[4\sqq\ell _{a}\ell ^{c}k_{b}k^{d}~\delta ^{abp_{1}q_{1}\dots p_{m}q_{m}}_{cdr_{1}s_{1}\dots r_{m}s_{m}}\left(q^{s_{1}}_{i}q^{j}_{q_{1}}\pounds _{\ell}\Theta ^{i}_{j}+\Theta ^{s_{1}}_{i}\Theta ^{i}_{q_{1}}\right)
	\nonumber
	\\
	&\quad \quad \times \sum _{t=0}^{m-1}\Mycomb[m]{t+1}~4^{t+1}\left(~^{(D-2)}R^{r_{2}s_{2}}_{p_{2}q_{2}}
	\dots ~^{(D-2)}R^{r_{m-t}s_{m-t}}_{p_{m-t}q_{m-t}}\right)
	\left(\Theta ^{r_{m-t+1}}_{p_{m-t+1}}\Psi ^{s_{m-t+1}}_{q_{m-t+1}}\dots \Theta ^{r_{m}}_{p_{m}}\Psi ^{s_{m}}_{q_{m}}\right)\Psi ^{r_{1}}_{p_{1}}\Bigg]
	\nonumber
	\\
	&+\delta \Bigg[4\sqq\kappa \ell _{a}\ell ^{c}k_{b}k^{d}\delta ^{abp_{1}q_{1}\dots p_{m}q_{m}}_{cdr_{1}s_{1}\dots r_{m}s_{m}}
	\sum _{t=0}^{m}\Mycomb[m]{t}~4^{t}\left(~^{(D-2)}R^{r_{1}s_{1}}_{p_{1}q_{1}}
	\dots ~^{(D-2)}R^{r_{m-t}s_{m-t}}_{p_{m-t}q_{m-t}}\right) \nn \\
	&\qquad \qquad \qquad \qquad \qquad \qquad \qquad \qquad \qquad \quad~ \times\left(\Theta ^{r_{m-t+1}}_{p_{m-t+1}}\Psi ^{s_{m-t+1}}_{q_{m-t+1}}\dots \Theta ^{r_{m}}_{p_{m}}\Psi ^{s_{m}}_{q_{m}}\right)\Bigg]+\dots~,
\end{align}
where all the terms which do not have variations of the derivatives of the metric along $k^a$ have been represented by the dots. From the above decomposition, we can straight away identify the boundary term for $m^{\textrm {th}}$ order \LL gravity [the above decomposition is for $(m+1)^{\textrm {th}}$ order] that will cancel the variations of non-tangential derivatives of the metric (derivatives in directions not tangent to the boundary) appearing in the boundary variation of the action. 

Grouping terms and arranging tensors according to the order of indices of the determinant tensor, we can write the expression for the null boundary term for $m^{\textrm {th}}$ order \LL gravity as
\begin{align}\label{BTLLF-pre}
\sqq&\mathcal{B}^{(m)}_{\textrm{\tiny LL}}[\ell_c]
\nn
\\
=&4\sqq\frac{m}{2^m}\delta ^{abp_{1}q_{1}p_{2}q_{2}\dots p_{m-1}q_{m-1}}_{cdr_{1}s_{1}r_{2}s_{2}\dots r_{m-1}s_{m-1}} \ell _{a}\ell ^{c}k^{d}
\nn 
\\
&\times \Bigg\{-\left[2\left(q^{i}_{b}k^{p}\nabla _{i}\ell _{p}\right)+\kappa k_b\right] \nn\\
&\qquad \phantom{+} \sum _{t=0}^{m-1}\Mycomb[m-1]{t}~4^{t}\left(~^{(D-2)}R^{r_{1}s_{1}}_{p_{1}q_{1}}
\dots ~^{(D-2)}R^{r_{m-t-1}s_{m-t-1}}_{p_{m-t-1}q_{m-t-1}}\right)\left(\Theta ^{r_{m-t}}_{p_{m-t}}\Psi ^{s_{m-t}}_{q_{m-t}}\dots \Theta ^{r_{m-1}}_{p_{m-1}}\Psi ^{s_{m-1}}_{q_{m-1}}\right)
\nonumber
\\
&\quad \quad+
  \sum _{t=0}^{m-2}  4^{t+1} \Big\{\left(m-1\right)k_{b}\left(q^{i}_{p_{1}}k^{p}\nabla _{i}\ell _{p}\right)\left[\left(q^{jr_{1}}k^{q}\nabla _{j}\ell _{q}\right)\Theta ^{s_{1}}_{q_{1}}+2 D^{r_{1}}\Theta ^{s_{1}}_{q_{1}}\right]\Mycomb[m-2]{t}\nn\\
  &\qquad \qquad \qquad \qquad \qquad \qquad \quad \quad+ \Psi ^{r_{1}}_{p_{1}}\left[k_{b}\left(q^{s_{1}}_{i}q^{j}_{q_{1}}\pounds _{\ell}\Theta ^{i}_{j}+\Theta ^{s_{1}}_{i}\Theta ^{i}_{q_{1}}\right)-2D_{b}\Theta ^{s_{1}}_{q_{1}}\right]\Mycomb[m-1]{t+1}\Big\} \nn\\
&\qquad\qquad\quad\phantom{+}\times  \left[\left(~^{(D-2)}R^{r_{2}s_{2}}_{p_{2}q_{2}}
\dots ~^{(D-2)}R^{r_{m-t-1}s_{m-t-1}}_{p_{m-t-1}q_{m-t-1}}\right)
\left(\Theta ^{r_{m-t}}_{p_{m-t}}\Psi ^{s_{m-t}}_{q_{m-t}}\dots \Theta ^{r_{m-1}}_{p_{m-1}}\Psi ^{s_{m-1}}_{q_{m-1}}\right)\right]
\nonumber
\\
&\quad \quad+\left(m-1\right)k_{b}~\Psi ^{r_{1}}_{p_{1}}\left(D^{r_{2}}\Theta ^{s_{1}}_{q_{1}}\right)\left(D_{p_{2}}\Theta ^{s_{2}}_{q_{2}}\right)
\nonumber
\\
& \quad \quad \phantom{+}~
 \sum _{t=0}^{m-3}\Mycomb[m-2]{t+1}~4^{t+2}\left(~^{(D-2)}R^{r_{3}s_{3}}_{p_{3}q_{3}}
\dots ~^{(D-2)}R^{r_{m-t-1}s_{m-t-1}}_{p_{m-t-1}q_{m-t-1}}\right)  \left(\Theta ^{r_{m-t}}_{p_{m-t}}\Psi ^{s_{m-t}}_{q_{m-t}}\dots \Theta ^{r_{m-1}}_{p_{m-1}}\Psi ^{s_{m-1}}_{q_{m-1}}\right)
 \Bigg\}.
\end{align}
We can simplify this result further. Let us examine the determinant tensor present:
\begin{align}
\delta ^{abp_{1}q_{1}p_{2}q_{2}\dots p_{m-1}q_{m-1}}_{cdr_{1}s_{1}r_{2}s_{2}\dots r_{m-1}s_{m-1}} 
&=-\delta ^{abp_{1}q_{1}p_{2}q_{2}\dots p_{m-1}q_{m-1}}_{dcr_{1}s_{1}r_{2}s_{2}\dots r_{m-1}s_{m-1}} \nn\\
&= -{\mathrm{det}} \left[ \begin{array}{c|ccccc}
\delta^a_d & \delta^a_{c} & \delta^a_{r_1} & \delta^a_{s_1} &\dots & \delta^a_{s_{m-1}} 
\\
\hline
\\
\delta^{b}_d &  &  & & & 
\\
 &  &  & & & 
\\
\delta^{p_{1}}_d &  &  & & & 
\\
\vdots & & & &\delta ^{bp_{1}q_{1}p_{2}q_{2}\dots p_{m-1}q_{m-1}}_{cr_{1}s_{1}r_{2}s_{2}\dots r_{m-1}s_{m-1}} & 
\\
\delta^{q_{m-1}}_d & & & & & \vphantom{\bigg{|}} \end{array} 
\right] \,~.\label{determinant_tensor-1}
\end{align}
Expanding the determinant along the first row, we can see that only the contribution from $\df^a_d$ survives in \ref{BTLLF-pre}, since all the other delta functions in the first row contract $\ell_a$ with such tensors as to give zero. Thus, effectively,
\begin{align}
\delta ^{abp_{1}q_{1}p_{2}q_{2}\dots p_{m-1}q_{m-1}}_{cdr_{1}s_{1}r_{2}s_{2}\dots r_{m-1}s_{m-1}}&\rightarrow-\delta^a_d \delta ^{bp_{1}q_{1}p_{2}q_{2}\dots p_{m-1}q_{m-1}}_{cr_{1}s_{1}r_{2}s_{2}\dots r_{m-1}s_{m-1}} \nn\\
&=-\delta^a_d\times{\mathrm{det}} \left[ \begin{array}{c|cccc}
\delta^b_c &  \delta^b_{r_1} & \delta^b_{s_1} &\dots & \delta^b_{s_{m-1}} 
\\
\hline
\\
\delta^{p_1}_c  &  & & & 
\\
&  &  & & 
\\
\delta^{q_{1}}_c &  &  & &  
\\
\vdots & & & \delta ^{p_{1}q_{1}p_{2}q_{2}\dots p_{m-1}q_{m-1}}_{r_{1}s_{1}r_{2}s_{2}\dots r_{m-1}s_{m-1}} & 
\\
\delta^{q_{m-1}}_c  & & & & \vphantom{\bigg{|}} \end{array} 
\right]~.
\end{align}
By making a similar argument but by expanding along the first column and looking at contractions of $\ell_c$, we can reduce this determinant further down to give
\begin{align}
\delta ^{abp_{1}q_{1}p_{2}q_{2}\dots p_{m-1}q_{m-1}}_{cdr_{1}s_{1}r_{2}s_{2}\dots r_{m-1}s_{m-1}}\rightarrow-\delta^a_d \delta^b_c\delta ^{p_{1}q_{1}p_{2}q_{2}\dots p_{m-1}q_{m-1}}_{r_{1}s_{1}r_{2}s_{2}\dots r_{m-1}s_{m-1}}~.
\end{align}
Substituting this back in \ref{BTLLF-pre}, we obtain
\begin{align}\label{BTLLF}
\sqq&\mathcal{B}^{(m)}_{\textrm{\tiny LL}}[\ell_c]
\nn
\\
=&4\sqq\frac{m}{2^m} \delta ^{p_{1}q_{1}p_{2}q_{2}\dots p_{m-1}q_{m-1}}_{r_{1}s_{1}r_{2}s_{2}\dots r_{m-1}s_{m-1}} 
\nn 
\\
&\times \Bigg\{\kappa \sum _{t=0}^{m-1}\Mycomb[m-1]{t}~4^{t}\left(~^{(D-2)}R^{r_{1}s_{1}}_{p_{1}q_{1}}
\dots ~^{(D-2)}R^{r_{m-t-1}s_{m-t-1}}_{p_{m-t-1}q_{m-t-1}}\right)\left(\Theta ^{r_{m-t}}_{p_{m-t}}\Psi ^{s_{m-t}}_{q_{m-t}}\dots \Theta ^{r_{m-1}}_{p_{m-1}}\Psi ^{s_{m-1}}_{q_{m-1}}\right)
\nonumber
\\
&\quad \quad-
\sum _{t=0}^{m-2}  4^{t+1} \Big\{\left(m-1\right)\left(q^{i}_{p_{1}}k^{p}\nabla _{i}\ell _{p}\right)\left[\left(q^{jr_{1}}k^{q}\nabla _{j}\ell _{q}\right)\Theta ^{s_{1}}_{q_{1}}+2 D^{r_{1}}\Theta ^{s_{1}}_{q_{1}}\right]\Mycomb[m-2]{t}\nn\\
&\qquad \qquad \qquad \qquad \qquad \qquad \qquad \quad \quad+ \Psi ^{r_{1}}_{p_{1}}\left[\left(q^{s_{1}}_{i}q^{j}_{q_{1}}\pounds _{\ell}\Theta ^{i}_{j}+\Theta ^{s_{1}}_{i}\Theta ^{i}_{q_{1}}\right)\right]\Mycomb[m-1]{t+1}\Big\} \nn\\
&\qquad\qquad\quad\phantom{+}\times  \left[\left(~^{(D-2)}R^{r_{2}s_{2}}_{p_{2}q_{2}}
\dots ~^{(D-2)}R^{r_{m-t-1}s_{m-t-1}}_{p_{m-t-1}q_{m-t-1}}\right)
\left(\Theta ^{r_{m-t}}_{p_{m-t}}\Psi ^{s_{m-t}}_{q_{m-t}}\dots \Theta ^{r_{m-1}}_{p_{m-1}}\Psi ^{s_{m-1}}_{q_{m-1}}\right)\right]
\nonumber
\\
&\quad \quad-\left(m-1\right)~\Psi ^{r_{1}}_{p_{1}}\left(D^{r_{2}}\Theta ^{s_{1}}_{q_{1}}\right)\left(D_{p_{2}}\Theta ^{s_{2}}_{q_{2}}\right)
\nonumber
\\
& \quad \quad \phantom{+}~
\sum _{t=0}^{m-3}\Mycomb[m-2]{t+1}~4^{t+2}\left(~^{(D-2)}R^{r_{3}s_{3}}_{p_{3}q_{3}}
\dots ~^{(D-2)}R^{r_{m-t-1}s_{m-t-1}}_{p_{m-t-1}q_{m-t-1}}\right)  \left(\Theta ^{r_{m-t}}_{p_{m-t}}\Psi ^{s_{m-t}}_{q_{m-t}}\dots \Theta ^{r_{m-1}}_{p_{m-1}}\Psi ^{s_{m-1}}_{q_{m-1}}\right)
\Bigg\}.
\end{align}
The object $-q^{i}_{j}k^{p}\nabla _{i}\ell _{p}$ appearing in these boundary terms is in fact the H\'a\'{\j}i\v{c}ek 1-form $\Omega_j$ \cite{hajivcek1973exact,Gourgoulhon:2005ng,Corichi:2016eoe}. The expression for the null boundary term as presented in \ref{BTLLF} is for $m^{\textrm th}$ order \LL gravity, but one can write down the boundary term for general \LL gravity by simply incorporating a summation over the \LL order $m$ with appropriate coefficients $c_{m}$. Thus, just by considering the variation of the \LL gravitational action on a null boundary, we have been able to determine the relevant null boundary term for \LL gravity. 

Note that we have not commented on the normal being future-directed or past-directed. This is because the normal is not a priori future-directed or past-directed in our conventions. In the conventions of \cite{Lehner:2016vdi}, where the null normal is defined to be future-directed, the boundary term expression will pick up an overall negative sign. This is explained in detail in \ref{app:surf_elem_conven}.

In \cite{Cano:2018aqi}, the late time growth of complexity for charged black holes in Lanczos-Lovelock gravity was calculated with the assumption that the null boundary terms are polynomials in the intrinsic and extrinsic geometric quantities of the null boundary. It was shown that such polynomial boundary terms do not contribute to the action at late times. Now that we have derived the complete expression for the boundary term, we can check whether their assumption is valid. The intrinsic and extrinsic geometric quantities of the null boundary that they used were taken from \cite{Vega:2011ue}. Comparing with our notation, we see that the correspondence is as follows: $\hat{\mathcal{R}}^{AB}_{CD}$ corresponds to our $~^{(D-2)}R^{ab}_{cd}$, $\mathcal{B}_{AB}$ to $\Theta_{ab}$, $\mathcal{K}_{ab}$ to $\Psi_{ab}$, $\omega_A$ to the H\'a\'{\j}i\v{c}ek 1-form $\Omega_a$ and $\kappa$ to our $\kappa$, up to overall signs and factors. But our boundary term does not appear to be completely polynomial in these objects, since it also contains derivatives of $\Theta_{ab}$. Hence, we are not sure if the conclusions in \cite{Cano:2018aqi} are still valid. This matter requires further exploration.

To check the correctness of this result, we shall evaluate this expression for the values of $m=1$ and $m=2$, which are the cases of \EH gravity and \GB gravity, respectively. (The case of $m=0$ gives zero, which is expected since the cosmological constant term does not need a boundary term.)
\subsection{Obtaining \EH and \GB null boundary terms from the general \LL result} \label{sec:LL_to_EH_GB}

In this section, we shall consider two specific cases of the general result in \ref{BTLLF}: one being the case of \EH action, for which the null boundary term is available in the literature \cite{Neiman:2012fx,Parattu:2015gga,Lehner:2016vdi}; and the other being \GB gravity, for which we have determined the null boundary term in \ref{sec:GB}. 
\subsubsection{$m=1$ Einstein-Hilbert case}

Consider substituting $m=1$ in \ref{BTLLF}. Of the three sums present in the equation, only the first one will survive. The last sum can be summarily dropped because of the $(m-1)$ factor. The second sum runs over positive integer values of $t$ from $0$ to $m-2$, which in this case would be $0$ to $-1$. This is to be interpreted as the sum not being able to run at all, giving a zero contribution. The first sum runs from $0$ to $m-1$, and thus will have only the single $t=0$ term in the present case. The determinant tensor has the indices $p_i, q_i, r_i$ and $s_i$ with $i$ running from $1$ to $m-1$, which is $1$ to $0$ for $m=1$. This means that the indices $p_i, q_i, r_i$ and $s_i$ will not be present, the determinant tensor in \ref{BTLLF} will not be present and all the $^{(D-2)}R^{r_{i}s_{i}}_{p_{i}q_{i}}$ and $\Theta ^{r_{i}}_{p_{i}}\Psi ^{s_{i}}_{q_{i}}$ will not be present. This does not mean that the determinant tensor has to be replaced by zero. Rather, we should replace it with unity. This is clearer if we look at the previous form in \ref{BTLLF-pre}, where the determinant tensor would reduce from $\delta ^{abp_{1}q_{1}p_{2}q_{2}\dots p_{m-1}q_{m-1}}_{cdr_{1}s_{1}r_{2}s_{2}\dots r_{m-1}s_{m-1}}$ to just $\delta ^{ab}_{cd}$. Further, $\Mycomb[m-1]{t}=\Mycomb[0]{0}=1$, using the usual formula for $\Mycomb[n]{k}$ with $0!=1$. Thus, the $t=0$ term reads
\begin{align}
\sqq\mathcal{B}^{(1)}_{\textrm{\tiny LL}}[\ell_c]
=2\sqq\kappa ~,
\end{align}
which matches with the expression for the null boundary term of the \EH action available in the literature \cite{Parattu:2015gga,Lehner:2016vdi}. Our definitions and conventions follow \cite{Parattu:2015gga}. The difference in choosing the direction of the null normal leads to the difference of a minus sign from \cite{Lehner:2016vdi}, as explained in \ref{app:surf_elem_conven}. We have assumed $\ell_a=A\nabla_a \phi$ and assumed that all $\phi=\textrm{constant}$ surfaces are null surfaces. Thus, our result can be directly compared with the result in Appendix G in the arXiv version of \cite{Parattu:2015gga}, but after imposing the additional assumption of $\ell^a \ell_a=0$ everywhere. The expression for the boundary term provided there is $2\left(\sqg/A\right)\left(\kappa+\Theta\right)$. From \ref{app:g_in_q}, $\left(\sqg/A\right)=\sqq$ and hence this reduces to $2\sqq\left(\kappa+\Theta\right)$ in our case. The $2\sqq\Theta$ term contains only derivatives of the metric along the null boundary, so it may be dropped, as has been done in \cite{Lehner:2016vdi}, if our aim is just to cancel off derivatives along $k^a$ from the boundary variation. So we are left with just $2\sqq\kappa$, which is our result above. [On the other hand, the unified boundary term in \cite{Parattu:2016trq,Jubb:2016qzt} reduces to $2\sqq\left(\kappa+\Theta\right)$ in the null case. The question of which of the boundary terms, the bare $2\sqq\Theta$ or the whole $2\sqq\left(\kappa+\Theta\right)$, is the correct boundary term requires further exploration. The derivation of the boundary term in this work has been performed with the aim of eliminating derivatives of the metric in directions off the null boundary only.] The additional assumption of $\ell^a \ell_a=0$ everywhere that we have assumed in this work just changes the expression for $\kappa$ from $\kappa = \ell^a \partial_a \ln A -\left(k^a/2\right) \partial_a\left(\ell^b \ell_b\right)$ to $\kappa = \ell^a \partial_a \ln A$.   
\subsubsection{$m=2$ Gauss-Bonnet case}

For the case of \GB gravity, we substitute $m=2$ in \ref{BTLLF}. The last sum in \ref{BTLLF} does not contribute because the sum goes over the invalid range of $0$ to $-1$. The other two sums do contribute. The first sum goes over the two values $t=0$ and $t=1$ and the second sum just consists of the $t=0$ term. From the first sum, we obtain
\begin{align}
\sqq\mathcal{B}^{(2,1)}_{\textrm{\tiny LL}}[\ell_c]
&=2\sqq\delta ^{p_{1}q_{1}}_{r_{1}s_{1}}
\kappa 
\left(~^{(D-2)}R^{r_{1}s_{1}}_{p_{1}q_{1}}+4\Theta ^{r_{1}}_{p_{1}}\Psi ^{s_{1}}_{q_{1}} \right) \nn\\
&=2\sqq \kappa \left(\delta ^{p_{1}}_{r_{1}}\delta ^{q_{1}}_{s_{1}}-\delta ^{p_{1}}_{s_{1}}\delta ^{q_{1}}_{r_{1}}\right) 
\left(~^{(D-2)}R^{r_{1}s_{1}}_{p_{1}q_{1}}+4\Theta ^{r_{1}}_{p_{1}}\Psi ^{s_{1}}_{q_{1}} \right) \nn\\
&=2\sqq\kappa \left(2~^{(D-2)}R+4\Theta \Psi -4\Theta _{ab}\Psi ^{ab}\right)~. \label{GB_1}
\end{align}
Moving over to the second sum, we have
\begin{align}
\sqq\mathcal{B}^{(2,2)}_{\textrm{\tiny LL}}[\ell_c]
=&2\sqq\delta ^{p_{1}q_{1}}_{r_{1}s_{1}}
\bigg\{-4\left(q^{i}_{p_{1}}k^{j}\nabla _{i}\ell _{j}\right)
\left[\left(q^{kr_{1}}k^{q}\nabla _{k}\ell _{q}\right)\Theta ^{s_{1}}_{q_{1}}
+2D^{r_{1}}\Theta ^{s_{1}}_{q_{1}}\right]
\nonumber
\\
&\qquad \qquad \qquad \qquad \qquad \quad ~~ -4  \Psi ^{r_{1}}_{p_{1}}
\left(q^{s_{1}}_{i}q^{j}_{q_{1}}\pounds _{\ell}\Theta ^{i}_{j}+\Theta ^{s_{1}}_{i}\Theta ^{i}_{q_{1}}\right)\bigg\}
\nonumber
\\
=& -8\sqq\left(\delta ^{p_{1}}_{r_{1}}\delta ^{q_{1}}_{s_{1}}-\delta ^{p_{1}}_{s_{1}}\delta ^{q_{1}}_{r_{1}}\right)\left(q^{i}_{p_{1}}k^{j}\nabla _{i}\ell _{j}\right)
\left[\left(q^{kr_{1}}k^{q}\nabla _{k}\ell _{q}\right)\Theta ^{s_{1}}_{q_{1}}
+2D^{r_{1}}\Theta ^{s_{1}}_{q_{1}}\right]
\nonumber
\\
&-8\sqq\left(\delta ^{p_{1}}_{r_{1}}\delta ^{q_{1}}_{s_{1}}-\delta ^{p_{1}}_{s_{1}}\delta ^{q_{1}}_{r_{1}}\right) \Psi ^{r_{1}}_{p_{1}}
\left(q^{s_{1}}_{i}q^{j}_{q_{1}}\pounds _{\ell}\Theta ^{i}_{j}+\Theta ^{s_{1}}_{i}\Theta ^{i}_{q_{1}}\right)\nn\\
=& -8\sqq\left[q^{ki}\left(k^{j}\nabla _{i}\ell _{j}\right)\left(k^{q}\nabla_{k}\ell _{q}\right)\Theta
+2\left(k^{j}\nabla _{i}\ell _{j}\right)D^{i}\Theta\right.\nn\\
&\left.
\qquad \quad~-\left(k^{j}\nabla _{i}\ell _{j}\right)\left(k^{q}\nabla _{k}\ell _{q}\right)\Theta^{ik}
-2\left(k^{j}\nabla _{i}\ell _{j}\right)D^{r}\Theta ^{i}_{r}\right]
\nonumber
\\
&-8\sqq \left[\Psi \left(q^{j}_{i}\pounds _{\ell}\Theta ^{i}_{j}+\Theta ^{j}_{i}\Theta ^{i}_{j}\right)-
\left(\Psi ^{j}_{i}\pounds _{\ell}\Theta ^{i}_{j}+\Psi ^{q}_{p}\Theta ^{p}_{i}\Theta ^{i}_{q}\right)\right]~.
\label{GB_2}
\end{align}
In order to compare this with \ref{BTGB}, we just need to make one more simplification. We have
\begin{align}
q^{j}_{i}\pounds _{\ell}\Theta ^{i}_{j} = \pounds _{\ell}\Theta-\Theta ^{i}_{j}\pounds _{\ell}q^{j}_{i}
&=\ell^a \nabla_a \Theta-\Theta ^{i}_{j}\left(\ell^a \nabla_a q^{j}_{i}+q^{j}_{a}\nabla_i \ell^a-q^{a}_{i}\nabla_a\ell^j\right) \nn\\
&=\ell^a \nabla_a \Theta-\left[\Theta ^{i}_{j}\ell^a \nabla_a \left(k^{j}\ell_{i}+\ell^{j}k_{i}\right)+\Theta ^{i}_{j}\Theta ^{j}_{i}-\Theta ^{i}_{j}\Theta ^{j}_{i}\right] \nn\\
&=\ell^a \nabla_a \Theta=\frac{d\Theta}{d\lambda}~, \label{simpl_qLT}
\end{align}
where $\lambda$ was defined below \ref{null_Raych}. Using \ref{simpl_qLT} and combining \ref{GB_1} and \ref{GB_2}, the complete $m=2$ null boundary term is
\begin{align}
\sqq\mathcal{B}^{(2)}_{\textrm{\tiny LL}}[\ell_c]
=&2\sqq\kappa \left(2~^{(D-2)}R+4\Theta \Psi -4\Theta _{ab}\Psi ^{ab}\right)\nn\\
& -8\sqq\left[q^{ki}\left(k^{j}\nabla _{i}\ell _{j}\right)\left(k^{q}\nabla_{k}\ell _{q}\right)\Theta
+2\left(k^{j}\nabla _{i}\ell _{j}\right)D^{i}\Theta\right.\nn\\
&\left.
\qquad \quad~-\left(k^{j}\nabla _{i}\ell _{j}\right)\left(k^{q}\nabla _{k}\ell _{q}\right)\Theta^{ik}
-2\left(k^{j}\nabla _{i}\ell _{j}\right)D^{r}\Theta ^{i}_{r}\right]
\nonumber
\\
&-8\sqq \left[ \Psi\frac{d\Theta}{d\lambda}+\Psi\Theta ^{j}_{i}\Theta ^{i}_{j}-
\Psi ^{j}_{i}\pounds _{\ell}\Theta ^{i}_{j}-\Psi ^{q}_{p}\Theta ^{p}_{i}\Theta ^{i}_{q}\right]~,
\end{align}
which can be seen to coincide with \ref{BTGB}, the null boundary term for \GB gravity we had derived earlier. 

Thus, the null boundary term presented in \ref{BTLLF} leads to the null boundary terms for both \EH gravity and \GB gravity for appropriate values of $m$. 
\section{Conclusion} \label{sec:conclusion}

\LL models of gravity, which includes general relativity, have actions that are not well-posed, and need boundary terms. Even though the boundary terms for \LL theories in the context of non-null surfaces have been known for some time, the null boundary terms were not available in the literature. In this work, we have filled this gap by providing appropriate null boundary terms. The framework and the assumptions that we have used are provided in \ref{sec:Proj_Riem_Intro}. The following are the assumptions that we have imposed:
\begin{inparaenum}[i)]
\item $B=1 \, ;$
\item $\ell^2=0$ everywhere, i.e. the null boundary is part of a null foliation;
\item $\df \ell^2=0$ and $\df k^2=0$ everywhere.
\end{inparaenum}
These assumptions have been made for simplifying the calculations. But we shall shortly argue that the results are not expected to change even if we remove these assumptions, enforcing the last two only on the null surface. We hope to show this explicitly in a future work.
 
We first derived a null boundary term for Gauss-Bonnet gravity in \ref{sec:GB}. This boundary term is the following integral over the null boundary:
\begin{align}\label{BTGB-concl}
\int_{\partial \mathcal{M}} &d\psi d^{d-2}z~\sqq~\mathcal{B}_{\textrm{\tiny GB}}\nn\\
=&\int_{\partial \mathcal{M}} d\psi d^{d-2}z~ 4\sqq\bigg[\kappa\left(~^{(D-2)}R-2 \Theta ^{p}_{q}\Psi ^{q}_{p}+2 \Theta \Psi\right) \nn\\
&\qquad \qquad \qquad \quad \quad~+2\Psi ^{p}_{q}\pounds _{\ell}\Theta ^{q}_{p}-2\Psi \ell^a \nabla_a \Theta +2\Theta ^{a}_{c}\Theta ^{c}_{b}\Psi ^{b}_{a} -2\Psi \Theta ^{a}_{b}\Theta ^{b}_{a}
\nonumber
\\
&\qquad \qquad \qquad \quad \quad~ -4\left(D^{m}\Theta -D^{a}\Theta ^{m}_{a}\right)\left(k^{q}\nabla _{m}\ell _{q}\right)
-2\left(k^{q}\nabla _{m}\ell _{q}\right)\left(k^{p}\nabla _{a}\ell_{p}\right)\left(\Theta q^{ma}
-\Theta ^{ma}\right)\bigg]~,
\end{align}
where $\psi$ is the scalar whose level surfaces give the auxiliary null foliation with $k_a=\nabla_a \psi$, $d^{d-2}z$ stands for $dz^1 dz^2\dots dz^{\left(D-2\right)}$ with the $z^A$ standing for the $D-2$ coordinates on the codimension-$2$ surfaces orthogonal to $k^a$ on the null boundary (see \ref{app:null_surf_elem} and \ref{app:g_in_q}). The definitions of the null surface objects involved are given in \ref{sec:Proj_Riem_Intro}. 

Then, in \ref{sec:gen_LL}, we derived the null boundary term for $m^{\textrm{th}}$ order \LL as
\begin{align}\label{BTLLF-concl}
\int_{\partial \mathcal{M}} &d\psi d^{d-2}z~\sqq \mathcal{B}^{(m)}_{\textrm{\tiny LL}}[\ell_c]
\nn
\\
=&\int_{\partial \mathcal{M}} d\psi d^{d-2}z~4\sqq\frac{m}{2^m} \delta ^{p_{1}q_{1}p_{2}q_{2}\dots p_{m-1}q_{m-1}}_{r_{1}s_{1}r_{2}s_{2}\dots r_{m-1}s_{m-1}} 
\nn 
\\
&\times \Bigg\{\kappa \sum _{t=0}^{m-1}\Mycomb[m-1]{t}~4^{t}\left(~^{(D-2)}R^{r_{1}s_{1}}_{p_{1}q_{1}}
\dots ~^{(D-2)}R^{r_{m-t-1}s_{m-t-1}}_{p_{m-t-1}q_{m-t-1}}\right)\left(\Theta ^{r_{m-t}}_{p_{m-t}}\Psi ^{s_{m-t}}_{q_{m-t}}\dots \Theta ^{r_{m-1}}_{p_{m-1}}\Psi ^{s_{m-1}}_{q_{m-1}}\right)
\nonumber
\\
&\quad \quad-
\sum _{t=0}^{m-2}  4^{t+1} \Big\{\left(m-1\right)\left(q^{i}_{p_{1}}k^{p}\nabla _{i}\ell _{p}\right)\left[\left(q^{jr_{1}}k^{q}\nabla _{j}\ell _{q}\right)\Theta ^{s_{1}}_{q_{1}}+2 D^{r_{1}}\Theta ^{s_{1}}_{q_{1}}\right]\Mycomb[m-2]{t}\nn\\
&\qquad \qquad \qquad \qquad \qquad \qquad \qquad \quad \quad+ \Psi ^{r_{1}}_{p_{1}}\left[\left(q^{s_{1}}_{i}q^{j}_{q_{1}}\pounds _{\ell}\Theta ^{i}_{j}+\Theta ^{s_{1}}_{i}\Theta ^{i}_{q_{1}}\right)\right]\Mycomb[m-1]{t+1}\Big\} \nn\\
&\qquad\qquad\quad\phantom{+}\times  \left[\left(~^{(D-2)}R^{r_{2}s_{2}}_{p_{2}q_{2}}
\dots ~^{(D-2)}R^{r_{m-t-1}s_{m-t-1}}_{p_{m-t-1}q_{m-t-1}}\right)
\left(\Theta ^{r_{m-t}}_{p_{m-t}}\Psi ^{s_{m-t}}_{q_{m-t}}\dots \Theta ^{r_{m-1}}_{p_{m-1}}\Psi ^{s_{m-1}}_{q_{m-1}}\right)\right]
\nonumber
\\
&\quad \quad-\left(m-1\right)~\Psi ^{r_{1}}_{p_{1}}\left(D^{r_{2}}\Theta ^{s_{1}}_{q_{1}}\right)\left(D_{p_{2}}\Theta ^{s_{2}}_{q_{2}}\right)
\nonumber
\\
& \quad \quad \phantom{+}~
\sum _{t=0}^{m-3}\Mycomb[m-2]{t+1}~4^{t+2}\left(~^{(D-2)}R^{r_{3}s_{3}}_{p_{3}q_{3}}
\dots ~^{(D-2)}R^{r_{m-t-1}s_{m-t-1}}_{p_{m-t-1}q_{m-t-1}}\right)  \left(\Theta ^{r_{m-t}}_{p_{m-t}}\Psi ^{s_{m-t}}_{q_{m-t}}\dots \Theta ^{r_{m-1}}_{p_{m-1}}\Psi ^{s_{m-1}}_{q_{m-1}}\right)
\Bigg\}.
\end{align}
with $-q^{i}_{j}k^{p}\nabla _{i}\ell _{p}$ being the H\'a\'{\j}i\v{c}ek 1-form $\Omega_j$ \cite{hajivcek1973exact,Gourgoulhon:2005ng,Corichi:2016eoe}. We have checked that this boundary term reduces to the corresponding expressions for general relativity and \GB in \ref{sec:LL_to_EH_GB}. These expressions are derived in the conventions specified in our earlier paper \cite{Parattu:2015gga}. To convert to the conventions followed in \cite{Lehner:2016vdi}, one just needs to append an overall minus sign to the boundary terms, as explained in \ref{app:surf_elem_conven}.

Let us briefly discuss whether the above results depend on the assumptions we had imposed. A review of the calculations suggests that lifting the assumption $B=1$ will only lead to the minor change of $\sqq$ being replaced by $B\sqq$ in the boundary terms. The remaining assumptions are $\ell^2=0$, $\df \ell^2=0$ and $\df k^2=0$ everywhere. But there are multiple reasons to expect that restricting these assumptions to be valid only on the null surface is enough to reproduce our results. Firstly, \cite{Lehner:2016vdi} reproduced the result in \cite{Parattu:2015gga} for the \EH null boundary term using a framework claimed to involve only tangential derivatives. We have followed \cite{Parattu:2015gga} here, but we expect the framework used in \cite{Lehner:2016vdi} to reproduce our results, just as in the \EH case. Then, the behaviour of $\ell^2$, $\df \ell^2$ and $\df k^2$ off the null surface should not matter. Secondly, our boundary terms contains all the intrinsic and extrinsic tensors associated with a null surface as given in \cite{Vega:2011ue}. Explicitly, $\gamma_{AB}$ corresponds to our $q_{ab}$, $\hat{\mathcal{R}}^{AB}_{CD}$ to our $~^{(D-2)}R^{ab}_{cd}$,  $\mathcal{B}_{AB}$ to $\Theta_{ab}$, $\mathcal{K}_{ab}$ to $\Psi_{ab}$, $\omega_A$ to the H\'a\'{\j}i\v{c}ek 1-form $\Omega_a$ and $\kappa$ to our $\kappa$, up to overall signs and factors. If these are indeed all the extrinsic and intrinsic tensors associated with a null surface, we do not expect any additional contribution to arise if restrict the assumptions of $\ell^2=0$, $\df \ell^2=0$ and $\df k^2=0$ to the null surface. Finally, we have done a rapid calculation of the Gauss-Bonnet boundary term lifting the assumption of $B=1$ and enforcing $\ell^2=0$, $\df \ell^2=0$ and $\df k^2=0$ only on the null surface. The result was in accordance with the above arguments. Thus, we have strong reasons to believe that our boundary terms are valid in a more general setting. We hope to present a careful calculation of the \GB and \LL boundary terms to verify our claims above, along with a unified treatment of spacelike, timelike and null boundary terms for \LL gravity, elsewhere.

The derivation presented here opens up several directions of exploration. We have not kept track of the terms with variations of the metric or variations of the surface derivatives of the metric on the boundary, unlike in our earlier papers \cite{Parattu:2015gga,Parattu:2016trq,Chakraborty:2017zep}. These terms can be partitioned into a Dirichlet variation term and a surface total derivative term. From the Dirichlet variation term, one can discover the degrees of freedom to be fixed on the null boundary for \LL models of gravity, the conjugate momenta corresponding to these degrees of freedom, the associated Poisson bracket structure and route for quantization. The surface total derivative term in the variational principle is also of importance, since the structure of corner terms associated with a null boundary can be derived from this term \cite{Lehner:2016vdi}. (One can also derive the corner terms from the boundary term by obtaining the corners as the limit of smooth portions of the boundary \cite{Hayward:1993my}. See \cite{Ruan:2017tkr} for a comparison of the three methods of finding corner terms, with the claim that they are equivalent.) These corner terms as well as the null boundary terms are important in checking the ``complexity=action'' conjecture for \LL theories (see \cite{Lehner:2016vdi} where the null boundary term for \EH action was used to check the conjecture for general relativity). This is perhaps the one of most current interest among the possible applications of our results. In \cite{Cano:2018ckq}, the \LL corner terms for non-null boundaries were derived, and it was argued that these terms can be extended in a straightforward manner to the case of corners on null boundaries. Our derivation here provides us with the resources to check this claim explicitly. The claim was further used to calculate the late time growth of complexity for charged black holes in \LL gravity in \cite{Cano:2018aqi}. With the assumption that the null boundary terms are polynomials in extrinsic and intrinsic geometric quantities of the null surface, it is demonstrated that they do not contribute to the action. As our result explicitly demonstrates, there are also non-polynomial terms present in the null boundary terms for \LL gravity. Thus, it will be interesting to check whether the null boundary term contribution indeed vanishes at late times. Also, these results can presumably be extended to scenarios where the null boundary term is of importance, now that we have derived the appropriate null boundary terms. There are many other applications that we can imagine, and we hope the community puts our results to fruitful use.
\section*{Acknowledgements}

Research of SC is funded by the INSPIRE Faculty Fellowship (Reg. No. DST/INSPIRE/04/2018/000893) from Department of Science and Technology, Government of India. Research of KP is supported by the SERB-NPDF grant (No. PDF/2017/002782), from DST, Government of India, the DGAPA postdoctoral fellowship from UNAM and the Fondecyt Postdoctoral fellowship (No. 3180421) from the Government of Chile. KP would like to thank Perimeter Institute for kind hospitality during a stay, discussions in which duration led to this project. SC thanks IIT Gandhinagar and Albert Einstein Institute, Golm, where parts of this work were done, for warm hospitality. SC and KP would like to thank T. Padmanabhan, K. Lochan, Dean Carmi and Pratik Rath for useful discussions. We would also like to thank the referee for his/her comments which have helped to improve the manuscript.
\appendix

\labelformat{section}{Appendix #1}
\labelformat{subsection}{Appendix #1}
\labelformat{subsubsection}{Appendix #1}
\section*{Appendices}
\section{Projections of the Riemann tensor near a null surface}\label{app:R_decomp_null}

In this appendix, we shall derive the expressions of the projections of the Riemann tensor near a null surface, which will be necessary for our work. We shall start by introducing the relevant geometrical quantities near the null surface.

Let us consider a null surface in a $D$-dimensional spacetime represented by $\phi(x)=\phi_0$, for some scalar function $\phi(x)$ and some $\phi_0=\textrm{constant}$. Other $\phi=\textrm{constant}$ surfaces need not be null, i.e., the null surface under consideration need not be the member of a null foliation. (We shall assume in the main text that all $\phi=\textrm{constant}$ surfaces are null. But the results in this appendix are derived without this assumption.) A null normal to the surface will have the form $\ell_a=A(x)\nabla_{a}\phi(x)$, for some other scalar function $A(x)$. Unlike the case for a non-null surface, the normal to a null surface cannot be normalized to unity. Hence, without a criterion for choosing the scaling, we consider all values of $A$ to be equally valid. This is the reason we talk about `a' null normal rather than `the' null normal. (Note that, while we may always \textit{choose} the normal vector field as $\ell_a=\nabla_{a}\phi$, it is not possible, in general, to start with an $\ell_a=A\nabla_{a}\phi$ and choose a different scalar field $\phi'$ to obtain $\ell_a=\nabla_{a}\phi'$. This is proved in \ref{app:need-of-A}. We want our results to be valid for any choice of the null normal and hence we will work with a general $A$.) 
\subsection{Induced metric}

The \textit{induced metric} of a hypersurface embedded in spacetime is derived from the metric of the spacetime by restricting its action to vectors tangent to the hypersurface. Although, strictly speaking, the induced metric acts on the lower-dimensional tangent space of the hypersurface, we can construct an `induced metric' for the full manifold---a covariant $2$-tensor acting on the tangent space of the whole manifold---by demanding that it annihilates the components of a vector normal to the hypersurface and coincides with the induced metric in its action on the components tangential to the hypersuface. We also take it to be a projector. This object is also called the \textit{orthogonal projector} on the hypersurface. For a non-null hypersurface, the orthogonal projector is
\begin{align}
	h_{ab}=g_{ab}-\epsilon n_a n_b~,
\end{align}
where $n_a$ is the normalized normal and $\epsilon$ is $-1$ or $+1$ depending on whether the surface is timelike or spacelike. (The timelike case is more often found in textbooks, see \cite{gravitation, gourgoulhon20123+}.) In this paper, we shall use the term `induced metric' to mean the orthogonal projector. If we have occasion to use the lower-dimensional version, we shall make it amply clear.      

Since one cannot construct a normalized normal for the null surface, it is not possible to define an induced metric in the above manner. The object $i_{ab}=g_{ab}+\lambda \ell_a \ell_b$ would not work for any finite $\lambda$ since this is not a projector:
\begin{equation}
	i^a_b i^b_c= \left(\df^a_b+\lambda \ell^a \ell_b\right) \left(\df^b_c+\lambda \ell^b \ell_c\right)=\df^a_c+2\lambda \ell^a \ell_c \neq i^a_c~.
\end{equation} 
The standard solution to the above problem is as follows: we choose an auxiliary null vector $k^a$ satisfying the relation $k^a\ell_a=-1$ (\cite{Carter:1979,Poisson}, also \sap). Then, the induced metric on the null hypersurface is chosen as the object
\begin{equation}\label{q_def}
	q_{ab}\equiv g_{ab}+\ell_a k_b+k_a \ell_b~.
\end{equation}
Once can check that this is a projector. Note that it is orthogonal to both $\ell^a$ and $k^a$, and hence its restriction to the null surface is a codimension-$2$ metric operating on the codimension-2 vector space orthogonal to $k^a$ and tangent to the null surface. In the case where these codimension-$2$ vector spaces may be stitched together to form the tangent spaces of codimension-2 surfaces lying on the null surface, it is the metric on those codimension-2 surfaces. Following \cite{Wald}, a metric is a symmetric and non-degenerate tensor of type $(0,2)$. In the null case, we cannot define a non-degenerate metric on the null surface because of the presence of a null direction on the surface. But we can have a non-degenerate metric when we eliminate the null direction and focus on codimension-$2$ surfaces on the null surface. We will soon choose $k^a$ to be hypersurface-orthogonal in order to define connection and curvature on the null surface without introducing torsion.
\subsection{Induced connection and covariant derivative} \label{appsubsec:ind_met}

For an arbitrary tensor $T^{ab\dots}_{cd\dots}$ that gives zero on contracting any index, up or down, with $\bm{\ell}$ or $\mathbf{k}$, the intrinsic covariant derivative $D$ on these codimension-$2$ surfaces may be defined as
\begin{equation}\label{toy_examp}
	D_aT^{bc\dots}_{de\dots}\equiv q^i_a q^b_j q^c_k  q_d^l q_e^m \dots \nabla_{i}T^{jk\dots}_{lm\dots}~,
\end{equation} 
analogous to the non-null case \cite{gravitation}. Here $\nabla$ is the usual covariant derivative of the parent manifold. To see whether the covariant derivative operator $D$ defined in \ref{toy_examp} is indeed a valid one, let us check each of the five criteria given in \cite{Wald} when operating on tensors orthogonal to $\bm{\ell}$ or $\mathbf{k}$. These five conditions are
\begin{enumerate}
	
	\item Linearity: This covariant derivative operator $D$ is certainly linear.
	
	\item Leibnitz rule: $D$ satisfies the Leibnitz rule, as can be easily checked using its definition.
	
	\item Commutativity with contraction: The operator $D$ also commutes with contraction, i.e., contraction with delta function of the space orthogonal to $k^a$ on the null surface. In working with indices of the whole manifold, this delta function is represented by $q^a_b=\delta^a_b + k^a \ell_b +\ell^a k_b$, where the extra terms are precisely to get rid of the components of the tensor not along the codimension-2 surfaces. [This can be easily seen using the canonical null basis and its dual basis (\sap), suitably extended to $D$ dimensions.] But note that we have already restricted ourselves to tensors orthogonal to $\bm{\ell}$ or $\mathbf{k}$. For such tensors, $\df^a_b$ and $q^a_b$ are equivalent in considering commutativity with contraction.
	
	\item Consistency with the notion of tangent vectors as directional derivatives on scalar fields: For a vector $t^a$ tangent to the codimension-$2$ surfaces on the null surface (i.e., such that $\ell_{a}t^{a}=0$ and $t_{a}k^{a}=0$), we have, for a scalar field $f$, the following relation:
	\begin{align}
		t^a D_a f= t^a q_a^b \nabla_b f= t^a \nabla_a f= t^a \partial_a f~.
	\end{align}
	Thus, $D_{a}$ indeed acts as a directional derivative when restricted to directions orthogonal to both $\ell^a$ and $k^a$.
	
	\item Torsion free: For all scalars $f$, we should have $D_a D_b f= D_b D_a f$. In the present context, we have
	\begin{align}
		D_a D_b f &=q^c_a q^d_b \nabla_c \left(D_d f\right) =q^c_a q^d_b \nabla_c \left(q^e_d \nabla_e f\right)
		=q^c_a q^d_b q^e_d\nabla_c \nabla_e f + q^c_a q^d_b \nabla_e f\nabla_c q^e_d \nn \\
		&=q^c_a q^e_b\nabla_c \nabla_e f + q^c_a q^d_b \nabla_e f\nabla_c \left(k^e \ell_d+\ell^e k_d\right) \nn \\
		&=q^c_a q^e_b\nabla_c \nabla_e f + q^c_a q^d_b \nabla_e f \left(k^e \nabla_c\ell_d+\ell^e \nabla_c k_d\right)~.	
	\end{align}
	Thus,
	\begin{align}
		D_a D_b f-D_b D_a f&= q^c_a q^d_b \left[k^e\nabla_e f \left(\nabla_c\ell_d-\nabla_d\ell_c\right)+\left(\ell^e\nabla_e f\right) \left(\nabla_c k_d-\nabla_d k_c\right)\right] 
		\nn
		\\
		&= k^e\nabla_e f \left(\Theta_{ab}-\Theta_{ba}\right)+\left(\ell^e\nabla_e f\right) q^c_a q^d_b \left(\nabla_c k_d-\nabla_d k_c\right)~,
	\end{align}
	where we have used the expression for the second fundamental form on the null surface,
	\begin{equation}\label{theta_def}
		\Theta_{cd}\equiv q^m_c q^k_d \nabla_m \ell_k~.
	\end{equation}
	In an identical manner, it is possible to define the corresponding quantity associated with the auxiliary null vector $k^a$ as
	\begin{equation}\label{psi_def}
		\Psi_{cd}\equiv q^m_c q^n_d \nabla_m k_n~.
	\end{equation}
	Using $\ell_a=A \nabla_a \phi$, one can prove that $\Theta_{ab}$ is symmetric (\sap). So we obtain
	\begin{align}
		D_a D_b f-D_b D_a f&= \ell^e\nabla_e f \left(\Psi_{cd}-\Psi_{dc}\right)~.
	\end{align}
	This will be zero for arbitrary $f$ only if $\Psi_{ab}$ is also symmetric. This can be achieved if we take $k_a$ to be hypersurface-orthogonal, i.e.,
	\begin{equation}
		k_a=B \nabla_a \psi~,
	\end{equation}
	with $B$ and $\psi$ being two scalar functions. (The proof that $\Theta_{cd}$ is symmetric automatically tells us that $\Psi_{cd}$ would also be symmetric, since $q_{ab}$ is symmetric with respect to an interchange of $\ell$ and $k$.) Since we do not intend to deal with torsion in this paper, we shall make the assumption that $k_a$ is also hypersurface-orthogonal. In this case, there are two foliations, represented by the scalars $\phi$ and $\psi$ with normals $\ell_a$ and $k_a$, respectively. A particular member of the $\phi$-foliation, say $\phi=\phi_0$, is null while the auxiliary foliation due to the scalar $\psi$ is taken to be composed of all null surfaces. In fact, the auxiliary foliation can be constructed by sending out null rays along $k^a$ from our fiducial null surface (see the description of Gaussian null coordinates \cite{Moncrief:1983,Friedrich:1998wq,Racz:2007pv,Morales} in Appendix C of \cite{Parattu:2015gga}). The object $q_{ab}$ then represents the induced metric on the codimension-$2$ surfaces formed by the intersection of the members of the $\psi$-foliation with the $\phi=\phi_0$ null surface.
	
\end{enumerate}

\subsection{Intrinsic (Gaussian) curvature}

Using the above definition of the induced covariant derivative and any vector $X^a$ satisfying $X^a\ell_a=0$ and $X^ak_a=0$, we can define the intrinsic Riemann tensor on the codimension-$2$ surfaces orthogonal to $k^a$ on the null surface, following \cite{Wald} and in analogy with the non-null case discussed in \cite{gravitation,gourgoulhon20123+}, as
\begin{equation}\label{R_d-2_def}
	~^{(D-2)}R_{abcd}X^b=D_c D_d X_a - D_d D_c X_a~.
\end{equation}
Note that what is defined above is the $D$-dimensional object. The components corresponding to the directions on the codimension-$2$ surfaces give the intrinsic curvature of those surfaces. See \cite{Poisson} for a formulation where these components are separated out and written using an appropriate set of basis vectors.
\vspace*{1cm}\\
We refer the reader to Appendix A.3 of the arxiv version of \cite{Parattu:2015gga} for more on the geometrical quantities and identities related to a null surface. Let us now carry out the derivation of the various identities relating the curvature tensor in $D$ spacetime dimensions to the curvature tensor on the codimension-2 surfaces that lie on the null surface and are orthogonal to $k^a$.

Before starting on the derivations of the projections necessary for our calculation, let us first compare the null case with the case of projections near a non-null hypersurface. For a non-null hypersuface, there are three relations--- the Gauss, the Codazzi-Mainardi and the Ricci relations---corresponding to none, one and two indices being projected on the normal, respectively, and the rest being projected on the hypersurface \cite{Baumgarte:2010ndz,gourgoulhon20123+,gravitation}. In the null case, we are dealing with the projector $q^a_b$, which is, in actuality, a projector on codimension-$2$ surfaces, the ones orthogonal to $k^a$ on the null surface, masquerading as a projector on the codimension-$1$ null surface. What this entails is the involvement of two normals, $\ell^a$ and $k^a$, rather than one. Let us enumerate the possibilities of projections with $\ell^a$, $k^a$ and $q^a_b$. The analogue of the Gauss relation would be the contraction of all indices of the Riemann tensor with $q^a_b$'s,
 just having to remember that we are projecting on the codimension-$2$ surfaces. (If we were just dealing with the codimension-$1$ null surface, even a projection on $\ell^a$ would be tangential to the null surface.) For the rest of the relations to follow, we shall assume that the remaining indices are contracted with $q^a_b$'s. The Codazzi-Mainardi relation is now two in number, one for contracting a single index with $\ell^a$ and another for contracting it with $k^a$. Moving on to the Ricci relation, the options of contracting two indices with the normals are four. One could contract two of the indices that are not antisymmetric among themselves with two $\ell^a$'s or two $k^a$'s, or one could take one $\ell^a$ and one $k^a$ and contract it either with two indices that are either antisymmetric among themselves or are not. In this case, we have the options of choosing three or four indices to be contracted with the normals. For the case of three, choose the case of two $\ell^a$'s or two $k^a$'s from the 
Ricci case and contract one of the two remaining indices (doesn't matter which one) with the other normal. Thus, there are two cases for three contractions with the normal. Finally, the completely normal case of contraction with four normals would require two $\ell^a$'s and two $k^a$'s, with the two of a pair occupying indices that are not antisymmetric. All such cases are the same, giving just one such completely normal case of contraction with four normals.

Thus, there are ten contractions of the Riemann tensor possible with the objects $\ell^a$, $k^a$ and $q^a_b$. Out of these ten, we are only going to consider the three obtained without using $k^a$, since these are the ones we will require for our calculations of the boundary term. 
\subsection{Various projections of the Riemann tensor}
\label{Projection_Riem_App}

\subsubsection{Identity involving $q^p_a q^q_b q^r_c q^s_d R_{pqrs}$: the null Gauss relation}

The first identity corresponds to the complete projection of the curvature tensor on the null hypersurface using the induced metric associated with the co-dimension two surfaces. For this purpose, we start with \ref{R_d-2_def} and expand according to \ref{toy_examp} to obtain
\begin{align}
	~^{(D-2)}R_{abcd}X^b&=\left\{q^m_c q^k_d q^l_a \nabla_m \left(q^i_k q^j_l \nabla_i X_j\right)\right\} - \left\{c \leftrightarrow d\right\} \nn 
	\\
	&=\left\{q^m_c q^i_d q^j_a \nabla_m \nabla_i X_j+q^m_c q^k_d q^j_a \nabla_i X_j \nabla_m q^i_k+q^m_c q^i_d q^l_a \nabla_i X_j\nabla_m  q^j_l \right\} - \left\{c \leftrightarrow d\right\}~,
	\label{qR-1}
\end{align}
where $\left\{c \leftrightarrow d\right\}$ represents the expression obtained by interchanging $c$ and $d$ in the expression inside the first set of curly brackets. The second term in the above expression may be rewritten as follows:
\begin{align}
	q^m_c q^k_d q^j_a \nabla_i X_j \nabla_m q^i_k&=q^m_c q^k_d q^j_a \nabla_i X_j \nabla_m \left(\ell^ik_k+k^i \ell_k\right) 
	\nn 
	\\
	&=q^m_c q^k_d q^j_a \nabla_i X_j  \left(\ell^i \nabla_m k_k+k^i \nabla_m \ell_k\right)
	\nn \\
	&=q^j_a \left[\Psi_{cd}   \left(\bm{\ell}.\nabla\right)X_j + \Theta_{cd} \left(\mathbf{k}.\nabla\right)X_j\right], \label{temp2}
\end{align}
where we have used \ref{theta_def} and \ref{psi_def}.
Proceeding in an identical manner, the third term in \ref{qR-1} may be rewritten as
\begin{align}
	q^m_c q^i_d q^l_a \nabla_i X_j\nabla_m  q^j_l
	&=q^m_c q^i_d q^l_a \nabla_i X_j \left(\ell^j\nabla_m k_l+k^j\nabla_m \ell_l\right)
	\nn
	\\
	&=q^i_d \left[\Psi_{ca}   \ell^j\nabla_iX_j + \Theta_{ca} k^j\nabla_iX_j\right]~.
	\label{temp3}
\end{align}
Substituting \ref{temp2} and \ref{temp3} in \ref{qR-1}, we obtain
\begin{align}
	~^{(D-2)}R_{abcd}X^b=&\left\{q^m_c q^i_d q^j_a \nabla_m \nabla_i X_j+q^j_a \left[\Psi_{cd}   \left(\bm{\ell}.\nabla\right)X_j + \Theta_{cd} \left(\mathbf{k}.\nabla\right)X_j\right]  +q^i_d \left[\Psi_{ca}   \ell^j\nabla_iX_j + \Theta_{ca} k^j\nabla_iX_j\right] \right\}\nn 
	\\
	& - \left\{c \leftrightarrow d\right\} 
	\nn 
	\\
	=& q^m_c q^i_d q^j_a \left(\nabla_m \nabla_i-\nabla_i \nabla_m\right) X_j 
	\nn 
	\\
	& +q^j_a \left[\left(\Psi_{cd}-\Psi_{dc}\right) \left(\bm{\ell}.\nabla\right)X_j + \left(\Theta_{cd}-\Theta_{dc}\right) \left(\mathbf{k}.\nabla\right)X_j\right] 
	\nn 
	\\
	& +\left[q^i_d \left(\Psi_{ca}   \ell^j\nabla_iX_j + \Theta_{ca} k^j\nabla_iX_j\right)\right]- \left[c \leftrightarrow d\right]
	\nn 
	\\
	=& q^m_c q^i_d q^j_a R_{jnmi}X^n 
	\nn 
	\\
	& +\left[q^i_d \left(-\Psi_{ca}   X_j\nabla_i \ell^j -\Theta_{ca} X_j\nabla_i k^j\right) \right]- \left[c \leftrightarrow d\right]
	~,
\end{align}
where, as in \ref{appsubsec:ind_met}, we have used the symmetry of $\Theta_{ab}$ and $\Psi_{ab}$. Next we would like to remove the arbitrary vector $X^a$ from both sides. Remember that $X^a$ here is a vector that is orthogonal to $\ell^a$ and $k^a$, i.e., a vector tangential to the codimension-$2$ surfaces on the null surface orthogonal to $k^a$. The above identity is not valid if $X^a$ is replaced by a general vector in the $D$-dimensional spacetime. Thus, we are not allowed to just cancel $X^a$ from both sides. But, given any vector $Y^a$ in the $D$-dimensional spacetime, we can construct a vector $X^a$ orthogonal to $\ell^a$ and $k^a$ as $X^a=q^a_b Y^b$. Thus, the above identity can be written in terms of an arbitrary $D$-dimensional vector $Y^a$ as 
\begin{align}
	~^{(D-2)}R_{abcd}q^b_e Y^e =& q^m_c q^i_d q^j_a q^n_e R_{jnmi} Y^e +
	\left[q^i_d \left(-\Psi_{ca}  \nabla_i \ell^j - \Theta_{ca} \nabla_i k^j\right)q_{je}Y^e \right]- \left[c \leftrightarrow d\right] \nn\\
	=& q^m_c q^i_d q^j_a q^n_e R_{jnmi} Y^e +\left[ \left(-\Psi_{ca} \Theta_{de}  - \Theta_{ca}\Psi_{de}\right) Y^e\right]- \left[c \leftrightarrow d\right]
	~. 
	\label{temp4}
\end{align}
Dropping the arbitrary vector $Y^e$, we obtain
\begin{align}
	~^{(D-2)}R_{abcd}q^b_e = q^m_c q^i_d q^j_a q^n_e R_{jnmi} -\left[ \left(\Psi_{ca} \Theta_{de}+\Theta_{ca}\Psi_{de}\right)- \left(c \leftrightarrow d\right)\right]
	~. 
	\label{temp4-1}
\end{align}
Now, we make the assumption that $~^{(D-2)}R_{abcd}$ is orthogonal to $\ell^a$ and $k^a$ on the index $b$. (It is already orthogonal to these vectors on the other indices, as is obvious from the right-hand side of the above equation.) Looking back at \ref{R_d-2_def}, we see that this equation only tells us the behaviour of $~^{(D-2)}R_{abcd}$ when the index $b$ is contracted with vectors orthogonal to $\ell^a$ and $k^a$. It does not tell us anything about the contraction of this index with $\ell^a$ or $k^a$. So we make the simplest assumption, not distinguishing $b$ in this respect from the other indices. With this assumption, only the $\delta ^{b}_{e}$ part of $q^b_e$ in \ref{temp4-1} survives and we obtain the null version of the \textit{Gauss relation} \cite{gourgoulhon20123+}:
\begin{align}
	^{(D-2)}R_{abcd}  =&q^j_a q^n_b q^m_c q^i_d  R_{jnmi} -\left[ \left(\Psi_{ca} \Theta_{db}+\Theta_{ca}\Psi_{db}\right)- \left(c \leftrightarrow d\right)\right]~, \label{Gauss-Codazzi-0}
\end{align}
or
\begin{align}
	q^j_a q^n_b q^m_c q^i_d  R_{jnmi}  =& ^{(D-2)}R_{abcd}+\left[ \left(\Psi_{ca} \Theta_{db}+\Theta_{ca}\Psi_{db}\right)- \left(c \leftrightarrow d\right)\right]~. \label{Gauss-Codazzi-alt}
\end{align}
Using \ref{Gauss-Codazzi-0}, it can be checked that $^{(D-2)}R_{abcd}$ satisfies all the algebraic identities (the identities without any covariant derivatives) of the Riemann tensor.
\subsubsection{Identity involving $\ell^p q^q_b q^r_c q^s_d R_{pqrs}$: the null Codazzi-Mainardi relation}

The next identity we derive will have one of the indices of the Riemann tensor contracted with the null normal $\ell^a$ and the other three projected on the null surface using $q^a_b$. Note that we need to derive only the case of the first index contracted with the null normal, since an instance of the null normal contracted with another index can be converted to this case using the symmetries of the Riemann tensor. We start with 
\begin{align}
	\ell^a q^b_p q^c_q q^d_r R_{abcd}&=q^b_p q^c_q q^d_r\left(\nabla_d \nabla_c \ell_b-\nabla_c \nabla_d \ell_b\right)
	=\left(q^b_p q^c_q q^d_r \nabla_d \nabla_c \ell_b\right)-\left(q \leftrightarrow r \right) 
	\nn 
	\\
	&=\left(q^d_r q^b_m q^m_p q^c_n q^n_q  \nabla_d \nabla_c \ell_b\right)-\left(q \leftrightarrow r \right) 
	\nn 
	\\
	&=\left[q^d_r  q^m_p  q^n_q  \nabla_d \left(q^b_m q^c_n \nabla_c \ell_b\right)-q^d_r  q^m_p  q^n_q  \nabla_d \left(q^b_m q^c_n \right)\nabla_c \ell_b\right]-\left[q \leftrightarrow r \right]
	\nn 
	\\
	&=\left[D_r \Theta_{pq}-q^d_r  q^b_p  q^n_q  \nabla_d \left( q^c_n \right)\nabla_c \ell_b-  q^d_r  q^m_p q^c_q \nabla_d \left(q^b_m  \right)\nabla_c \ell_b\right]-\left[q \leftrightarrow r \right] 
	\nn 
	\\
	&=\left[D_r \Theta_{pq}-q^d_r  q^b_p  q^n_q   \left( k^c \nabla_d\ell_n +\ell^c \nabla_d k_n\right)\nabla_c \ell_b-  q^d_r  q^m_p q^c_q  \left(k^b\nabla_d \ell_m +  \ell^b\nabla_d k_m  \right)\nabla_c \ell_b\right]
	\nn
	\\
	&\phantom{=}-\left[q \leftrightarrow r \right] 
	\nn 
	\\
	&=\left[D_r \Theta_{pq}-q^b_p \left( k^c \Theta_{rq} +\ell^c \Psi_{rq}\right)\nabla_c \ell_b-  q^c_q  \left(k^b\Theta_{rp}+\ell^b\Psi_{rp}  \right)\nabla_c \ell_b\right]-\left[q \leftrightarrow r \right] 
	\nn 
	\\
	&=\left[D_r \Theta_{pq}-  q^c_q  \left(k^b\Theta_{rp}+\ell^b\Psi_{rp}  \right)\nabla_c \ell_b\right]-\left[q \leftrightarrow r \right], 
\end{align}
where, in the second-last step, we have used the fact that $\Theta_{rq}$ and $\Psi_{rq}$ are symmetric. The expression $q^c_q \ell^b \nabla_c \ell_b=q^c_q  \left(\nabla_c \ell^2\right)/2=  \left(D_q \ell^2\right)/2$ is zero at the null surface, since this is a derivative along the codimension-2 surfaces orthogonal to $k^a$ on the null surface and $\ell^2$ is zero all over the null surface. Another way to see this is by substituting $\ell_b=A\nabla_b \phi$ and simplifying:
\begin{align}
	q^c_q \ell^b \nabla_c \ell_b&=q^c_q \ell^b \nabla_c \left(A\nabla_b \phi\right)=q^c_q \ell^b \ell_b\partial_c \ln A + q^c_q \ell^b  A\nabla_c \nabla_b \phi=q^c_q \ell^b  A\nabla_b \nabla_c \phi=q^c_q \ell^b  A\nabla_b \left(\frac{\ell_c}{A}\right)
	\nn 
	\\
	&=q^c_q \ell^b  \nabla_b\ell_c=q^c_q \kappa \ell_c=0~, 
	\label{qldl_zero}
\end{align}
where we have used $\ell^b\ell_b=0$ and the formula 
\begin{align}\label{rel_def_kappa}
\ell^a\nabla_a \ell_b=\kappa \ell_b~,
\end{align}
with $\kappa$ called the non-affinity coefficient (\sap). Thus, we obtain
\begin{align}
	\ell^a q^b_p q^c_q q^d_r R_{abcd}&=\left(D_r \Theta_{pq}-  q^c_q  k^b\Theta_{rp}\nabla_c \ell_b\right)-\left(q \leftrightarrow r \right) 
	\nn
	\\
	\Rightarrow \ell^a q^b_p q^c_q q^d_r R_{abcd}&=D_r \Theta_{pq}-D_q \Theta_{pr}+  \left(q^c_r  \Theta_{qp}-  q^c_q  \Theta_{rp}\right)k^b\nabla_c \ell_b~, \label{R_3q}
\end{align}
which is the null version of the \textit{Codazzi-Mainardi relation} \cite{gourgoulhon20123+}. (The name of Mainardi is often absent in physics literature.)
\subsubsection{Identities involving $q^p_a \ell^q  q^r_b \ell^s R_{pqrs}$: the null Ricci relation}

The last expression we shall derive is of the decomposition $q^a_p \ell^b  q^c_q \ell^d R_{abcd}$. This is the remaining projection of the Riemann tensor using only $q^a_b$ and $\ell^a$, since the symmetries of the Riemann tensor ensure that contracting more than two indices with the null normal will give zero. In our case, though, we also have the vector $k^a$ and other projections can be carried out with $k^a$ in the mix, as we have indicated before. But these are not required for our purpose and so we shall not be deriving them here. 

We start by decomposing $\nabla_a\ell_b$:
\begin{align}
	\nabla_a\ell_b=& \left(q^c_a-k^c \ell_a-\ell^c k_a\right)\left(q^d_b-k^d \ell_b-\ell^d k_b\right)\nabla_c\ell_d 
	\nn 
	\\
	=&q^c_aq^d_b\nabla_c\ell_d-q^c_a k^d \ell_b\nabla_c\ell_d-q^c_a\ell^d k_b\nabla_c\ell_d-k^c \ell_aq^d_b\nabla_c\ell_d+k^c \ell_ak^d \ell_b\nabla_c\ell_d 
	\nn 
	\\
	&+k^c \ell_a\ell^d k_b\nabla_c\ell_d -\ell^c k_aq^d_b\nabla_c\ell_d +\ell^c k_ak^d \ell_b\nabla_c\ell_d +\ell^c k_a\ell^d k_b\nabla_c\ell_d 
	\nn 
	\\
	=&\Theta_{ab}-q^c_a k^d \ell_b\nabla_c\ell_d-k^c \ell_aq^d_b\nabla_c\ell_d+k^c \ell_ak^d \ell_b\nabla_c\ell_d 
	\nn 
	\\
	&+k^c \ell_a\ell^d k_b\nabla_c\ell_d -\kappa k_a \ell_b \nn\\
	=&\Theta_{ab}-q^c_a k^d \ell_b\nabla_c\ell_d-k^c \ell_aq^d_b\nabla_c\ell_d+k^c \ell_ak^d \ell_b\nabla_c\ell_d -\tilde{\kappa}\ell_a k_b -\kappa k_a \ell_b ~, \label{temp5}
\end{align}
where we have used \ref{qldl_zero}, \ref{rel_def_kappa}  and the fact that $\ell^c \ell^d \nabla_c\ell_d=\ell^c \partial_c \left(\ell_d\ell^d/2\right)=0$ since the derivative is along the null surface where $\ell_d\ell^d=0$ throughout. We have also used the definition (\sap)
\begin{equation}\label{def_ktilde}
\tilde{\kappa}\equiv -\frac{k^a}{2} \partial_a\left(\ell^b \ell_b\right)= -k^a \ell^b \nabla_a \ell_b~.
\end{equation}
The object $\tilde{\kappa}$ may be called the ``nullness gradient" since it measures how $\ell^a \ell_a$ varies as one moves away from the null surface. This is related to $\kappa$ by the relation
\begin{equation}\label{rel_kappa_kappa_t-1}
	\kappa = \ell^a \partial_a \ln A + \tilde{\kappa}~.
\end{equation}
Note that when $A$ does not vary in the null direction $\ell^a$, we shall have
\begin{equation}\label{kappa_eq_kappa_t}
	\kappa = \tilde{\kappa}~.
\end{equation}
Regrouping the terms, it is possible to write down \ref{temp5} in a nice form, resulting in
\begin{equation}
	\nabla_a\ell_b=\Theta_{ab}-\left(k^d\nabla_c\ell_d \right)  q^c_a  \ell_b- \left(k^c\nabla_c\ell_d\right) \ell_a q^d_b+ \left(k^c  k^d \nabla_c\ell_d\right) \ell_a \ell_b -\tilde{\kappa}\ell_a k_b -\kappa k_a \ell_b~. \label{dl_Theta}
\end{equation}
This equation has $\nabla_a\ell_b$ decomposed along $q^a_b$, $\ell_a$ and $k_a$. Using \ref{dl_Theta}, the projection $q^a_p \ell^b  q^c_q \ell^d R_{abcd}$ can be decomposed as follows:
\begin{align}
	&R_{abcd}q^a_p \ell^b q^c_q \ell^d =q^a_p  q^c_q \ell^d \left(\nabla_c \nabla_d-\nabla_d \nabla_c \right) \ell_a 
	\nn 
	\\
	&= q^a_p  q^c_q \ell^d\nabla_c \left[\Theta_{da}-\left(k^i\nabla_j\ell_i \right)  q^j_d  \ell_a- \left(k^i\nabla_i\ell_j\right) \ell_d q^j_a+ \left(k^i  k^j \nabla_i\ell_j\right) \ell_d \ell_a -\tilde{\kappa}\ell_d k_a -\kappa k_d \ell_a\right] 
	\nn 
	\\
	&\phantom{=}-q^a_p  q^c_q \ell^d\nabla_d \left[\Theta_{ca}-\left(k^i\nabla_j\ell_i \right)  q^j_c  \ell_a- \left(k^i\nabla_i\ell_j\right) \ell_c q^j_a+ \left(k^i  k^j \nabla_i\ell_j\right) \ell_c \ell_a -\tilde{\kappa}\ell_c k_a -\kappa k_c \ell_a\right]  
	\nn 
	\\
	&= q^a_p  q^c_q \ell^d\nabla_c \left[\Theta_{da}- \left(k^i\nabla_i\ell_j\right) \ell_d q^j_a  -\kappa k_d \ell_a\right] 
	\nn 
	\\
	&\phantom{=}-q^a_p  q^c_q \ell^d\nabla_d \left[\Theta_{ca}-\left(k^i\nabla_j\ell_i \right)  q^j_c  \ell_a- \left(k^i\nabla_i\ell_j\right) \ell_c q^j_a\right]  
	\nn 
	\\
	&=q^a_p  q^c_q \ell^d \left(\nabla_c \Theta_{da}-\nabla_d \Theta_{ca}\right)- \left(k^i\nabla_i\ell_j\right) q^a_p  q^c_q q^j_a \ell^d\nabla_c\ell_d +\kappa  q^a_p  q^c_q \nabla_c \ell_a 
	\nn 
	\\
	&\phantom{=}+ \left(k^i\nabla_j\ell_i \right)  q^j_cq^a_p  q^c_q \ell^d\nabla_d \ell_a+q^a_p  q^c_q \left(k^i\nabla_i\ell_j\right)  q^j_a \ell^d\nabla_d \ell_c 
	\nn 
	\\
	&=q^a_p  q^c_q \ell^d \left(\nabla_c \Theta_{da}-\nabla_d \Theta_{ca}\right) +\kappa  q^a_p  q^c_q \nabla_c \ell_a
	=q^a_p  q^c_q \ell^d \left(\nabla_c \Theta_{da}-\nabla_d \Theta_{ca}\right) +\kappa \Theta_{pq}~,
\end{align}
where, in the third step, some terms have dropped out because when we act the covariant derivative on them and use the Leibniz's rule for the action of a derivative on a product, we see that the resulting terms are zero when contracted with the factors outside. We have also used \ref{qldl_zero} and \ref{rel_def_kappa}. We have thus obtained the following simple relation:
\begin{equation}
	R_{abcd}q^a_p \ell^b q^c_q \ell^d = q^a_p  q^c_q \ell^d \left(\nabla_c \Theta_{da}-\nabla_d \Theta_{ca}\right) +\kappa \Theta_{pq}~. \label{qlqlR-1}
\end{equation}
A little more manipulation leads to another nice form: 
\begin{align}
	R_{abcd}q^a_p \ell^b q^c_q \ell^d &= q^a_p  q^c_q \ell^d \left(\nabla_c \Theta_{da}-\nabla_d \Theta_{ca}\right) +\kappa \Theta_{pq} 
	\nn 
	\\
	&= q^a_p  q^c_q  \nabla_c \left(\ell^d\Theta_{da}\right)- q^c_q \Theta_{dp} \nabla_c \ell^d -q^a_p  q^c_q \ell^d\nabla_d \Theta_{ca} +\kappa \Theta_{pq} 
	\nn 
	\\
	\Rightarrow R_{abcd}q^a_p \ell^b q^c_q \ell^d&=- \Theta_{pd}  \Theta^d_{q} -q^a_p  q^c_q \ell^d\nabla_d \Theta_{ca} +\kappa \Theta_{pq}~. \label{qlqlR-2} 
\end{align}
 It turns out that one can write the above expression in terms of the Lie derivative of $\Theta_{ca}$ along $\ell^{a}$, given in terms of covariant derivatives by the following formula:
\begin{equation}
	\pounds_{\ell} \Theta_{ca} = \ell^b \nabla_{b} \Theta_{ca}+ \Theta_{ba} \nabla_{c} \ell^b + \Theta_{cb} \nabla_{a} \ell^b~.
\end{equation}
Substituting for the covariant derivative of $\Theta_{ca}$ along $\ell^a$ by the corresponding Lie derivative, \ref{qlqlR-2} becomes
\begin{align}
	R_{abcd}q^a_p \ell^b q^c_q \ell^d&=- \Theta_{pd}  \Theta^d_{q} -q^a_p  q^c_q \left(\pounds_{\ell} \Theta_{ca}- \Theta_{ba} \nabla_{c} \ell^b - \Theta_{cb} \nabla_{a} \ell^b\right) +\kappa \Theta_{pq} 
	\nn 
	\\
	&=- \Theta_{pd}  \Theta^d_{q}  -q^a_p  q^c_q\pounds_{\ell} \Theta_{ca}+\Theta_{bp} \Theta^b_{q} +\Theta_{qb} \Theta^b_{p} +\kappa \Theta_{pq} 
	\nn 
	\\
	\Rightarrow R_{abcd}q^a_p \ell^b q^c_q \ell^d&= -q^a_p  q^c_q\pounds_{\ell} \Theta_{ca}+\Theta_{pd}  \Theta^d_{q} +\kappa \Theta_{pq}~. \label{qlqlR-3} 
\end{align}
We would like to use a particular form of the above equation where one index on $\Theta$ is up and the other is down. So we take \ref{qlqlR-3} and manipulate it as follows:
\begin{align}
	R_{abcd}q^a_p \ell^b q^c_q \ell^d &= -q^a_p  q^c_q\pounds_{\ell} \Theta_{ca}+\Theta_{pd}  \Theta^d_{q} +\kappa \Theta_{pq} 
	\nn 
	\\
	&= -q^a_p  q^c_q\pounds_{\ell} \left(g_{ce}\Theta^e_{a}\right)+\Theta_{pd}  \Theta^d_{q} +\kappa \Theta_{pq} 
	\nn 
	\\
	&= -q^a_p  q^c_q\left(\Theta^e_{a}\pounds_{\ell} g_{ce}+ g_{ce}\pounds_{\ell} \Theta^e_{a}\right)+\Theta_{pd}  \Theta^d_{q} +\kappa \Theta_{pq} 
	\nn 
	\\
	&= -q^a_p  q^c_q\left[\Theta^e_{a}\left(\nabla_c \ell_e + \nabla_e \ell_c\right)+ g_{ce}\pounds_{\ell} \Theta^e_{a}\right]+\Theta_{pd}  \Theta^d_{q} +\kappa \Theta_{pq} 
	\nn 
	\\
	&= -q^a_p  q^c_q\left[2\Theta^e_{a}\Theta_{ce}+ g_{ce}\pounds_{\ell} \Theta^e_{a}\right]+\Theta_{pd}  \Theta^d_{q} +\kappa \Theta_{pq} 
	\nn 
	\\
	&= -2\Theta^e_{p}\Theta_{qe}- q^a_pq_{qe}\pounds_{\ell} \Theta^e_{a}+\Theta_{pd}  \Theta^d_{q} +\kappa \Theta_{pq} 
	\nn 
	\\
	\Rightarrow R_{abcd}q^a_p \ell^b q^c_q \ell^d&= - q^a_pq_{qe}\pounds_{\ell} \Theta^e_{a}-\Theta^e_{p}\Theta_{qe} +\kappa \Theta_{pq} ~, \label{qlqlR-5} 
\end{align}
where we have used the symmetry of $\Theta_{ab}$. This is the formula we will be using in our derivation of the boundary term. But for getting this into a form similar to the Ricci relation \cite{Baumgarte:2010ndz,gravitation}, we have to take the $q$'s inside the Lie derivative. We start with the formula for the action of the Lie derivative on $q^a_b$'s:
\begin{align}
	\pounds_{\ell} q^a_b &=\ell^c \nabla_c q^a_b - q^c_b \nabla_c \ell^a+q^a_c \nabla_b \ell^c  
	= \ell^c \nabla_c \left(\df^a_b+\ell^a k_b+k^a \ell_b\right)- q^c_b \nabla_c \ell^a+q^a_c \nabla_b \ell^c\nn\\ 
	&= \ell^c \nabla_c \left(\ell^a k_b+k^a \ell_b\right)- q^c_b \nabla_c \ell^a+q^a_c \nabla_b \ell^c~.  \label{lie_q}
\end{align}
Instead of the previous result, \ref{qlqlR-5}, we start with \ref{qlqlR-3}. Taking the $q$'s inside the Lie derivative in \ref{qlqlR-3}, we obtain
\begin{align}
	R_{abcd}q^a_p \ell^b q^c_q \ell^d&= -\pounds_{\ell} \Theta_{pq} + \left[\left(\Theta_{ca}q^c_q \pounds_{\ell}q^a_p\right)+\left(p\leftrightarrow q\right) \right]+\Theta_{pd}  \Theta^d_{q} +\kappa \Theta_{pq} \nn \\
	&= -\pounds_{\ell} \Theta_{pq}+\Theta_{pd}  \Theta^d_{q} +\kappa \Theta_{pq} + \left\{\Theta_{qa} \left[\ell^c \nabla_c \left(\ell^a k_p+k^a \ell_p\right)- q^c_p \nabla_c \ell^a+q^a_c \nabla_p \ell^c\right]\right\}+\left\{p \leftrightarrow q\right\}  \nn \\
	&= -\pounds_{\ell} \Theta_{pq}+\Theta_{pd}  \Theta^d_{q} +\kappa \Theta_{pq} +  \left\{\left[\Theta_{qa} \left(k_p\ell^c \nabla_c\ell^a+\ell_p\ell^c \nabla_ck^a\right)-\Theta_{qa} \Theta^a_{p}+ \Theta_{qc} \nabla_p \ell^c\right]\right\}\nn\\
	&\phantom{=}+\left\{p \leftrightarrow q\right\}  \nn \\
	&= -\pounds_{\ell} \Theta_{pq}+\kappa \Theta_{pq}-\Theta_{qa} \Theta^a_{p} +  \left\{\left[\Theta_{qa} \left(k_p\kappa \ell^a+\ell_p\ell^c \nabla_ck^a\right)+ \Theta_{qc}\left(q^a_p-\ell^a k_p-k^a \ell_p\right) \nabla_a \ell^c\right]\right\}\nn\\
	&\phantom{=}+\left\{p \leftrightarrow q\right\} \nn \\
	&= -\pounds_{\ell} \Theta_{pq}+\kappa \Theta_{pq}-\Theta_{qa} \Theta^a_{p} + \left\{\Theta_{qa} \ell_p\ell^c \nabla_ck^a+ \Theta_{qc}\Theta^c_{p} -\Theta_{qc} \ell_p k^a\nabla_a \ell^c\right\}+\left\{p \leftrightarrow q\right\}  \nn \\
	\Rightarrow R_{abcd}q^a_p \ell^b q^c_q \ell^d&= -\pounds_{\ell} \Theta_{pq}+\kappa \Theta_{pq}+\Theta_{qc}\Theta^c_{p} + \left(\Theta_{pa} \ell_q+\Theta_{qa} \ell_p\right)\left(\ell^c \nabla_ck^a- k^c\nabla_c \ell^a\right)   ~,
	\label{qlqlR-3-1} 
\end{align}
where we have used the symmetry of $\Theta_{ab}$ and \ref{rel_def_kappa}. Recognizing that $\ell^b \nabla_{b}k^a-k^b \nabla_{b}\ell^a=\pounds_{\ell}k^a$ (the Lie bracket $\left[\bm{\ell},\bm{k}\right]$), we can write the above equation as
\begin{align}
	R_{abcd}q^a_p \ell^b q^c_q \ell^d= -\pounds_{\ell} \Theta_{pq}+ \left(\Theta_{pa} \ell_q+\Theta_{qa} \ell_p\right)\pounds_{\ell}k^a+\kappa \Theta_{pq}+\Theta_{qc}\Theta^c_{p} ~. \label{qlqlR-3-2} 
\end{align}
\section{Why $\ell_a=\nabla_a \phi'$ is not the general case}\label{app:need-of-A}
If we have $\ell_a=A\nabla_a \phi$, is it possible to choose a different scalar field $\phi'$ to write $\ell_a=\nabla_a \phi'$? Here we demonstrate that this cannot be done in all cases. Of course, one could choose $A$ to be one to start with. But we wish to keep our results general and have hence worked with a general $A$.

Consider $\ell_a=A \nabla_a \phi$. Let us assume that this can be written as $\ell_a=\nabla_a \phi'$. Then we would have
\begin{equation}
\nabla_a \ell_b-\nabla_b \ell_a = \nabla_a \nabla_b \phi'-\nabla_b \nabla_a \phi'=0~.
\end{equation}  
Therefore, the tensor equation $\nabla_a \ell_b-\nabla_b \ell_a=0$ should be satisfied. But writing this using the scalar field $\phi$, we have
\begin{equation}
\nabla_a \ell_b-\nabla_b \ell_a = \nabla_a A \nabla_b \phi-\nabla_b A \nabla_a \phi =\partial_a A \partial_b \phi-\partial_b A \partial_a \phi=0~,
\end{equation} 
where the covariant derivatives are just partial derivatives since $A$ and $\phi$ are scalars. It is easy to prove that this equation cannot be satisfied for all choices of $A$ and $\phi$. For example, suppose we consider a four-dimensional spacetime with coordinates $\left(t,x,y,z\right)$. Take $\phi=t$. The $tx$-component in the above equation is then
\begin{equation}
\partial_t A \partial_x t-\partial_x A \partial_t t=0
\Rightarrow \partial_x A =0~.
\end{equation}
Similarly, we can prove that $\partial_y A =0$ and $\partial_z A =0$. Thus, in this case, we can write $\ell_a=\nabla_a \phi'$ only in the case where $A=A(t)$. If $A$ depends on any of the other coordinates, this will not be possible.
\section{Directed surface element on the null surface}\label{app:null_surf_elem}
In this appendix, we shall derive the expression for the directed surface element on a null surface. We shall follow the treatment in an appendix of the arxiv version of \cite{Parattu:2015gga}, which itself was inspired from \cite{Poisson}, and adapt it to our framework.

Let $y^\alpha=(y^1,y^2,\dots,y^{D-1})$ be any set of coordinates on a codimension-$1$ null boundary, with associated basis vectors $e^{a}_{\alpha}$. Then, the invariant directed surface element for the surface is given by \cite{Poisson}
\begin{eqnarray}
	d\Sigma_{a}= \epsilon_{a a_1 a_2\dots a_{D-1}}e^{a_1}_{1}e^{a_2}_{2}\dots e^{a_{D-1}}_{D-1} dy^{1}dy^{2}\dots dy^{D-1}~.
\end{eqnarray}
The contraction of $d\Sigma_{a}$ with any vector on the boundary surface, expressible as a linear combination of $e^{a}_{\alpha}$, is zero because of the antisymmetry of $\epsilon_{a a_1 a_2\dots a_{D-1}}$. Thus, if $\ell_a$ is a null normal to the surface, $d\Sigma_{a}$ must be proportional to $\ell_a$ and we may write 
\begin{equation} \label{surf_cov}
	\epsilon_{a a_1 a_2\dots a_{D-1}}e^{a_1}_{1}e^{a_2}_{2}\dots e^{a_{D-1}}_{D-1}=f \ell_{a}~,
\end{equation} 
for some scalar function $f$. Contracting with the auxiliary null vector $k^{a}$, we obtain
\begin{equation}\label{f}
	f=-\epsilon_{a a_1 a_2\dots a_{D-1}}k^a e^{a_1}_{1}e^{a_2}_{2}\dots e^{a_{D-1}}_{D-1}~.
\end{equation}
Thus, the directed surface element can be written as
\begin{eqnarray}\label{invar_surf_elem}
	d\Sigma_{a}= -\ell_{a}\epsilon_{b a_1 a_2\dots a_{D-1}}k^b e^{a_1}_{1}e^{a_2}_{2}\dots e^{a_{D-1}}_{D-1} dy^{1}dy^{2}\dots dy^{D-1}~. 
\end{eqnarray}
This is the unique directed surface element on the null surface. There are two ambiguities that one can think of: one being the change of the factor $A$ in the null normal $\ell_a=A \nabla_a \phi$, the second being the choice of the direction of $k^a$ (since there are many null directions to choose from---of all the null directions at a point on the null surface, only the directions along and opposite to $\ell^a$ are out of limits for a choice of $k^a$). It is easy to see that the choice of $A$ does not matter in the above surface element. A scaling of $\ell_a$ by changing $A$ will lead to a scaling of $k^a$ by the inverse factor to preserve the relation $\ell_ak^a=-1$. Going on to the second ambiguity, note first that $(k^a,e^{a_1}_{1},e^{a_2}_{2},\dots,e^{a_{D-1}}_{D-1})$ forms a basis in the $D$-dimensional spacetime. Any new choice of $k^a$, say $k'^a$, can be expanded in this basis as $k'^a=k^a + \mu_{\alpha} e^{a}_{\alpha}$ for some coefficients $\mu_{\alpha}$. (The coefficient of $k^a$ is fixed as one here to ensure $k'^a \ell_a=-1$. We are considering a fixed $\ell_a$ here, since we have already shown that the scalings, the only freedom in choosing $\ell_a$, do not matter.) But we have
\begin{align}
\epsilon_{a a_1 a_2\dots a_{D-1}}k'^a e^{a_1}_{1}e^{a_2}_{2}\dots e^{a_{D-1}}_{D-1}=\epsilon_{a a_1 a_2\dots a_{D-1}}k^a e^{a_1}_{1}e^{a_2}_{2}\dots e^{a_{D-1}}_{D-1}~,
\end{align}
since the antisymmetry of $\epsilon_{a a_1 a_2\dots a_{D-1}}$ kills the $\mu_{\alpha} e^{a}_{\alpha}$ part. Thus, we have proved that both the ambiguities we have mentioned do not matter, so that \ref{invar_surf_elem} is the invariant directed surface element on the null surface, independent of our choices for $\ell_a$ and $k^a$.

The invariant surface element in \ref{invar_surf_elem} is valid for any choice of coordinates $(y^1,y^2,\dots,y^{D-1})$ on the codimension-$1$ null boundary. We now consider choices of coordinates for the $D$-dimensional spacetime. Consider an arbitrary set of $D$ coordinates $\left(x^1,x^2,\dots,x^{D}\right)$. Then, denoting $k^{x^{1}}=k^1,k^{x^{2}}=k^2,$ etc., we have the scalar density $f$ defined in \ref{f} in this coordinate system as 
\begin{align}
f&=\sqg\begin{bmatrix}
k^1 &k^2 &k^3 &\dots &k^{D}\\
e_1^1 &e_1^2 &e_1^3 &\dots &e_1^{D}\\
e_2^1 &e_2^2 &e_2^3 &\dots &e_2^{D}\\
\vdots       &\vdots       &\vdots    &\ddots &\vdots\\
e_{D-1}^1 &e_{D-1}^2 &e_{D-1}^3 &\dots &e_{D-1}^{D}\\
\end{bmatrix} ~.
\end{align}
A convenient choice is the set of coordinates $(\phi,y^1,y^2,\dots,y^{D-1})$, with the null boundary being specified as a constant value of $\phi$. Then the null normal is $\ell_a=A \nabla_a \phi$ for some $A$. Also, we can write $e^a_{\alpha}=\partial x^a/\partial y^\alpha$. (The partial derivative is taken keeping the other coordinates constant. In particular, $\phi$ being constant makes sure that $\partial x^a/\partial y^\alpha$ at the boundary surface is tangential.) In this coordinate system,
\begin{align}
f&=\sqg\begin{bmatrix}
k^{\phi} &k^{y^1} &k^{y^2} &\dots &k^{y^{D-1}}\\
0 &1 &0 &\dots &0\\
0 &0 &1 &\dots &0\\
\vdots       &\vdots       &\vdots    &\ddots &\vdots\\
0 &0 &0 &\dots &1\\
\end{bmatrix} \nn\\
&=\sqg k^{\phi}=\sqg k^a \nabla_a \phi~.
\end{align}
Thus, in any coordinate system $(\phi,y^1,y^2,\dots,y^{D-1})$ with $\phi$ being constant on the null boundary and $(y^1,y^2,\dots,y^{D-1})$ being arbitrary, we have the surface element in \ref{invar_surf_elem} as
\begin{eqnarray}\label{invar_surf_elem-sp}
d\Sigma_{a}= -\sqg\ell_{a}\left(k^b \nabla_b \phi\right)dy^{1}dy^{2}\dots dy^{D-1}~. 
\end{eqnarray}
One could use the condition $\ell_a k^a=-1$ to write
\begin{align}
k^b \nabla_b \phi=\frac{k^b \ell_b}{A}=-\frac{1}{A}~,
\end{align}
so that the surface element becomes 
\begin{eqnarray}\label{invar_surf_elem-sp-1}
d\Sigma_{a}= \frac{\sqg}{A}\ell_{a}dy^{1}dy^{2}\dots dy^{D-1}~. 
\end{eqnarray}
This is a form that we had used in our previous paper \cite{Parattu:2015gga} as well as in the derivation in this paper. But perhaps \ref{invar_surf_elem-sp} is a more convenient form since it has the $\ell_a$, $k^a$ and the coordinate $\phi$, while \ref{invar_surf_elem-sp-1} has the secondary scalar $A$ that we obtain by expressing $\ell_a$ as $A\nabla_a \phi$.

In the next section, we discuss the case where $k_a$ is chosen as $k_a=B \nabla_a \Psi$ and the scalar $\Psi$ is chosen as one of the coordinates on the null boundary. This allows us to decompose the determinant of the metric $g$ in terms of the determinant $q$ of the metric on the constant surfaces of $\Psi$ on the null boundary.

Let us briefly digress to note that the derivation till here has not used $\ell_a \ell^a=0$ or $k_a k^a=0$, but only $\ell_a k^a =-1$. Thus, the derivation also works for a non-null surface with appropriate choices of $\ell_a$ and $k^a$. For example, a spacelike surface with a timelike normal would have the normalized normal satisfying $n_a n^a=-1$. We can thus choose $\ell_a=k_a=n_a$. Choosing the boundary surface to be a level surface of time $t$, we can write $n_a=-N \nabla_{a} t$, where $N$ is the normalization factor and the minus sign has been added so that $n^a$ will face in the direction of increasing $t$. Then, in a coordinate system $\left(t,y^1,y^2,\dots,y^{D-1}\right)$, the surface element in the form in \ref{invar_surf_elem-sp} would be
\begin{align}\label{invar_surf_elem-sp-nn}
d\Sigma_{a}&= -\sqg n_{a}\left(n^b \nabla_b t\right)dy^{1}dy^{2}\dots dy^{D-1} \nn \\
&= \sqg \frac{n_{a}}{N}dy^{1}dy^{2}\dots dy^{D-1} \nn \\
&= \sqrt{h} n_{a}dy^{1}dy^{2}\dots dy^{D-1} ~,
\end{align}
which is of the familiar form for a spacelike surface (see Appendix B in the arxiv version of \cite{Parattu:2015gga} for a derivation of $\sqg=N\sqrt{h}$). 

A popular convention is to have the normal vector to be future-directed for spacelike and null surfaces. For the null case, we have provided a comparison of this convention with our convention in \ref{app:surf_elem_conven}.
\section{Directed null surface element with $\phi$ and $\psi$ as coordinates}\label{app:g_in_q}
In this appendix, we shall derive the form of the directed null surface element in the case where the scalars $\phi$ and $\psi$, appearing in $\ell_a=A\nabla_a \phi$ and $k_a=B\nabla_a \phi$, are taken as coordinates. (We have further assumed $B=1$ in deriving the null boundary term, but let us work with a general $B$ for the time being.) For this, we need to derive a formula for the determinant of the metric $g$ in terms of the determinant of the metric on the codimension-$2$ surface, analogous to the formula $g=-N^2 h$ (\cite{gravitation}, see Appendix B in the arxiv version of \cite{Parattu:2015gga} for a derivation) connecting the determinant of the metric to the determinant of the induced metric on a spacelike surface. In the last part of \ref{app:null_surf_elem}, we had specialized to a coordinate system $(\phi,y^1,y^2,\dots,y^{D-1})$. Let us choose $\psi=y^1$. Then the codimension-$2$ surfaces on the null boundary given by constant values of $\psi$ are the surfaces whose induced metric we have represented as $q_{ab}$. Let the rest of the $D-2$ coordinates be $\left(z^1,z^2,\ldots,z^{D-2} \right)$. These are the coordinates that run over our codimension-$2$ surfaces at the intersection of the level surfaces of $\phi$ and $\psi$. While the small Latin letters $a,b,c, \ldots$ are used as indices to run over all the spacetime coordinates, we shall use the capital Latin letters $A,B,C,\ldots$ to run over only the coordinates $\left(z^1,z^2,\ldots,z^{D-2} \right)$ corresponding to the codimension-$2$ surfaces. Let $e_I=\partial/\partial z^I$, with $I\in \left\{1,2,\dots D-2 \right\}$ correspond to the coordinate vectors on the codimension-$2$ surfaces. We shall write
\begin{align}
q_{AB}\equiv g_{ab}\frac{\partial x^a}{\partial z^A} \frac{\partial x^b}{\partial z^B}
\end{align}
for the components of the metric on the codimension-$2$ surfaces. Then, the metric in such a coordinate system is given by
\begin{align}
g_{ab}=\begin{bmatrix}
        g_{\phi \phi}&g_{\phi \psi}&g_{\phi 1}&\dots &g_{\phi,D-2}\\
        g_{\phi \psi}&g_{\psi \psi}&g_{\psi 1}&\dots &g_{\psi,D-2}\\
        g_{\phi 1}   &g_{\psi 1}   &q_{11}    &\dots &q_{1,D-2}\\
        \vdots       &\vdots       &\vdots    &\ddots &\vdots\\
        g_{\phi,D-2} &g_{\psi,D-2}    &q_{1,D-2} &\dots &q_{D-2,D-2}
        \end{bmatrix}~,
\end{align}
and the inverse metric is
\begin{align}
g^{ab}=\begin{bmatrix}
g^{\phi \phi}&g^{\phi \psi}&g^{\phi 1}&\dots &g^{\phi,D-2}\\
g^{\phi \psi}&g^{\psi \psi}&g^{\psi 1}&\dots &g^{\psi,D-2}\\
g^{\phi 1}   &g^{\psi 1}   &q^{11}    &\dots &q^{1,D-2}\\
\vdots       &\vdots       &\vdots    &\ddots &\vdots\\
g^{\phi,D-2} &g^{\psi,D-2}    &q^{1,D-2} &\dots &q^{D-2,D-2}
\end{bmatrix}~.
\end{align}
From Appendix A.2 in the arxiv version of \cite{Parattu:2015gga}, we have the result that (originally proved there for four dimensions, but the proof can be easily seen to be valid for $D$ dimensions)
\begin{align}\label{g_in_q_1}
g=\frac{q}{g^{\phi \phi}g^{\psi \psi}-\left(g^{\phi \psi}\right)^2}~.
\end{align}
At the null boundary, the constraints $g^{ab}\ell_a \ell_b=0$, $g^{ab}k_a k_b=0$ and $g^{ab}\ell_a k_b=-1$ mean the following conditions on the inverse metric, respectively:
\begin{align}
g^{\phi \phi}=0,\quad g^{\psi \psi}=0,\quad AB g^{\phi \psi}=-1~.  
\end{align} 
Substituting this in \ref{g_in_q_1}, we obtain
\begin{align}\label{g_in_q}
g=-A^2B^2q~.
\end{align}
Taking the square root,
\begin{align}\label{sqg_in_q}
\sqg=AB\sqq~.
\end{align}
which, under our assumption of $B=1$, becomes
\begin{align}\label{sqg_in_q-B1}
\sqg=A\sqq~.
\end{align}
Substituting this result in the surface element in \ref{invar_surf_elem-sp-1}, we obtain
\begin{eqnarray}\label{dse-B_1}
d \Sigma_a = \sqq \ell_{a} d\psi dz^{1}\dots dz^{D-2}\equiv \sqq \ell_{a} d\psi  d^{D-2}z~,
\end{eqnarray} 
where we have set $(y^1,y^2,\dots,y^{D-1})=(\psi,z^1,\dots,z^{D-2})$.
\section{Translating our results to the future-directed normal convention} \label{app:surf_elem_conven}

In this appendix, we clarify how our convention of choosing the null normal compares with the popular convention of choosing the normal vector to be future-directed (see, e.g., \cite{Lehner:2016vdi}) and specify how our results can be translated to the other convention.

Our conventions for the normal follow those in our earlier paper \cite{Parattu:2015gga}. The expression obtained after converting the total divergence on the boundary to a surface integral is given in \ref{ST_1} as
\begin{equation}
ST = \int_{\mathcal{\partial V}} d^{D-1}x ~2 \left(\frac{\sqrt{-g}}{A}\right) \ell_j P^{ibjd}\nabla_b \left(\delta g_{di}\right)~,\label{ST-11}
\end{equation} 
where $\ell_a$ is defined as $\ell_a\equiv A \nabla_a \phi$. But the factor $A$ cancels off and we are left with the object $\nabla_a \phi$. This is what we obtain if the boundary surface $\phi= \textrm{constant}$ we are considering is such that $\phi$ increases as we move through it from inside the integration volume to the exterior. Thus, the normal is a priori neither future-directed nor past-directed in the conventions of our paper. For any specific example, the nature of the normal would depend on how the scalar $\phi$ is defined. 

To make connection with the results in \cite{Lehner:2016vdi}, we first make the additional assumption that $\phi$ increases towards the future. Combining this with our assumption that $\phi$ increases as we move from inside the integration volume to the exterior through the boundary surface, moving out through the boundary surface will involve moving forward in time. Since we are working in the $(-,+,+,+)$ convention, the vector corresponding to the normal $\nabla_a \phi$ would be past-directed (and inward-directed) at this surface. We can make it future-directed (and outward-directed) by adding an extra minus sign to the definition of the normal vector, as has been done in \cite{Lehner:2016vdi}. Then, the boundary integral becomes
\begin{equation}
ST = -\int_{\mathcal{\partial V}} d^{D-1}x ~2 \left(\frac{\sqrt{-g}}{A}\right) \ell_j P^{ibjd}\nabla_b \left(\delta g_{di}\right)~,\label{ST-1}
\end{equation} 
where there is an extra minus sign occurring outside because the definition of $\ell_j$ now has an extra minus sign compared to the definition in our paper. If we work in this convention, we will get the same expressions for the boundary terms as in our paper except for overall minus signs. 

In the Einstein-Hilbert case, this is why our boundary term differs from that in \cite{Lehner:2016vdi} by a minus sign.

\section{Deriving the null Raychaudhuri equation from the null projections of the Riemann tensor} \label{app:null_raych}
In this appendix, we shall derive the null Raychaudhuri equation \cite{Poisson} with non-affine parametrisation \cite{Chakraborty:2015hna} using the projections of the Riemann tensor near a null surface derived in \ref{app:R_decomp_null}. We start with
\begin{align}
R^{a}_{c}\ell _{a}\ell ^{c} =R_{abcd}\ell^a \ell^c g^{bd}=R_{abcd}\ell^a \ell^c \left(q^{bd}-k^b\ell^d -\ell^b k^d \right)= R_{abcd}\ell^a \ell^c q^{bd}~. \label{Rll}
\end{align}
We take \ref{qlqlR-5},
\begin{align}
R_{abcd}q^a_p \ell^b q^c_q \ell^d= - q^a_pq_{qe}\pounds_{\ell} \Theta^e_{a}-\Theta^e_{p}\Theta_{qe} +\kappa \Theta_{pq}~,
\end{align}
and contract with $g^{pq}$ to obtain
\begin{align}
R_{abcd}q^{ac} \ell^b  \ell^d&= - q^a_{e}\pounds_{\ell} \Theta^e_{a}-\Theta^e_{p}\Theta^p_{e} +\kappa \Theta \nn\\
&= - \pounds_{\ell}\left(q^a_{e} \Theta^e_{a}\right)+\Theta^e_{a}\pounds_{\ell}q^a_{e} -\Theta^e_{p}\Theta^p_{e} +\kappa \Theta \nn\\
&= - \pounds_{\ell}\left(\Theta\right)+\Theta^e_{a}\pounds_{\ell}q^a_{e} -\Theta^e_{p}\Theta^p_{e} +\kappa \Theta \nn \\
&= - \ell^a \nabla_a \Theta +\Theta^e_{a}\pounds_{\ell}q^a_{e} -\Theta^e_{p}\Theta^p_{e} +\kappa \Theta \nn \\
&= - \ell^a \nabla_a \Theta +\Theta^e_{a}\left(- q^c_e \nabla_c \ell^a+q^a_c \nabla_e \ell^c\right) -\Theta^e_{p}\Theta^p_{e} +\kappa \Theta \nn \\
&= - \ell^a \nabla_a \Theta -\Theta^c_{a}  \Theta_c^a + \Theta^e_{a} \Theta^a_e -\Theta^e_{p}\Theta^p_{e} +\kappa \Theta \nn \\
&= - \frac{d\Theta}{d \lambda} -\Theta^e_{p}\Theta^p_{e} +\kappa \Theta ~, \label{null_raych_app}
\end{align}
where we have defined the parameter $\lambda$ varying along the vector $\ell^a$ such that $\ell^a \nabla_a \lambda=1$ (or, equivalently, $\ell^a = \partial x^a/ \partial \lambda$) and used the result for the Lie derivative of $q^a_e$ from \ref{lie_q}:
\begin{align}
\pounds_{\ell} q^a_b = \ell^c \nabla_c \left(\ell^a k_b+k^a \ell_b\right)- q^c_b \nabla_c \ell^a+q^a_c \nabla_b \ell^c~. 
\end{align}
From \ref{Rll} and \ref{null_raych_app}, we get the null Raychaudhuri equation:
\begin{align}
\frac{d\Theta}{d \lambda}= -R^{a}_{c}\ell _{a}\ell ^{c} -\Theta^a_{b}\Theta_a^{b} +\kappa \Theta ~. \label{null_raych_app_std}
\end{align}

\bibliography{References}

\bibliographystyle{./utphys1}

\end{document}